%% file: nn_su3_slaph_firstlook.tex
\documentclass[aps,prd,twocolumn,tightenlines,preprintnumbers,showpacs,superscriptaddress,notitlepage,nofootinbib,eqsecnum,floatfix,longbibliography,10pt]{revtex4-1}
\input{preamble}
\input{def}

\input{affiliations}

\newcount\hour \newcount\hourminute \newcount\minute
\hour=\time \divide \hour by 60
\hourminute=\hour \multiply \hourminute by 60
\minute=\time \advance \minute by -\hourminute

\begin{document}

\title{Two-nucleon S-wave interactions at the SU(3) flavor-symmetric point with $m_{ud}\approx m_s^{\rm phys}$:\\
a first lattice QCD calculation with the stochastic Laplacian Heaviside method}

\author{Ben H\"{o}rz}
\affiliation{\lblnsd}

\author{Dean~Howarth}
\affiliation{\llnl}
\affiliation{\lblnsd}

\author{Enrico~Rinaldi}
\affiliation{\arithmer}
\affiliation{\ithems}

\author{Andrew~Hanlon}
\affiliation{\mainz}

\author{Chia~Cheng~Chang~(\begin{CJK*}{UTF8}{bsmi}\mbox{張家丞}\end{CJK*})}
\affiliation{\ithems}
\affiliation{\ucb}
\affiliation{\lblnsd}

\author{Christopher K\"{o}rber}
\affiliation{\bochum}
\affiliation{\ucb}
\affiliation{\lblnsd}

\author{Evan~Berkowitz}
\affiliation{\mcfp}

\author{John~Bulava}
\affiliation{\cpthree}

\author{M.A.~Clark}
\affiliation{\nvidia}

\author{Wayne~Tai~Lee}
\affiliation{\columbia}

\author{Colin~Morningstar}
\affiliation{\cmu}

\author{Amy~Nicholson}
\affiliation{\unc}
\affiliation{\lblnsd}

\author{Pavlos~Vranas}
\affiliation{\llnl}
\affiliation{\lblnsd}

\author{Andr\'{e}~Walker-Loud}
\affiliation{\lblnsd}
\affiliation{\ucb}
\affiliation{\llnl}

%\date{\mydate}

\begin{abstract}
We report on the first application of the stochastic Laplacian Heaviside method for computing multiparticle interactions with lattice QCD to the two-nucleon system.
Like the Laplacian Heaviside method, this method allows for the construction of interpolating operators which can be used to construct a set of positive definite two-nucleon correlation functions, unlike nearly all other applications of lattice QCD to two nucleons in the literature.
It also allows for a variational analysis in which optimal linear combinations of the interpolating operators are formed that couple predominantly to the eigenstates of the system.
Utilizing such methods has become of paramount importance to help resolve the discrepancy in the literature on whether two nucleons in either isospin channel form a bound state at pion masses heavier than physical, with the discrepancy persisting even in the SU(3)-flavor-symmetric point with all quark masses near the physical strange quark mass.
This is the first in a series of papers aimed at resolving this discrepancy.
In the present work, we employ the stochastic Laplacian Heaviside method without a hexaquark operator in the basis at a lattice spacing of $a\approx0.086$~fm, lattice volume of $L=48a\approx4.1$~fm and pion mass $m_\pi\approx714$~MeV.
With this setup, the observed spectrum of two-nucleon energy levels strongly disfavors the presence of a bound state in either the deuteron or dineutron channel.
\end{abstract}

\preprint{LLNL-JRNL-813871, RIKEN-iTHEMS-Report-20, MITP/20-055}

\maketitle
%%%%%%%%%%%%%%%%%%%%%%%%%%%%%%%%%
%%%%%%%%%%%%%%%%%%%%%%%%%%%%%%%%%
%%%%%%%%%%%%%%%%%%%%%%%%%%%%%%%%%
%\tableofcontents
%----------------------------------------------------------
%    INTRODUCTION
\section{Introduction \label{sec:intro}}

Quantum Chromodynamics (QCD), the fundamental theory of nuclear strong interactions, encodes the interactions of nearly massless quarks and massless gluons which are confined into protons and neutrons, the nucleons, with a mass of $\mathrm{O}(1)$~GeV.  These nucleons, which form the basis of matter, have a residual strong interaction that leads to the formation of nuclei with binding energies that are typically two orders of magnitude smaller than this confinement scale: in the case of the deuteron, the smallest nucleus made of one proton and one neutron, with a binding energy of $B_d\approx2.2$~MeV, the residual interaction is a part-per-mille.
In the case of the dineutron system, the residual interaction is smaller still, leading to a barely unbound system.

The nonperturbative nature of QCD at these low energies, combined with these disparate energy scales, severely complicates our ability to understand the emergence of nuclear physics directly from the Standard Model (SM) of particle physics.
The only model-independent and systematically improvable method for computing the properties and interactions of nucleons directly from QCD is lattice QCD (LQCD), the Euclidean spacetime formulation of QCD on a finite and discrete grid, or lattice.
For a discussion of the importance, challenges and prospects of connecting our understanding of nuclear physics to the SM through a coupling of LQCD and effective theories, see the recent review articles in Refs.~\cite{Drischler:2019xuo,Tews:2020hgp,Cirigliano:2020yhp}.

The first application of LQCD to the two-nucleon system was in the quenched approximation (infinitely massive $\bar{q}q$ virtual pairs) 25 years ago~\cite{Fukugita:1994ve}.
The next calculation was performed in 2006 using dynamical quarks with pion masses ranging from $350\lesssim m_\pi\lesssim 600$~MeV~\cite{Beane:2006mx}.  Since that time, there has been a measured growth in the application of LQCD to systems with two or more baryons,
with the first calculation of a bound two-baryon system appearing in 2010~\cite{Beane:2010hg,Inoue:2010es}.
However, significant challenges, most notably the exponentially bad signal-to-noise (StoN) ratio~\cite{Lepage:1989hd}, have prevented substantive progress:
some 15 years later, even with all the growth in computing power and algorithmic advances, there are still no computations of two-nucleon systems utilizing the \luscher method~\cite{Luscher:1986pf,Luscher:1990ux} with pion masses lighter than $m_\pi\approx300$~MeV~\cite{Yamazaki:2015asa}.

However, we have seen the emergence of LQCD calculations of light nuclei at $m_\pi\approx800$~MeV (up to $A=4$)~\cite{Beane:2009gs,Beane:2012vq} which have been used to match to a pion-less effective field theory of few-nucleon interactions and used to calibrate and predict nuclei up to $A=6$~\cite{Barnea:2013uqa}.
We have also seen the development of a new method which first constructs a two-nucleon potential, known as the HAL QCD potential~\cite{Ishii:2006ec,Aoki:2009ji,HALQCD:2012aa,Aoki:2012tk,Aoki:2012bb}, from which a Schr{\"o}dinger equation is solved and can then be used to predict the scattering phase shifts.
If the HAL QCD method can be demonstrated to numerically agree with the \luscher method, it offers a promising alternative for computing the interactions of baryons from LQCD.

However, there is controversy in the literature concerning the aforementioned agreement and, in turn, there is a discrepancy on whether or not two nucleons form a bound state at medium and heavy pion masses.
In short, most calculations of two-nucleons that utilize the \luscher method observe the presence of (deeply) bound deuteron and dineutron systems at pion masses larger than $\approx 300$~MeV~\cite{Beane:2012vq,Yamazaki:2012hi,Beane:2013br,Berkowitz:2015eaa,Orginos:2015aya,Wagman:2017tmp}---with the exception of the Mainz group which found a bound dineutron to be unlikely at $m_\pi \approx 960$~MeV~\cite{Francis:2018qch}---while the HAL QCD Collaboration, utilizing their potential method, concludes that there are no bound states in either channel~\cite{HALQCD:2012aa,Inoue:2011ai}.
For a more detailed discussion of the controversy, see the recent review in Ref.~\cite{Drischler:2019xuo}.

Some have found it tempting to think this disagreement is a demonstration that the HAL QCD method has uncontrolled systematic uncertainties. However, while the \luscher method provides a rigorous mapping between the finite-volume energy spectrum and the infinite-volume scattering amplitudes, there are potential unresolved systematic uncertainties in the application of the method, particularly in properly identifying the multiparticle energy spectrum. All applications that observe the existence of bound two-nucleon systems rely upon a local hexaquark creation operator at the source and dilute, two-nucleon momentum-space annihilation operators at the sink.
The HAL QCD Collaboration has shown that the extracted spectrum in many of these cases does not pass basic consistency checks, demonstrating that there are larger systematic uncertainties than have been reported~\cite{Iritani:2017rlk}.
Combined with the StoN challenges and the very small elastic scattering energy gaps, this has led HAL QCD to speculate the calculations which observe bound states have been misled by ``false plateaux'' in the effective masses of the system, which can arise with non-positive-definite correlation functions~\cite{Iritani:2016jie}.\footnote{This has been challenged, but not conclusively demonstrated to be wrong~\cite{Yamazaki:2017euu,Yamazaki:2017jfh,Beane:2017edf}.}

These non-positive definite correlation functions require the assumption that the overlap onto the eigenstates of the system are dominated by a single interpolating field constructed from the projection of each nucleon individually onto a state of definite momentum at the sink side.
Consider the center of mass (CoM) for simplicity, such that
\begin{equation}
\langle NN(q)| \sim \sum_{\mathbf{x},\mathbf{y}}
    c(\mathbf{p}) \langle 0 |  e^{i\mathbf{p}\cdot\mathbf{x}} e^{-i\mathbf{p}\cdot\mathbf{y}}
    N(\mathbf{x}) N(\mathbf{y})\, ,
\end{equation}
where $q$ is the relative interacting momentum, which is determined through the \luscher quantization condition; $\mathbf{p}$ is given by a noninteracting plane-wave momentum mode allowed in the finite periodic volume, $\mathbf{p} = \frac{2\pi}{L}\mathbf{n}$ with $\mathbf{n}$ vector of integers; and $c(\mathbf{p})$ is a weight which can be chosen arbitrarily if a single momentum mode $\mathbf{p}$ dominates the overlap (in practice, the existing calculations have chosen $c(\mathbf{p})=1$).
For weakly interacting systems, such as $I=2\ \pi\pi$ scattering, this type of interpolating fields works reasonably well as demonstrated by the consistency between the results from NPLQCD~\cite{Beane:2011sc} and HadSpec~\cite{Dudek:2010ew} which utilized this simplistic operator and a full variational basis, respectively.
For strongly interacting systems, such as the two-nucleon system, the results in the literature are insufficient to draw a conclusion one way or the other as to how well this simplistic basis of interpolating fields couples cleanly to the spectrum.

In contrast, with a variational basis of interpolating fields, one is not restricted to this assumption and instead utilizes a linear combination of creation and annihilation operators
\begin{equation}
\langle NN(q)| \sim \sum_{\mathbf{p}} \sum_{\mathbf{x},\mathbf{y}}
    c(\mathbf{p}) \langle 0 |  e^{i\mathbf{p}\cdot\mathbf{x}} e^{-i\mathbf{p}\cdot\mathbf{y}}
    N(\mathbf{x}) N(\mathbf{y})\, ,
\end{equation}
where now the $c(\mathbf{p})$ coefficients are determined through a diagonalization of the set of interpolating fields used and constrained by the numerical values of the correlation function.
Even with the variational basis, experience shows that it is still necessary to have a large basis of operators which provide sufficient overlap onto the various states of the system.  For example, in the $I=1\ \pi\pi$ system, one must include operators that look both like local $\rho$ operators ($\bar{q}\g_\mu q$) as well as displaced two-pion operators to obtain a spectrum that is consistent with the expected $\rho$ resonance~\cite{Dudek:2012xn}.  A similar study of the negative parity nucleon found that a nonlocal $N\pi$ operator is required~\cite{Lang:2012db}.
In the case of the two-nucleon system, it could be that the hexaquark operator is important for coupling to a deeply bound state, as speculated in Ref.~\cite{Berkowitz:2015eaa}.\footnote{A recent study showed the use of hexaquark operators, at both the source and sink, gives effective energies above threshold (and even above effective energies utilizing two-baryon interpolators at the sink) in the dineutron channel, suggesting a hexaquark operator may not be necessary to accurately extract the spectrum~\cite{Francis:2018qch}.}

In the present work we take a first step towards trying to resolve this discrepancy by performing the first LQCD calculation of the two-nucleon systems in both isospin channels using a positive-definite correlation matrix with a variational basis of operators.
The first application of a variational basis to two-baryon systems was applied recently by the Mainz group to the H-dibaryon and dineutron systems~\cite{Francis:2018qch,Hanlon:2018yfv} in which significant tension with the local hexaquark results from NPLQCD was observed~\cite{Beane:2010hg,Beane:2012vq}, although some tension with the HAL QCD potential method exists as well~\cite{Inoue:2010es,Inoue:2011ai}.
One possible explanation for this is the use of only two dynamical quark flavors (with a ``quenched'' strange quark) from Mainz, giving rise to potentially large systematic effects in the determination of the binding energy as compared to HAL QCD and NPLQCD. Furthermore, all calculations were performed with a single lattice spacing with different lattice actions.

For the present work, we focus on the two two-nucleon channels with total isospin $I=0$ and $1$, which we refer to as the deuteron and dineutron channel, respectively.
We utilize the stochastic Laplacian Heaviside method~\cite{Morningstar:2011ka}, which is a stochastic variant of the distillation method~\cite{Peardon:2009gh}.  We will summarize the method in \secref{sec:slaph}.
As discussed in \secref{sec:lattice}, our calculation strongly disfavors a bound state in either the deuteron or dineutron channel.
We discuss the implications of this work as well as the limitations and provide an outlook in \secref{sec:discussion}.
In order for the broader community to have confidence in the application of LQCD to nuclear physics, it is of paramount importance to resolve the issue underlying the contradictory results in the literature.

%----------------------------------------------------------
%    sLapH
\section{Stochastic Laplacian Heaviside Method \label{sec:slaph}}

A successful computation of two-nucleon energies relies heavily on the construction of optimal operators. Unfortunately, this leads to several sources of computational difficulty. First, the six valence quarks present in two-nucleon correlation functions give rise to a large number of Wick contractions. Next, to
maximize overlap onto the individual finite-volume two-nucleon states, both nucleon interpolating operators should be projected onto a definite spatial momentum at the source and sink.
Finally, two-nucleon interpolators which transform irreducibly under the finite-volume remnant of rotational symmetry require a summation over two-nucleon momentum combinations which transform among themselves under the little group.

The considerations above necessitate a flexible and efficient treatment of
all-to-all quark propagation in which the quark propagator is determined
between all spatial lattice points. The stochastic Laplacian Heaviside (LapH) method, which is employed here, has been  successful for
two-meson~\cite{Bulava:2016mks,Brett:2018jqw} and meson-baryon~\cite{Andersen:2017una} correlators. This method enables a particular choice of quark smearing in which the quark fields are projected onto the space spanned by the lowest $N_{\rm ev}$ eigenmodes of the gauge-covariant
three-dimensional laplace operator~\cite{Peardon:2009gh}. Stochastic estimators with $N_r$ noise sources are then introduced in this $N_{\rm ev}\times N_{\rm spin}$ dimensional LapH subspace rather than the entire spatial lattice, significantly improving the variance~\cite{Morningstar:2011ka}.

The stochastic estimators are improved by `dilution'~\cite{Foley:2005ac}, in which each stochastic field
is partitioned into $N_{\rm dil}$ fields, each of which has support on a unique
subset of the LapH subspace. For this work we employ $N_{\rm ev} = 384$,
full spin dilution, and 12 LapH eigenvector projectors, so $N_{\rm dil} =
4 \times 12 = 48$.%
\footnote{In practice only the upper two spin components are used for the computation of states propagating forward in time, reducing the effective $N_\mathrm{dil}$ by a factor two in the correlator construction.}
To ensure unbiased estimates of products of quark
propagators, independent stochastic fields are required for each valence quark line, so that estimates for the quark propagators are given by
\begin{gather}\label{e:qprop}
Q_{a\alpha,b\beta}(x,y) = \lim_{N_r \rightarrow \infty} \frac{1}{N_r} \sum_{r,d}\phi^{(r,d)}_{a\alpha}(x) \rho_{b\beta}(y)^{(r,d)*}
\end{gather}
where $(r,d)$ denote the noise and dilution indices respectively, $\rho(y)$ is a stochastic combination of LapH eigenvectors and $\phi = Q\rho$ is the result of a linear system solve, which are solved efficiently in GPU accelerated nodes with QUDA~\cite{Clark:2009wm,Babich:2011np}.

The computation of two-nucleon correlation functions is also simplified with
stochastic LapH. Each two-nucleon interpolator trasnforming irreducibly is given by
\begin{gather}\label{e:ops}
{\cal O}^{II_3\Lambda\lambda}_{\ell}(\boldsymbol{P}) = \sum_{\boldsymbol{p}_1\boldsymbol{p}_2} c_{\ell_1\ell_2}^{II_3\Lambda\lambda} \, N_{\ell_1} \, N_{\ell_2}
\end{gather}
with definite isospin $(I,I_3)$, little group irrep $\Lambda$, irrep row $\lambda$, and total momentum $\boldsymbol{P}$. The additional identifier $\ell$ distinguishes multiple linearly independent operators of this type, while the labels
$\ell_{1,2}$ for the single nucleon operators denote the individual $I_3$ and
momenta $\boldsymbol{p}_{1,2}$. Note that all terms have $\boldsymbol{p}_1 + \boldsymbol{p}_2 = \boldsymbol{P}$, with the $\boldsymbol{p}_{1,2}$ in different terms related via little group transformations. The coefficients $c_{\ell_1\ell_2}^{II_3\Lambda\lambda}$ are determined according to Ref.~\cite{Morningstar:2013bda} and are available upon request.

When using stochastic LapH estimates for quark propagators, temporal
correlators factorize into `source' and `sink' functions which depend on the
fields in \eqnref{e:qprop} at a given Euclidean time separation. For single nucleons, these fields are
\begin{equation}
\label{e:barfunc}
\begin{aligned}
	\Phi^{(i_1,i_2,i_3)}_\ell(\boldsymbol{p},t)
	=& \\
	c^{(\Lambda,\lambda)}_{\alpha\beta\gamma} \, \epsilon_{abc}& \, \sum_{\boldsymbol{x}} {\rm e}^{i\boldsymbol{p}\cdot \boldsymbol{x}}
	\phi^{(i_1)}_{a\alpha}(x)
	\phi^{(i_2)}_{b\beta}(x)
	\phi^{(i_3)}_{c\gamma}(x)
\end{aligned}
\end{equation}
and $\Omega^{(i_1,i_2,i_3)}_{\ell}(\boldsymbol{p},t)$ (in which the $\phi(x)$ are replaced with $\rho(x)$),
where we have used the shorthand $i_k = (r_k,d_k)$ to combine the noise and dilution indices.

The rank-three tensors of \eqnref{e:barfunc} are contracted over the $i_k$ to project onto definite $(I,I_3)$ and treat all Wick contractions for each of the terms in \eqnref{e:ops}.
To produce an unbiased stochastic estimate, each of the six valence
quark lines in a two nucleon correlation function require a different $r_k$.
However, for a given set of stochastic sources, each permutation and combination of six $r_k$ produces a new (in principle correlated) estimate. However, even using the minimal number of noise sources but moderately increasing the number
of permutations results in a scaling of the statistical errors consistent with independent measurements~\cite{Bulava:2017stw}. We use the maximal number of permutations of 6 noise sources, accounting for nucleon-level symmetries, giving a total of 180 permutations. Further details of algorithmic improvements and the optimization of our developed code are given in Appendix~\ref{app:code}.

%----------------------------------------------------------
%    lattice calculation
\section{Lattice Calculation \label{sec:lattice}}

We employ an isotropic clover-Wilson action with $N_f=2+1$ dynamical fermions that matches the setup being used by the CLS Collaboration~\cite{Bruno:2014jqa}.
We have generated the new \texttt{C103} ensemble with $m_u=m_d=m_s\approx m_s^{\rm phys}$, using the \texttt{openQCD} code~\cite{openQCD} on the BlueGene/Q machine at LLNL (Vulcan).
The lattice spacing is $a\approx0.086$~fm~\cite{Bruno:2016plf} with a lattice extent $V = 48^3\times96$, periodic boundary conditions in space and thermal boundary conditions in Euclidean time.
The C103 ensemble has 4 thermalized replicas (streams) of about $400$ configurations, and each replica is started from different thermalized configurations and with different random seeds.
Each configuration is saved after 2 HMC trajectories of length $\tau = 2$ in molecular dynamic time units.
The bare parameters of the lattice action are provided in \tabref{tab:lattice_action}.
The present computation of two-nucleon correlation functions uses 4 time-sources (cfr. $t$ in \eqnref{e:barfunc}) on 802 configurations spanning two of the replicas, for a total of 3208 time-sources.

%-------------------------------------------------------------------------------
\begin{table}
\caption{\label{tab:lattice_action}
Bare parameters for the lattice action of the C103 ensemble.
}
\begin{ruledtabular}
\begin{tabular}{cccccc}
ens.& $\beta$& $V$& $c_0$& $\kappa_{u,d}=\kappa_s$&$c_{sw}$\\
\hline
C103& 3.4& $96\times48^3$& 1.66& 0.136497611186012& 1.986246
\end{tabular}
\end{ruledtabular}
\end{table}
%-------------------------------------------------------------------------------

%----------------------------------------------------------
%    lattice parameters
\subsection{Correlation Functions \label{sec:corr_analysis}}

At low temperatures the spectral decomposition of a two-point correlation function is given by
\begin{equation}\label{eq:C2pt}
C_{ij}(t) = \sum_n z_{i,n} \tilde{z}^\dagger_{j,n} e^{-E_n t}\, ,
\end{equation}
where $z_{i,n} = \langle \O| O_i |n\rangle$ is the overlap of the $n^{th}$ enegy eigenstate onto the vacuum through the annihilation operator $O_i$.
If the creation and annihilation operators come from a Hermitian-conjugate basis, then this correlation function is positive definite such that all $z_{i,n} \tilde{z}^\dagger_{i,n} = |z_{i,n}|^2 \geq 0$. This simple fact greatly simplifies the analysis of excited state contamination to the ground state contribution in \eqnref{eq:C2pt}.
Specifically it eliminates the possibility of having a false plateau which could be generated by opposite sign contributions to \eqnref{eq:C2pt} from the lowest lying states in the spectrum.

For single-hadron correlation functions, a calculation which uses local point or gaussian-smeared (Wuppertal~\cite{Gusken:1989ad,Alexandrou:1990dq}) quark sources for the hadron creation operator while using momentum-space annihilation operators is still positive definite since translation invariance ensures that, up to a multiplicative constant arising from the Fourier transform, the creation and annihilation operators are still Hermitian conjugate to each other.
If the annihilation operator of a two-hadron correlation function was constructed with a single total-momentum Fourier transform, then it would also be positive definite for the same reason, but it is well known that such operators do not provide enough control over the eigenstates of the system to reliably extract the multitude of energy levels corresponding to the two hadrons interacting at different values of relative momentum.
Therefore, two-hadron correlation functions are typically computed with each of the two final-state hadrons separately Fourier transformed to a particular final-state momentum.  Unfortunately, such annihilation operators are no longer Hermitian conjugates of the spatially local creation operators, and thus the correlation functions lose their positive definite quality.

The sLapH (and LapH) methods allow for the construction of Hermitian-conjugate pairs of creation and annihilation operators in which each hadron at the source and sink can be separately Fourier transformed.  The advantage is twofold: a volume-averaging effect at the source as well as the sink, improving the stochastic precision and a positive definite matrix of correlation functions.

Another well-known feature of two-nucleon calculations is that a ratio of correlation functions constructed as
\begin{equation}\label{eq:ratio}
R(t) = \frac{C_{NN}(t)}{C_N(t) C_N(t)}\, ,
\end{equation}
provides the best way to estimate the interaction energy.
The stochastic correlation between the two-nucleon and single-nucleon correlation functions, $C_{NN}$ and $C_N$, is very strong, and the ratio $R$ benefits from a large cancellation of the single-hadron inelastic excited states: the effective mass of this ratio correlation function $R$ yields a precise estimate of the interaction energy.
However, prior to a time separation when the single-hadron correlation function has relaxed to the ground state, this ratio correlation function can be susceptible to false plateaus: the Taylor expansion of the single-hadron correlators in the denominator leads to opposite sign contributions to the ratio correlation function which are precisely the kind of corrections that can lead to false plateaus.

To describe this feature more precisely, suppose the single-nucleon correlation function was described by just the ground state and a single excited state
\begin{equation}\label{eq:n_simple}
C_N(t) = A_0 e^{-E_0 t} + A_1 e^{-E_1 t}\, .
\end{equation}
In this simplistic model, the two-nucleon correlation function would be given by
\begin{align}\label{eq:nn_simple}
C_{NN}(t) &= \sum_q B_{00,q} e^{-(2E_0 +\D E_{00}(q)) t}
\nonumber\\&\phantom{=}+\sum_{\tilde{q}} B_{01,\tilde{q}} e^{-(E_0 + E_1 + \D E_{10}(\tilde{q})) t}
\nonumber\\&\phantom{=}+\sum_{q^\prime} B_{11,q^\prime} e^{-(2E_1 + \D E_{11}(q^\prime)) t}
\end{align}
where the sums over $q$, $\tilde{q}$ and $q^\prime$ run over the elastic scattering modes between two ground state nucleons, a ground and excited state, and between two excited states respectively, as allowed by the \luscher quantization condition.
The interaction energies $\D E_{00}$, $\D E_{10}$ and $\D E_{11}$ depend upon the relative momentum between the states ($q$) and are typically much smaller than the inelastic excited state energy $E_1 - E_0$ as these elastic scattering energies must vanish as $L\rightarrow\infty$ except in the case of a bound state.
The large-time behavior of the ratio correlation function is then approximated by
\begin{align}\label{eq:ratio_taylor}
R(t) &= b_{00,0} e^{-\D E_{00}(q_0)t}
    + b_{00,1} e^{-\D E_{00}(q_1)t}
\nonumber\\&\phantom{=}
    + b_{10,0} e^{-(E_1-E_0 + \D E_{10}(\tilde{q}_0) )t}
    -2 a_1 e^{-(E_1 - E_0) t}
\nonumber\\&\phantom{=}
    +\cdots
\end{align}
where the $b_{00,n}$, $b_{10,0}$ and $a_1$ are ratios of overlap factors.
The observed near-exact cancellation of inelastic excited states in the ratio manifests as near-exact cancellation between the $b_{10,0}$ and $-2a_1$ terms on the second line of \eqnref{eq:ratio_taylor}.
Such cancellations with opposite signs can lead to false plateaus early in Euclidean time before the time-separation in which the single-nucleon correlation function is saturated by the ground state.%
%   FOOTNOTE
\footnote{The false plateaus HAL QCD has speculated occur for the local source, momentum-space sink correlation functions are not from this early time interference, but rather from non-positive definite contributions from various elastic scattering states which pollute the correlation function at late time, up until $\mathrm{O}(4)$~fm~\cite{Iritani:2016jie}.}
%------------------

To avoid this problem, the NPLQCD Collaboration has long advocated that a sufficient amount of statistics should be used such that the interaction energies can be precisely determined without the need of relying upon the ratio correlation function, but rather the two-nucleon and single-nucleon correlation functions can be fit independently and $\D E_{00}(q)$ can be extracted under jackknife or bootstrap resampling of the ground state energies such determined~\cite{Beane:2009py,Beane:2010hg,Beane:2011iw}.

For many calculations, including the present one, the statistical precision is insufficient to achieve a multisigma determination of the interaction energy from fits to the two-nucleon and single-nucleon correlation functions separately.  A simple measure of the feasibility of such a strategy is whether one can use the ratio correlation function $R(t)$ only at sufficiently late times that the single-nucleon has plateaued and still achieve a convincing energy extraction of the interaction energy~\cite{Wagman:2017tmp}.
This is almost the case in the present calculation, but we require the use of a few time slices ($\mathrm{O}(0.17-0.35)$~fm) prior to the ground state saturation of the single-nucleon correlation functions.

The desire to leverage the positive definite nature of the two-nucleon correlation functions with sLapH, and that, with the present stochastic precision, we must rely upon values of the correlation function prior to the single-nucleon being saturated by just the ground state, motivates the following set of correlation functions and their parametrizations.
First, we factorize the spectral decomposition by pulling out the ground state contribution as a prefactor.
For a single nucleon of momentum $q$, we parametrize the correlation function as
\begin{equation}\label{eq:n_exp}
C_{N_q}(t) = z_{q,0}^2 e^{-E^q_0 t}\left(
    1 +  z_{q,n}^2 e^{-\D E^q_{n,0}t}
    \right)\, ,
\end{equation}
with an implicit sum over all excited states $n>0$.
The ground state energy is $E^q_0$ and
\begin{equation}
    \D E^q_{n,0} \equiv E^q_{n} - E^q_0\, .
\end{equation}
The ground state overlap factor is given by $z_{q,0}$, and the $z_{q,n}$ are the ratio of overlap factors of the $n^{th}$ state to the ground state which all satisfy the bound $z_{q,n} > 0$.

To take advantage of the positive definiteness of the two-nucleon correlation function and also the cancellation of excited states in the ratio correlation function, instead of fitting the two-nucleon correlator, we fit the ratio correlator but with the following functional form
\begin{equation}\label{eq:nn_ratio_exp}
R(t) = \frac{r_{0}^2 e^{-\D E^{NN}_{0} t}
        \left( 1 + r_{l}^2 e^{-\D E^{NN}_{l,0}t}
        \right)
    }{\left(1 +  z_{q,n}^2 e^{-\D E^q_{n,0}t} \right)
      \left(1 +  z_{p,m}^2 e^{-\D E^p_{m,0}t} \right)}\, ,
\end{equation}
with implicit summations over the $l$, $n$ and $m$ excited states.
The various new terms in this expression are
\begin{itemize}[leftmargin=*]
\item $\D E^{NN}_{0} = E^{NN}_0 - E_0^q - E_0^p$, the ground state interaction energy of interest for total momentum $\mathbf{P} = \mathbf{p}+\mathbf{q}$;

\item $\D E^{NN}_{l,0} = E^{NN}_l - E^{NN}_0$, the energy gap between the $l^{th}$ two-nucleon excited state and the two-nucleon ground state energy.
The $l^{th}$ energy gap can arise from either an elastic scattering state of the two ground state nucleons or when one or both nucleons are in an inelastic excited state;

\item $r_0^2 = (z^{NN}_0)^2 / (z_{q,0} z_{p,0})$, the ratio of the ground state two-nucleon overlap factor to the product of single-nucleon overlap factors;

\item $r_l^2 = (z^{NN}_l / z^{NN}_0)^2 >0$, the ratio of the $l^{th}$ two-nucleon overlap factors to the ground state two-nucleon overlap factor, which are all positive.

\end{itemize}
With this fit function, \eqnref{eq:nn_ratio_exp}, if an equal number of ``inelastic'' excited states are included in the numerator as in the denominator, as well as possibly extra ``elastic'' excited states, then the fit function can naturally capture the cancellation of the inelastic excited states from the single nucleon that are observed to also pollute the two-nucleon correlation functions at early times, without forcing this cancellation to be exact.
In the next section, we will demonstrate the stability of the analysis with respect to the time-range and number of states used in the analysis.

%----------------------------------------------------------
%    nn energies
\subsection{Energy spectrum \label{sec:nn_energies}}

\subsubsection{The pion}
\begin{figure}
\includegraphics[width=\columnwidth]{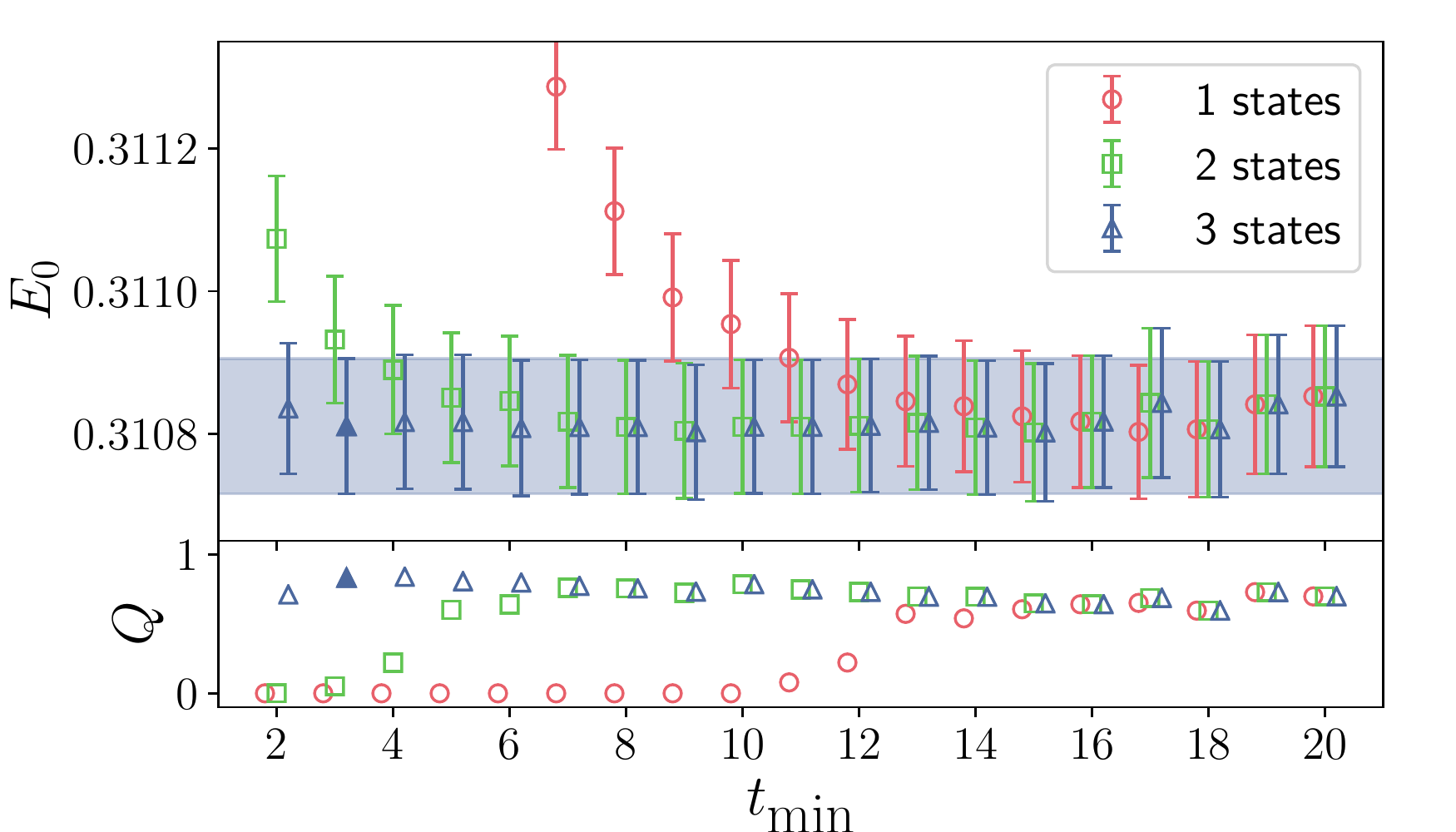}
\caption{\label{fig:pion}
A $t_{\rm min}$ stability plot of the pion ground state mass with different $N$ states in the fit function. See text for description.  The filled symbol at $t_{\rm min}=3$ is the chosen fit.
}
\end{figure}

We first look at the pion correlation function to estimate $m_{\pi}$.
A single operator was used to construct this correlation function which is fit to the cosh version of \eqnref{eq:C2pt} to take into account wrap-around effects
\begin{equation}
C_\pi(t) = \sum_n z_n z^\dagger_n \left( e^{-E_n t} + e^{-E_n (T-t)} \right)\, .
\end{equation}
\figref{fig:pion} shows an $N$-state stability plot of the ground state pion mass versus $t_{\rm min}$ with the chosen fit (given by the filled symbol) coming from $N=3$ states and $t_{\rm min}=3$.
The fits were performed with a Bayesian constrained analysis~\cite{Lepage:2001ym}, resulting in a determination of the pion mass in lattice units of
\begin{equation}\label{eq:m_pi}
m_\pi = 0.310810(95)\, .
\end{equation}

%-------------------------------------------------------------------------------
%-------------------------------------------------------------------------------

\subsubsection{Single nucleon analysis}

We then move on to study the mass of single nucleons.
The single-nucleon correlation functions were fit with \eqnref{eq:n_exp}, also using a Bayesian constrained analysis~\cite{Lepage:2001ym}.
The ground state energy prior is estimated from the long-time behavior of the effective mass and the ground state overlap factor is estimated from an effective overlap construction:
\begin{align}\label{eq:eff_mass_z}
m_{\rm eff}(t) &= \ln \left( \frac{C_{N_q}(t)}{C_{N_q}(t+1)} \right)\, ,
\nonumber\\
z_{N_q}^{\rm eff}(t) &= \left[ e^{m_{\rm eff}(t) t}\ C_{N_q}(t) \right]^{1/2}\, .
\end{align}
The prior central values are taken from the mean of these at a late reference time of $t=10$ and the prior widths are taken to be 10 times the uncertanties on the relative effective quantity at this time.
For the excited state energy splittings, we use a log-normal distributed prior such that the total energies are ordered.
The mean values of the priors are estimated at twice the pion mass with a width that comes down a little lower than the first $N\pi$ p-wave scattering state.  The central value of the $l^{th}$ state energy and the excited state overlap factors are then priored as
\begin{align}
E_{l}^{q} &= E_0^q + l \times \overline{\D E_q}\, ,
\nonumber\\
\ln(\D E_q) &= (\ln(2m_\pi), 0.7)\, ,
\nonumber\\
z_{q,l} &= (1.0, 1.0)
\end{align}
where $\overline{\D E_q}$ is the mean value and $l=1$ is the first excited state. We use the notation $(p_c, p_w)$ to represent a prior with central value $p_c$ and width $p_w$ assuming that its distribution is Normal unless the prior name is $\ln( \cdot)$, in which case a log-normal distribution is assumed.

\begin{figure}
\includegraphics[width=\columnwidth]{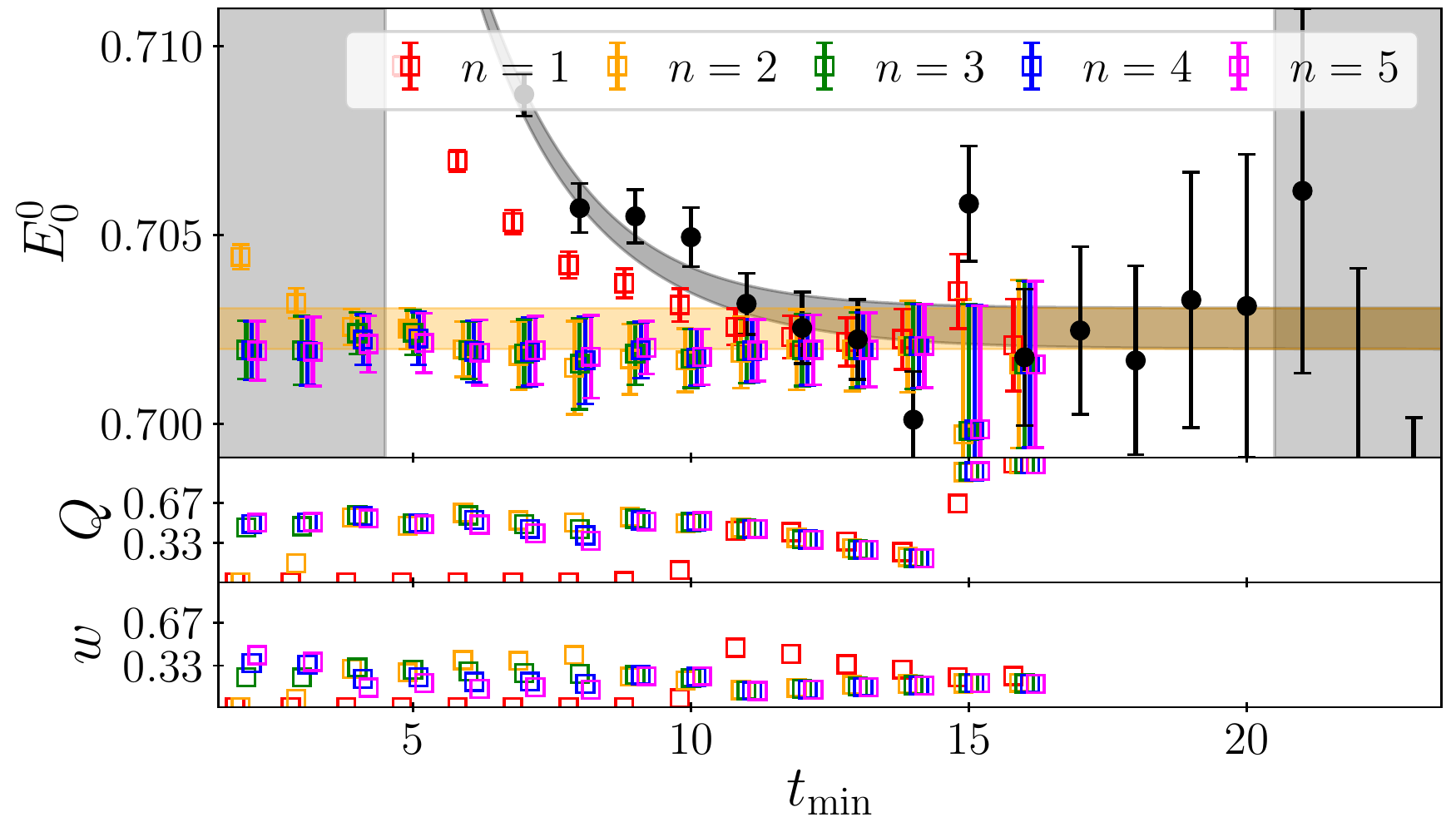}
\caption{\label{fig:single_nucleon_stability}
Stability of the single nucleon ground state at zero momentum.
The filled (black) circles are the effective mass data from the correlation function.
The open squares are the resulting ground state mass as a function of $t_{\rm min}$ and the number of states $n$ used in the analysis.
The filled square at $t_{\rm min}=5$ with $n=2$ is the chosen fit.
The vertical gray bands indicate time-regions excluded from this fit and the gray filled curve is the effective mass reconstructed from its posteriors and the horizontal band is the value of $m_N$.
}
\end{figure}

In \figref{fig:single_nucleon_stability}, we show the resulting ground state energy of the nucleon at rest versus $t_{\rm min}$ and the number of excited states.  It is sufficient to chose $n=3$ states (2 excited states) to fit the single nucleon as early as $t_{\rm min}=2$ to achieve an answer that is consistent with the general stability displayed.
We observe very similar stability of the ground state mass for all of the boosted single-nucleon correlation functions which are shown in the github repository accompanying this publication~\url{https://laphnn.github.io/nn_c103_qcotd_swave_only/}~\cite{laphnn_qcotd}.
In all cases, we observe an $n=3$ fit from $t_{\rm min}=2$ is in excellent agreement with the general stability of the ground state as well as $n=2$ with $t_{\rm min}=5$.  We use these two choices for our analysis and to explore systematics associated with the choice of the number of states and fit range.
We find in lattice units
\begin{equation}\label{eq:mn}
m_N = 0.70262(59)
\end{equation}

%-------------------------------------------------------------------------------
% single nucleon tmin plots

\subsubsection{Two-nucleon analysis}
To determine the two-nucleon eigenstates, a correlation matrix, $C_{ij} \equiv \langle \hat{O}_i(t) \hat{O}^{\dagger}_j(t) \rangle$, is formed from the set of operators, $\hat{O}_i(t)$, which have been projected onto a given $(\mathbf{P},\Lambda)$. Solutions to the following Generalized Eigenvalue Problem (GEVP),
\begin{eqnarray}
C(t_d) v_n(t_d,t_0) = \lambda_n C(t_0) v_n(t_d,t_0) \ ,
\end{eqnarray}
for given reference times $t_d,t_0$, may then be used to rotate the correlation matrices,
\begin{eqnarray}
\hat{C}_n(t) = \left( v_n(t_0,t_d), C(t) v_n(t_0,t_d)\right)
\end{eqnarray}
to a basis consisting of linear combinations of operators having optimal overlap (for the given basis) onto the eigenstates of the system.

From this set of correlation functions we form the ratio $R$ using \eqnref{eq:ratio} with single-nucleon correlators corresponding to momenta $\mathbf{p}_n$ of the nearest noninteracting energy level for a given state, $n$. This ratio is then fit to the functional form of \eqnref{eq:nn_ratio_exp} with similar Bayesian methods as the single-nucleon case. Priors for the various parameters are chosen as follows:
\begin{itemize}[leftmargin=*]

\item $\D E^{NN}_{0}$: similar to the single nucleon, these are estimated from the effective mass of the ratio correlation function at a reference time of $t=10$ with the prior mean estimated from the mean of the effective mass and a prior-width that is 10 times larger than the uncertainty of the effective mass;

\item $\D E^{NN}_{l,0}$: We add two towers of excited states, one corresponding to inelastic excited states with prior means and widths estimated as with the single nucleon inelastic excited states and a second tower with energy gaps estimated to arise from elastic scattering states.  Since we expect the GEVP to remove the low-lying elastic scattering excited states, the gap to the first excited state is estimated to be several levels above the ground state with a prior width that allows it to be as small as the first anticipated excited scattering state or as large as an inelastic single nucleon excited state.

\item $r_0^2$: As with the single nucleon, the ground state ratio overlap factor is estimated through an effective overlap factor of the ratio correlation function, \eqnref{eq:eff_mass_z};

\item $r_l^2 = (1.0, 1.0)$: Following from \eqnref{eq:nn_ratio_exp} and similar expectations as with the single nucleon excited states.

\end{itemize}

%----------------------------------------------------------
%    phase shift analysis
\subsection{Phase shift analysis \label{sec:nn_phaseshifts}}

The \luscher finite-volume formalism~\cite{Luscher:1986pf,Luscher:1990ux}, and its extension to moving frames~\cite{Rummukainen:1995vs,Kim:2005gf} and various generalizations~\cite{PhysRevD.83.114508,Fu:2011xz,Leskovec:2012gb,Hansen:2012tf,Gockeler:2012yj,Briceno:2013lba,Briceno:2014oea}, allows one to faithfully connect the finite-volume two-particle spectrum to the corresponding infinite-volume scattering phase shifts at the momenta associated with those energies. The reduced hypercubic symmetry of the lattices, however, mixes the partial waves associated with spherical symmetry in infinite volume. Thus, a system which has been projected onto the given irreps of the hypercubic group will, in general, have nonzero overlap with an infinite number of partial waves, and therefore a truncation in the partial waves considered is required. Fortunately, at low energies, we expect contributions from a partial wave, $l$, to fall off as $E^{-l}$, justifying a truncation to the lowest partial waves that couple to a given finite-volume irrep.

In this first look at NN interactions with sLapH (the present work), we ignore all partial-wave mixing induced by the finite, periodic volume and restrict ourselves to considering the $s$-wave interactions (a standard choice in the field so far for two-baryons).
We also restrict the energies considered to those below the $t$-channel cut, $q^* \leq m_\pi/2$ where $q^* $ is the magnitude of the momentum of each nucleon in the center-of-mass (CoM) frame.
We will relax these restrictions and assumptions in a forthcoming paper where we explore the partial wave mixing and energies up to the inelastic pion-production threshold.

For the low energies considered here, the K-matrix for each partial wave is expected to be well described by a smooth polynomial in $q^{*2}$, known as the effective range expansion (ERE)
\begin{eqnarray}\label{eq:ere}
q^* \cot\d(q^*) = -\frac{1}{a} + \frac{1}{2} r_0 q^{*2} + \frac{1}{6} r_1 q^{*4} + \cdots \ ,
\end{eqnarray}
where $\d(q^*)$ is the scattering phase shift, $a$ is the scattering length, $r_0$ is the effective range, and $r_n \ , n>0$ are higher-order shape parameters which give the short-distance details of the potential.  In terms of the potential, the convergence of the ERE is expected to be rapid for $q R \ll 1$, where $R$ is the range of the potential.

Under the assumption that partial wave mixing is negligible, the \luscher quantization condition provides a one-to-one mapping between the spectrum and $q^* \cot\d(q^*)$, which for the $s$-wave is
\begin{eqnarray}\label{eq:luscher}
q^*\cot\delta(q^*) = \frac{2}{\gamma L \sqrt{\pi}}
    Z_{00}^{\mathbf{d}}\left(1, \frac{q^{*2}L^2}{4\pi^2} \right) \ ,
\end{eqnarray}
where $\gamma$ is the ratio of the energy to the CoM energy, $\gamma = E/E^*$, and $Z_{00}^{\mathbf{d}}$ is a generalized zeta function defined in Ref.~\cite{Rummukainen:1995vs}, characterized by the boost vector
\begin{equation}\label{eq:boost_vec}
\mathbf{d} \equiv \frac{L}{2\pi} \mathbf{P}\, .
\end{equation}
The input values $q^*$ are derived starting from the lattice extracted energies, $E = \sqrt{ E^{*2} + |\mathbf{P}|^2}$, where $E^*$ is then related to $q^*$ via
\begin{equation}\label{eq:E_to_delta}
    E^* = 2\sqrt{q^{*2} + m_N^2}\, .
\end{equation}

There are several ways to proceed in fitting the numerical results to extract the ERE parameters; here, we discuss three.
In the first method, referred to as the determinant residual method, the \luscher quantization condition (truncated to some maximum partial wave and parameterized appropriately) is used directly to form the residuals of the $\chi^2$ function, which is subsequently minimized~\cite{Morningstar:2017spu}. Using the quantization condition directly in the fitting procedure is a natural way to include multiple partial waves. A convenient feature of this method, as opposed to methods that directly solve the quantization condition, is that the generalized zeta functions can all be computed once before the minimization process starts. However, one cannot avoid recomputing the covariance matrix each time the parameters are adjusted during the fit, since the model cannot be separated from the data.
In subsequent papers, we will explore this method in more detail when we consider the partial wave-mixing induced both by the finite volume as well as the physical mixing of the ${}^{3}S_1$--${}^{3}D_1$ waves in the deuteron channel.

The second method, which we refer to as the $q\cot\d$ method, has been common in the application to two-baryon systems under the truncated partial-wave expansion (also considered here).
First, one converts the energy levels, which are determined typically with Gaussian distributed noise, to values of the CoM momentum \eqnref{eq:E_to_delta} which are used to determine the phase shift values through \eqnref{eq:luscher}.
These values of the $q^* \cot \d(q^*)$ are then fit with the ERE \eqnref{eq:ere}, to determine the values of $a$, $r_0$ and other shape parameters that describe the low-energy interactions.

As is well known, the zeta functions appearing in \eqnref{eq:luscher} have nonlinear dependence upon $q^*$ in the typical range over which the momenta can be determined.
This transforms the roughly Gaussian distributed determination of $E$ (and hence $q^*$) into a highly asymmetric distribution of $q^* \cot \d(q^*)$.
Moreover, it is common to perform the ERE fit by treating $q^* \cot \d(q^*)$ data points as having uncertainties in both the $x$ and $y$ directions.
However, under the assumption of no partial wave mixing, the \luscher quantization condition, \eqnref{eq:luscher} provides a one-to-one mapping between the $x$ ($q^{*2}$) and $y$ ($q^* \cot \d(q^*)$) values, such that there is really only a single variable with uncertainty.
For sufficiently precise determinations of $q^*$ values such that a linear approximation to \eqnref{eq:luscher} describes the results, treating the pairs of $(q^{*2}, q^* \cot \d(q^*))$ points with correlated uncertainties is expected to faithfully reproduce the true uncertainty with the standard linear transformations for handling $x$ and $y$ uncertainty.
However, when the nonlinearity of \eqnref{eq:luscher} is important, this method can produce biased results.
See for example Ref.~\cite{Iritani:2017rlk} for a treatment that enforces this constraint from \eqnref{eq:luscher}.

We propose an alternative method that properly handles this nonlinear relationship, which we refer to as the spline/gradient method.
Consider a bootstrap (BS) resampling of the values of $(X_i,Y_i) = (q^{*2}_i, q^*_i \cot \d(q^*_i))$ pairs on irrep $i$.
For a given BS sample, one can define the squared distance between this point and the intersection of the ERE function with the $i^{th}$ irrep as the distance along the curve defined through \eqnref{eq:luscher} which we denote $Y_i=f(X_i)$ for convenience, with the distance given by
\begin{equation}
s_i(f_i^\prime, z_{i,\hat{\b}}, z_{i,bs}) = \left| \int_{z_{i,\hat{\b}}}^{z_{i,bs}}
    dx \sqrt{1 + f_i^\prime(X)^2} \right|\, ,
\end{equation}
where $f^\prime_i$ is the derivative of $f$ along the curve,
$z_i$ is a generalized coordinate along the curve,
$z_{i,\hat{\b}}$ is the location of intersection of the ERE function with the $i^{th}$ irrep and $z_{i,bs}$ is the coordinate of the $bs^{th}$ sample of irrep $i$.
These BS distances can be used to construct the objective function that penalizes data discrepancy between all irreps with the intersection of the ERE parameterization.
Then, an uncorrelated, unweighted least square penalty for BS sample $bs$ would be given by $\sum_i s_i(f_i^\prime, z_{i,\hat{\b}}, z_{i,bs})^2$.
To estimate the appropriate covariance, we leverage the Delta Method~\cite{bickel2001mathematical} that scales the covariance from $X$ using the gradient of $f(X)$.
For a normally distributed set of $X$ variables of mean $\mu_X$
\begin{equation*}
(\bar{X} - \mu_X) \Rightarrow N(0, \Sigma_X)\, ,
\end{equation*}
with $\Sigma_X$ is the covariance of the $X$ variables over the irreps,
the Delta Method states for a differentiable $f$, the distribution of $f(X)$ follows
\begin{equation*}
f(\bar{X}) - f(\mu_X) \Rightarrow
    N\left(0, \nabla f(\mu_X)^T \Sigma_X \nabla f(\mu_X) \right)\, ,
\end{equation*}
where $\nabla f(\mu_X)$ is the vector of $f^\prime_i$ over the irreps.
And thus, the correlated objective function we can minimize to estimate the ERE parameters ($\hat{\b}$), for a given BS sample is given by
\begin{equation}
\tilde{\chi}_{bs}^2 = \sum_{ij} s_i(f^\prime_i, z_{i,\hat{\b}}, z_{i,bs})
    W_{bs, ij}
    s_j(f^\prime_j, z_{j,\hat{\b}}, z_{j,bs})\, ,
\end{equation}
where the inverse ``covariance matrix'' for sample $bs$ is
\begin{equation}
W_{bs} = \left[
    \nabla f(X_{bs})^T \hat{\S}_X \nabla f(X_{bs})
    \right]^{-1}\, .
\end{equation}
While the variance of $X$ is fixed for each BS sample, the gradients are evaluated sample by sample.
To estimate the distance and gradient along the \luscher curve, a cubic spline is fit to each pair of values using the BS samples.
The ERE parameters are then estimated with the resulting BS distribution of $\hat{\b}_{BS}$.
For more detail, see \appref{app:qcot_analysis} and Ref.~\cite{steiner2018multidimensional}.

A third method we consider is essentially the ``spectrum method'' described in Ref.~\cite{Morningstar:2017spu}, which directly minimizes spectrum residuals, thus avoiding skewed $q\cot\d$ distributions.
This method is composed into two steps:
An outer step, effectively computing the spectrum as a function of ERE parameters and an inner root-finding step, solving for the value of $q^{*2}$ which satisfies the equation $\eqref{eq:ere}=\eqref{eq:luscher}$ for given $q \cot \delta$ parameterization.
This inner step performs a least-squares minimization for fixed ERE parameters that minimize the residual of the predicted $q^{*2}$ values with those determined from the spectrum.
The outer step computes a function $f(a,r_0,...;{\rm irrep},\mathbf{P},n)$ that returns the value of $q^{*2}$ for a given irrep at boost $\mathbf{P}$ of the $n^{th}$ principle correlator, which are compared against the spectrum by $q^{*2}_{{\rm irrep},\mathbf{P},n} = E^{*^2}/4 -m_N^2$ which is the numerical value of $q^{*2}$ for the same state.  We then minimize the $\chi^2$ with respect to the ERE parameters
\begin{equation}
\chi^2_{\rm spec} = \sum_{i,j} (f(\b;i) - q^{*2}_i)
    {\rm Cov}^{-1}_{q^{*2},ij}
    (f(\b;j) - q^{*2}_j)\, ,
\end{equation}
where $\b=\{a, r_0,\dots\}$ and $i,j$ are master indices running over the combinations of ${\rm irrep},\mathbf{P},n$. The covariance is constructed from the bootstrap distributions of $q^{*2}_i$ with respect to the bootstrap means $\overline{q^{*2}_{i}}$,
\begin{equation}
    {\rm Cov}_{q^{*2},ij} = \frac{1}{N_{\rm bs}} \sum_{bs}
        (q^{*2}_{i,bs} - \overline{q^{*2}_{i}}) (q^{*2}_{j,bs} - \overline{q^{*2}_{j}})\, .
\end{equation}
There are many other variants of extracting the physical parameters from the two-particle spectrum which are discussed in the literature, see for example Refs.~\cite{Bernard:2007cm,Beane:2012ey,Hall:2013qba,Wu:2014vma,Wilson:2014cna,Wilson:2015dqa,Dudek:2016cru,Guo:2016zos,Woss:2020cmp}.

As we will show in Secs.~\ref{sec:deuteron} and \ref{sec:dineutron}, of the two methods used in this work, the spectrum method is less susceptible to outliers, since the $q^{*2}$ values determined are bound to finite intervals and have a near Gaussian distribution following from their parent $E$ distributions, and therefore, the resulting uncertainty on the extracted ERE parameters is smaller.
This is in contrast to the values of $q\cot\d$ determined with \eqnref{eq:luscher} as these distributions become highly nonsymmetric and heavy tailed.
Nevertheless, the spline/gradient method we introduce reproduces the same values of the ERE parameters and is less susceptible to the heavy-tailed fluctuations than the more standard analysis of $q\cot\d$ values one finds in the literature.

%----------------------------------------------------------
% T1g plot
\begin{figure}
\begin{tabular}{c}
\includegraphics[width=\columnwidth]{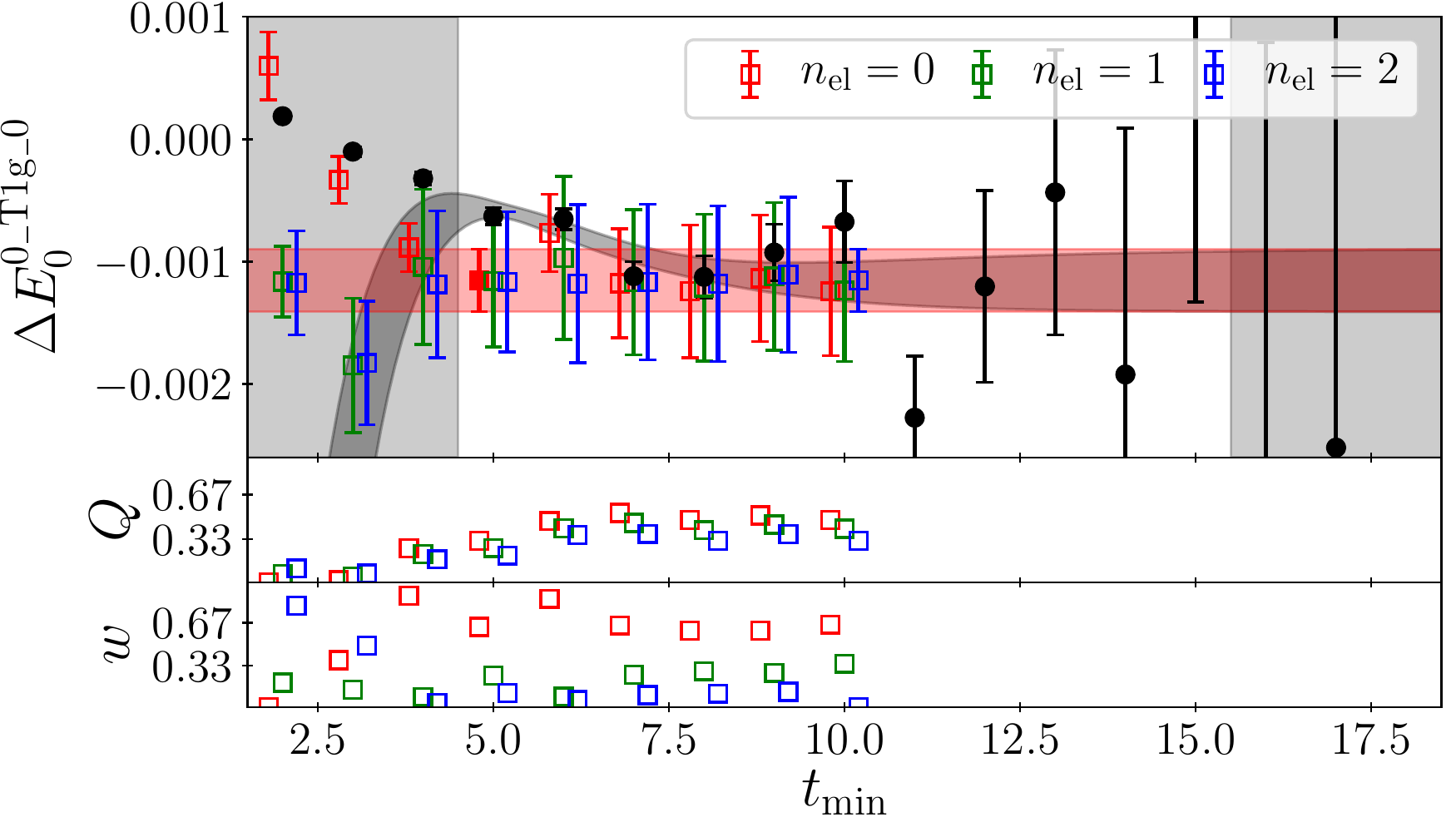}\\ (a)\\
\includegraphics[width=\columnwidth]{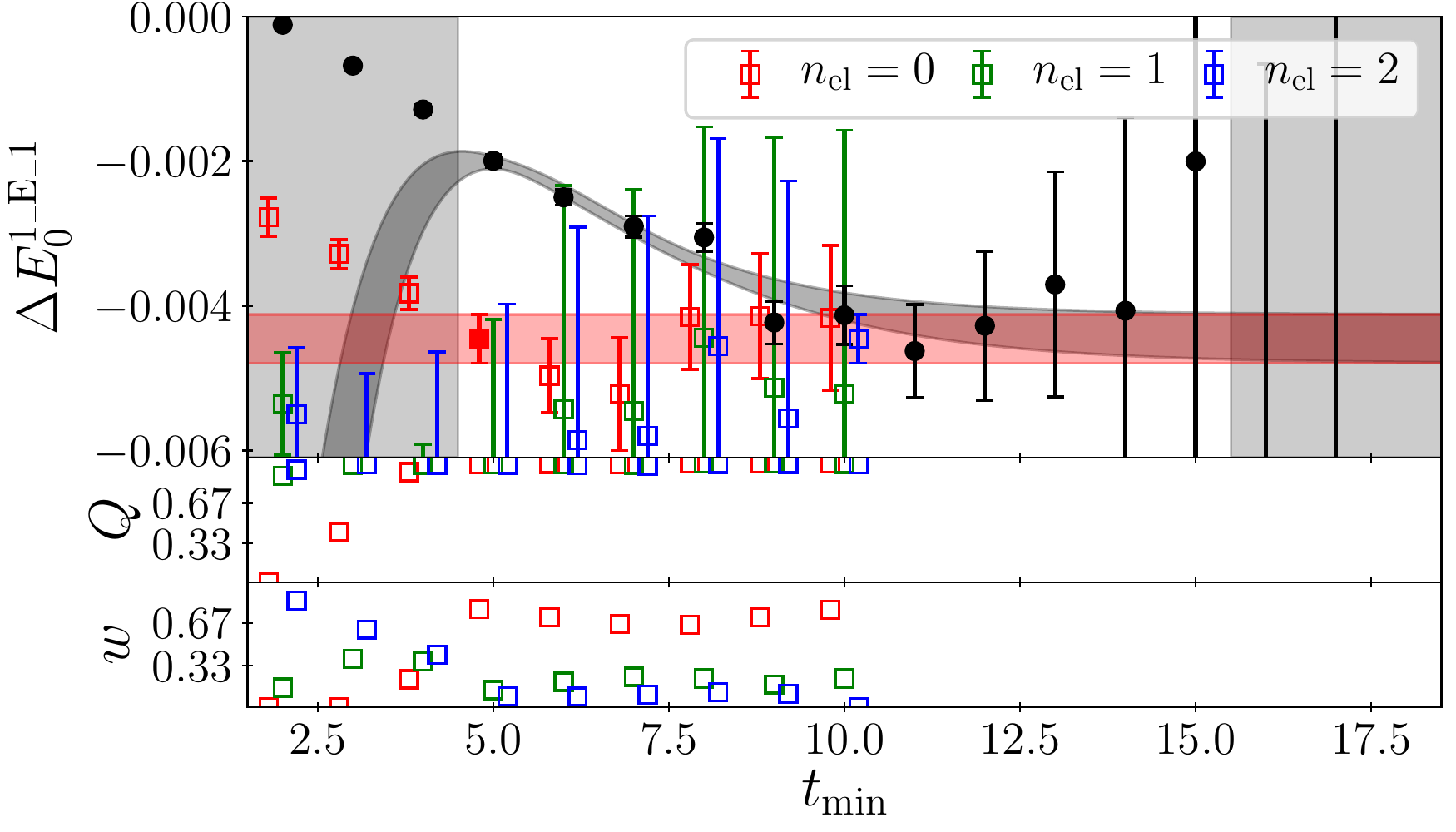}\\ (b)
\end{tabular}
\caption{\label{fig:deut_T1g}
Stability plot of the ground state energy in the T1g irrep with $\mathbf{d}=0$ (a) and the second principal correlator in the $E$ irrep with $\mathbf{d}=1$ (b).
The filled (black) circles are the effective mass of the ratio correlation function, \eqnref{eq:nn_ratio_exp}.
The open squares are the resulting $\D E_{g.s.}$ energy as a function of $t_{\rm min}$ and the number of ``elastic'' excited states used, see the text.
The filled square is the chosen fit.
The vertical gray bands indicate time-regions excluded from this fit, the gray curve is the effective mass reconstructed from its posteriors and the (red) horizontal band is the value of $\D E_{g.s.}$.
}
\end{figure}
%-------------------------------------------------------------------------------

%----------------------------------------------------------
%    deuteron
\subsubsection{Deuteron channel \label{sec:deuteron}}
%-------------------------------------------------------------------------------

%-------------------------------------------------------------------------------
%    qcotd deuteron
\begin{figure*}
\begin{tabular}{cc}
\includegraphics[width=0.49\textwidth]{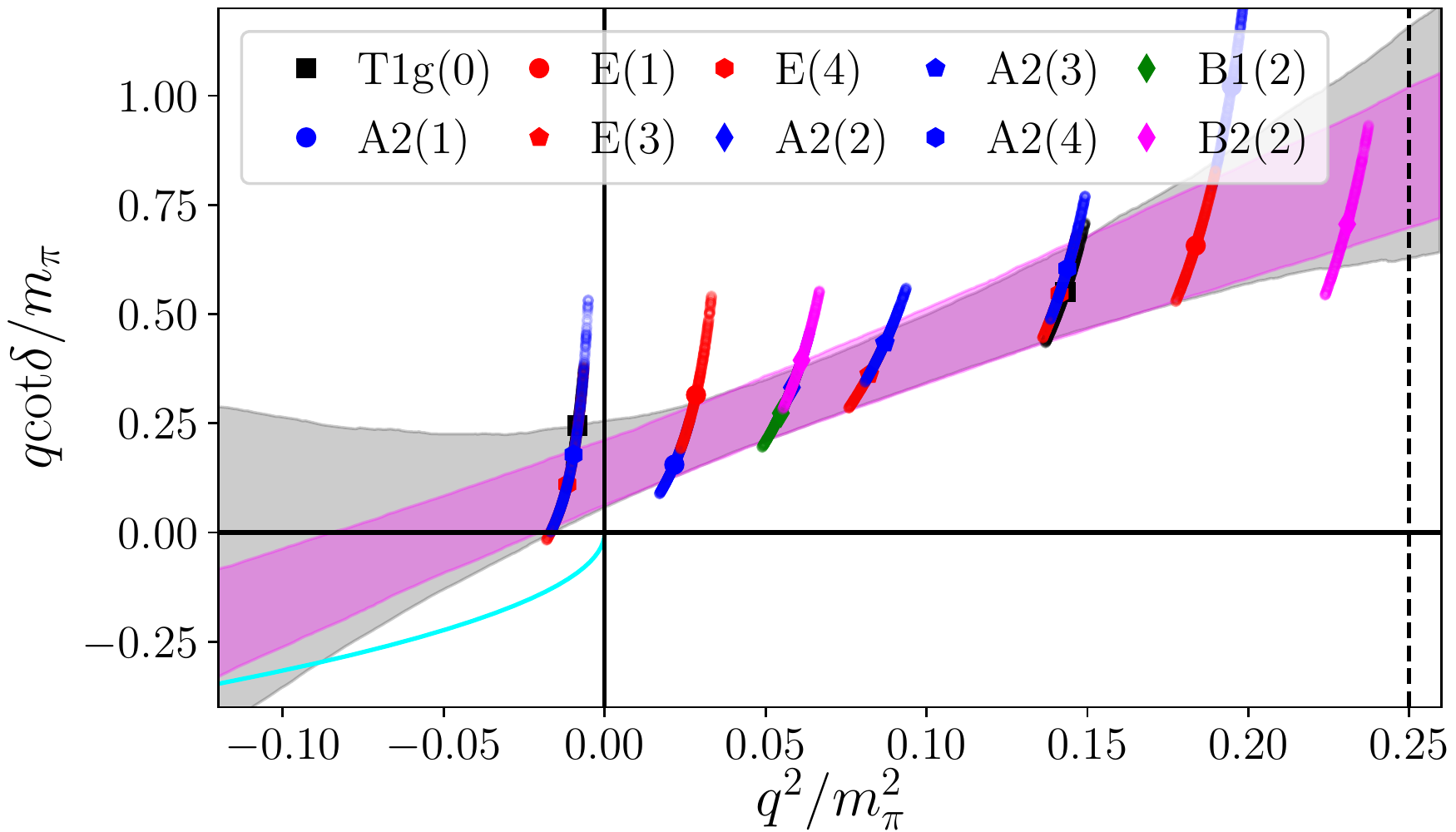}
&
\includegraphics[width=0.49\textwidth]{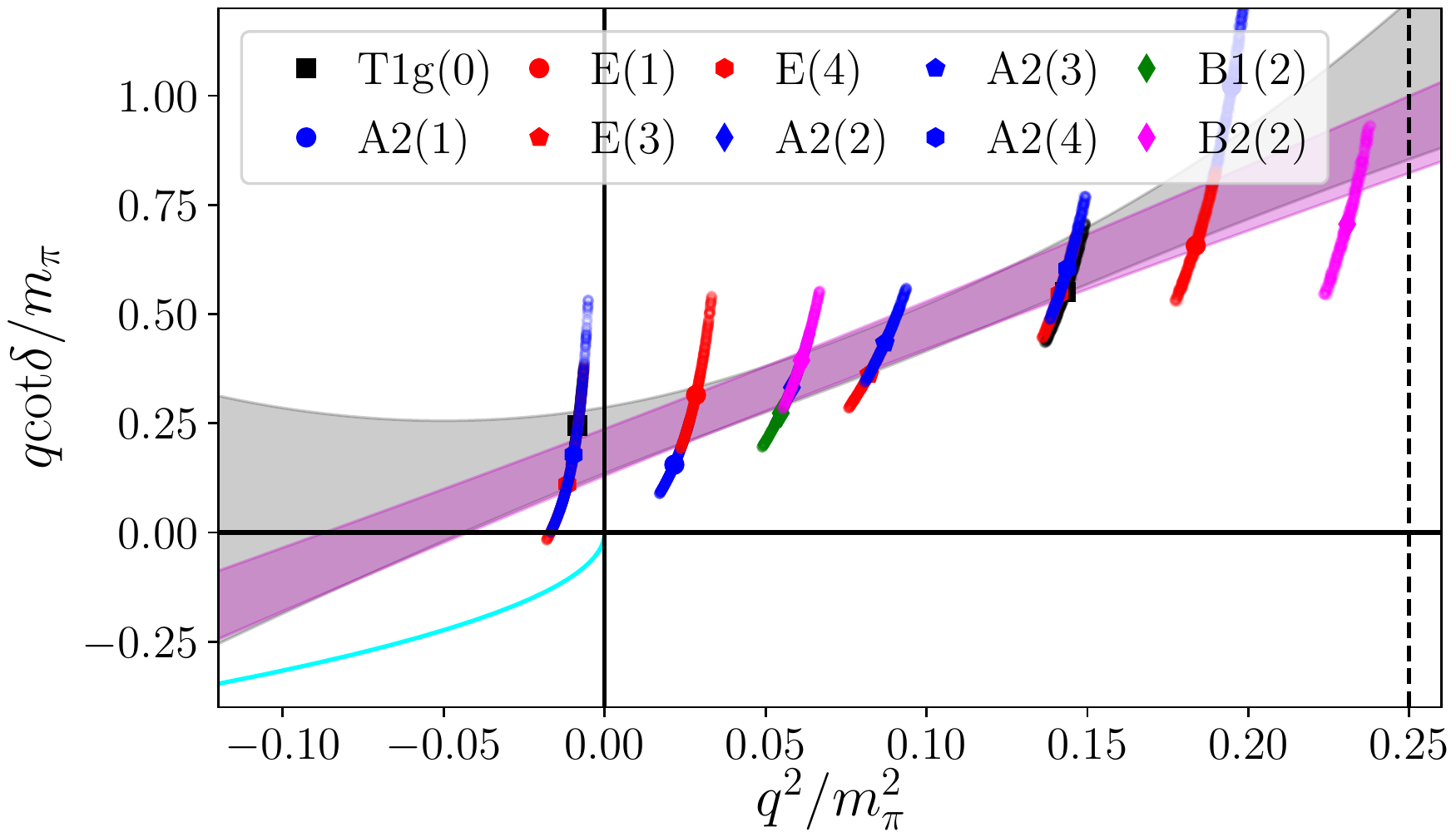}
\\
($q\cot\d$ analysis)& (spectrum analysis)
\end{tabular}
\caption{\label{fig:qcotd_deuteron}
Phase shift analysis, ignoring all partial wave mixing, in the irreps that overlap with the S-wave deuteron.
The spline/gradient method is shown on the left and the spectrum method on the right.
The smaller (magenta) band is the 1-sigma result of the NLO ($q^{*2}$) order fit while the larger (gray) band is the 1-sigma result of the NNLO ($q^{*4}$) order analysis.
The solid (cyan) line is the solution of $q\cot\d=iq$ where a bound state would occur if it were in the spectrum.
}
\end{figure*}
%-------------------------------------------------------------------------------

To extract results for the deuteron channel, we consider all irreps whose lowest partial-wave contribution corresponds to $s$-wave scattering of nucleons with isospin $I=0$ and spin $s=1$.
To determine the spectrum, we first perform a stability analysis of the two-nucleon correlation function as a function of $t_{\rm min}$ and the number of ``elastic'' excited states used above and beyond the $n=2$ states used for the single nucleon.
In \figref{fig:deut_T1g}, we show sample stability plots for fits to the $NN$ ratio correlation functions in two different irreps.
In all irreps, we find that the choices
\begin{itemize}[leftmargin=*]
\item N, \eqnref{eq:C2pt}: $N_{\rm states}=2$, $t=[5,20]$;
\item NN, \eqnref{eq:nn_ratio_exp}: $N_{\rm states}=2$, $n_{\rm el}=0$, $t=[5,15]$;
\end{itemize}
lead to an optimal, or near optimal fit as measured by three factors:
\begin{itemize}[leftmargin=*]
\item Good quality of fit, $Q$;
\item For a given $t_{\rm min}$, the highest weight $w=e^{\rm logGBF}$ as measured by the relative Bayes Factor, see Ref.~\cite{Jay:2020jkz} for further discussion on this point where fits with different amounts of data are also considered;
\item Consistency with the long time values of the effective mass of the ratio correlation function.
\end{itemize}
We opted to select the values of $t_{\rm min}=5$ and $n_{\rm el}=0$ to be the same for all irreps analyzed to minimize the chance of accidentally biassing the result through a more fine-grained optimization.

Stability plots for all irreps can be found with the git repository accompanying this publication \url{https://laphnn.github.io/nn_c103_qcotd_swave_only/}~\cite{laphnn_qcotd}.
In \appref{app:energy_levels} in \tabref{tab:deuteron}, we list the irreps and the resulting ground state energies of the two-nucleon system and corresponding boosted single nucleons as well as the processed values of $q^{*2}$ and $q^* \cot \d$ in $m_\pi$ units used in this analysis.

In \figref{fig:qcotd_deuteron} we show the resulting values of ($q^*$,$q^*\cot\d$) from all irreps along with an ERE fit using the spline/gradient method (left) and the spectrum method (right).
To cleanly display the correlated distributions of ($q^*$,$q^*\cot\d$) pairs, we bootstrap our energy results and show the resulting $68\%$ confidence intervals in the data.
We note that the results from different irreps agree nicely within their respective energy ranges. The purely $s$-wave contributions from each irrep are expected to be consistent with each other, with any discrepancies arising from mixing of higher partial waves. The smooth $q\cot\delta$ behavior taken from multiple irreps thus gives some confidence that mixing from higher partial waves is negligible within our errors.

We find that the fits to the ERE to $q^{*2}$, next-to-leading order (NLO) and $q^{*4}$, next-to-next-to leading order (NNLO) give consistent results for the phase shift within our energy range at our given uncertainties. This, coupled with the smooth behavior of the data, strongly indicates a convergence of the expansion within the energies considered. Our results for the effective range parameters are as follows:
{\small
\begin{equation}
\begin{array}{rrlll}
\hline\hline
\textrm{method}& \textrm{order}& {\quad m_\pi a}& {\ m_\pi r_0}& m_\pi^3 r_1\\
\hline
q\cot\d& \textrm{NLO}       &  -7.9(_{-6.8}^{+3.5})& 5.5(_{-1.1}^{+1.5})& -\\[2pt]
\textrm{spec}& \textrm{NLO} &  -5.5(1.6)& 5.82(71)& -\\[2pt]
q\cot\d& \textrm{NNLO}      & -7.6(_{-7.9}^{+3.9})& 5.3(_{-3.5}^{+2.4})& 2(_{-33}^{+56})\\
\textrm{spec}& \textrm{NNLO}& -4.7(1.7)& 4.2(2.3)& 29(37)\\
\hline
\end{array}
\end{equation}}
Using the NLO ERE expansion, one can solve a quadratic equation for solutions of
\begin{eqnarray}
    q\cot\delta = iq \ ,
\end{eqnarray}
resulting in the two solutions
\begin{equation}
\frac{q_\pm}{m_\pi} = \frac{i}{m_\pi r_0} \left(
    1 \pm \sqrt{1 - 2\frac{r_0}{a}}
    \right)\, .
\end{equation}
Taking the results from the more stable spectrum analysis, the plus solution is found to be
\begin{equation}
\frac{q_+}{m_\pi} = i\, 0.476(62)\, .
\end{equation}
In principle, this could correspond to a bound state solution.  However, this solution lies well outside the range where our results are constraining the amplitude (it is the crossing of our $q\cot\d$ and $-\sqrt{-q^2}$ at large, negative value of $q^2$).  However, this cannot be a physical bound state as the slope of the $q\cot\d$ curve is larger than the tangent of the $-iq$ curve at this crossing, as discussed in detail in Ref.~\cite{Iritani:2017rlk}.
The negative solution
\begin{equation}
\frac{q_-^{\rm deuteron}}{m_\pi} = -i\, 0.132(32)
\end{equation}
lies in the range of our results, is purely imaginary with a negative sign and thus corresponds to a virtual bound state.
This state is expected physically for an attractive interaction with a large, negative scattering length.  As the strength of the interaction increases, such that the system would form a bound state, the virtual bound state solution would move towards zero and become a positive imaginary solution which is the bound state.
There have been few previous identifications of virtual states with lattice QCD in the two-meson sector~\cite{Dudek:2014qha,Wilson:2019wfr}.

Such a bound state solution would have a positive scattering length, such that the intercept of $q\cot\d$ at threshold ($q^2=0$) would be negative.  This implies one should find negative values of $q\cot\d$ for small, positive $q^2$, which we do not find with our results.
Thus, the results of this computation strongly disfavor the existence of a bound deuteron at this pion mass, and with this particular action at finite lattice spacing.

These results are not sufficient to rule out a bound state in the system.
For example, the operator basis we have chosen, which does not include a hexaquark operator, may not have sufficient overlap with a bound state to correctly extract energy levels.
If this were the case, then all of our results would have to systematically shift downwards by several sigma with the inclusion of this otherwise missing operator.
In a forthcoming publication, we will investigate the impact of including such a hexaquark operator in the basis, which has yet to be included due to its numerical cost.

%----------------------------------------------------------
%    dineutron
\subsubsection{Dineutron channel \label{sec:dineutron}}

%    qcotd dineutron
\begin{figure*}
\begin{tabular}{cc}
\includegraphics[width=0.49\textwidth]{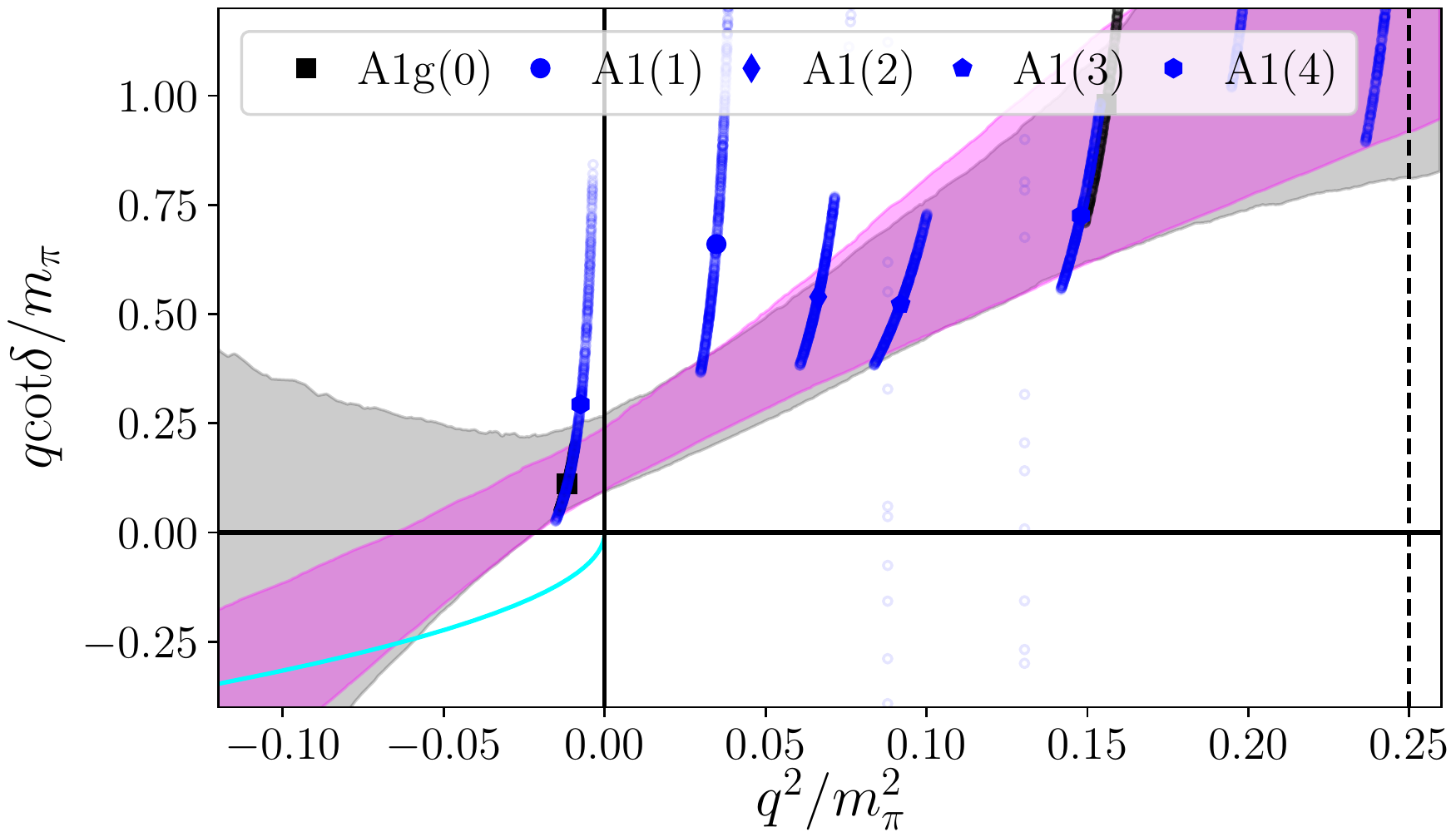}
&
\includegraphics[width=0.49\textwidth]{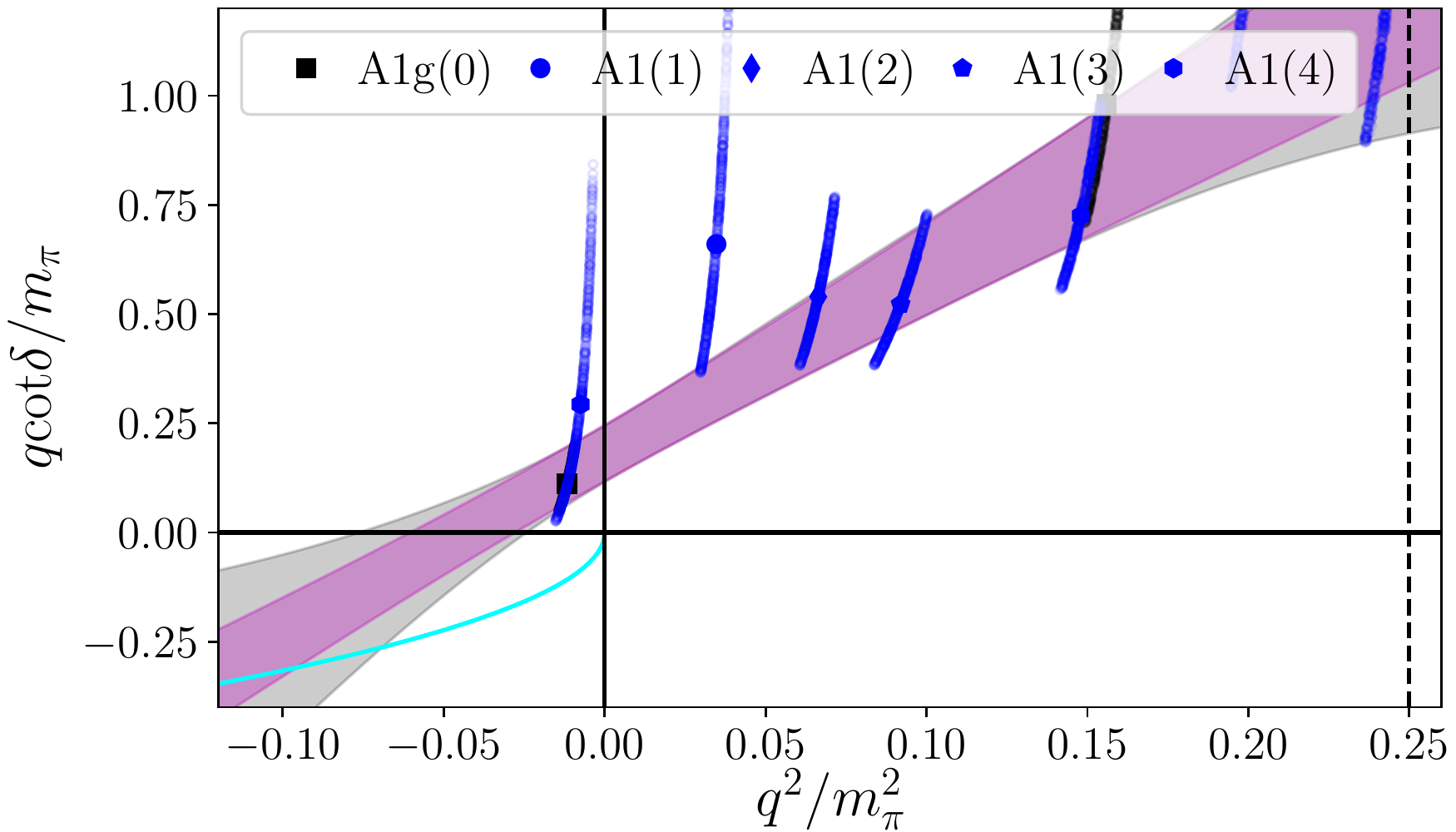}
\\
($q\cot\d$ analysis)& (spectrum analysis)
\end{tabular}
\caption{\label{fig:qcotd_dineutron}
Same as described in the caption of \figref{fig:qcotd_deuteron} except for the $s$-wave dineutron channel.
}
\end{figure*}
%-------------------------------------------------------------------------------

For the dineutron channel, we similarly chose all irreps corresponding to $I=1$, $s=0$, having overlap onto the $s$-wave as the leading contribution at low energies for values of $q^{*} < m_\pi/2$.
After performing a stability analysis, we also observe the same choice of $t_{\rm min}=5$ and $n_{\rm el.}=2$ provides an optimal or near-optimal fit for all irreps.
The irreps and resulting energies and processed values of $q^{*2}$ and $q^* \cot \d$ are given in \tabref{tab:dineutron} in \appref{app:energy_levels}.

The resulting values of $q^* \cot \d$ also suggest minimal partial wave-mixing and a smooth $q^{*2}$ dependence.
The ERE analysis with the two methods described above is displayed in \figref{fig:qcotd_dineutron} and yields the following parameters
{\small
\begin{equation}
\begin{array}{rrlll}
\hline\hline
\textrm{method}& \textrm{order}& {\quad m_\pi a}& {\ m_\pi r_0}& m_\pi^3 r_1\\
\hline
q\cot\d& \textrm{NLO}       &  -6.6(_{-2.6}^{+3.1})& 8.4(_{-2.3}^{+4.4})& -\\[2pt]
\textrm{spec}& \textrm{NLO} &  -5.5(2.0)& 8.4(1.5)& -\\[2pt]
q\cot\d& \textrm{NNLO}      & -6.3(_{3.0}^{+3.2})& 7.5(_{-5.5}^{+5.4})& 14(_{-85}^{+117})\\
\textrm{spec}& \textrm{NNLO}& -5.6(2.0)& 8.7(2.6)& -5(45)\\
\hline
\end{array}
\end{equation}}
Similar to the deuteron, the results are consistent with no bound state and a virtual bound state at
\begin{equation}
\frac{q_-^{\rm dineutron}}{m_\pi} = -i\, 0.121(32)\, .
\end{equation}
Taken together, our results, while not conclusive, strongly disfavor the presence of a bound state in either the deuteron or dineutron channel.

%----------------------------------------------------------
%    Discussion
\section{Discussion and Outlook \label{sec:discussion}}

We have presented the first lattice QCD calculation of two-nucleon systems using the stochastic Laplacian Heaviside method~\cite{Morningstar:2011ka}.
There are only two such two-baryon calculations using a variational operator basis in the literature, the other being an application to the H-dibaryon system and the dineutron system~\cite{Francis:2018qch,Hanlon:2018yfv} using the more common ``distillation'' method~\cite{Peardon:2009gh}.
In this work and Refs.~\cite{Francis:2018qch,Hanlon:2018yfv}, the pion mass is rather heavy. In our case it is set approximately equal to the physical strange quark mass resulting in $m_\pi\approx714$~MeV, and in Ref.~\cite{Francis:2018qch} it corresponds to $m_\pi\approx960$~MeV in an $N_f=2$ calculation.\footnote{Ref.~\cite{Francis:2018qch} also performed calculations at pion masses as low as $m_\pi \approx 436$~Mev, but reliable fits to the phase shift were unattainable.}

Even at the SU(3) flavor-symmetric point with a heavy pion mass, the pion is the lightest propagating degree of freedom emerging from QCD.
It is therefore natural to measure other length scales with respect to the pion mass and possibly natural to expect that the range $R$ of the potential would approximately be given by $m_\pi^{-1}$.
While the scattering length can take on any value (with the unitary limit, $a\rightarrow\infty$, being the crossover between BEC and BCS like systems), the effective range is typically the size of the potential.

In the present work, we have found that in both the dineutron and deuteron channels, the effective range is $r_0 m_\pi\approx5-9$, which is an unusually large value.
The delta-nucleon mass splitting is another small energy scale, and it is found that the splitting decreases with increasing pion mass at a mild rate such that it is $m_\D - m_N\approx200$~MeV at the SU(3) flavor-symmetric point near the physical strange quark mass~\cite{WalkerLoud:2008bp}.
While this is a small energy scale compared to $m_\pi$, it is not clear how this translates into a range of the two-nucleon potential, but one should keep in mind that QCD does naturally produce such an energy scale. NPLQCD similarly found large values of the effective range in their calculations at $m_\pi\approx800$~MeV~\cite{Beane:2013br,Wagman:2017tmp}, though not quite as large.

Causality and unitarity can be used to place a bound on the size of the effective range in terms of the range of the potential~\cite{Wigner:1955zz} with corrections arising from a finite scattering length~\cite{Phillips:1996ae}%
% FOOTNOTE ---------------------------------------------------------------------
\footnote{This formula is derived for a finite range potential, $V(r)=0$ for $r>R$.  Incoprorating corrections from a Yukawa tail was found to slightly reduce the lower bound on $R$ for the physical $^1{\rm S}_0$ channel~\cite{Scaldeferri:1996nx}.}
\begin{equation}\label{eq:r_causality}
r_0 \leq 2 \left[ R - \frac{R^2}{a} + \frac{R^3}{3a^2} \right]\, .
\end{equation}
Since we do not know the range of the potential, $R$, as it is dynamically generated by QCD (and it is not an observable), we can invert this relation and use our determination of the effective range to place a lower bound on $R$.
Using the NLO ERE parameters from the spectrum fit of the deuteron, the real solution of \eqnref{eq:r_causality} provides the limit
\begin{align*}
    &m_\pi R \gtrsim 2.0\, ,&
    &R \gtrsim 0.55 \textrm{ fm}\, ,&
\end{align*}
which is roughly the same or larger than the size of the nucleon: as the pion mass increases, the pion cloud of the nucleon shrinks till the size of the nucleon roughly corresponds to a size $r_N\approx \L_{\rm QCD}^{-1}$, similar to this value.
Perhaps the range of the potential is set by the nucleons coming ``into contact'' with each other.

\subsection{Comparing with the literature}

Several groups~\cite{Beane:2012vq,Yamazaki:2012hi,Beane:2013br,Berkowitz:2015eaa,Orginos:2015aya,Wagman:2017tmp,Francis:2018qch} have used the \luscher method to compute the scattering phase shifts of the two-nucleon systems, deuteron and dineutron, at pion masses larger than 300~MeV. In all cases, except the Mainz group, they have found (deeply) bound states with a reasonable degree of certainty. However, the HAL QCD Collaboration~\cite{HALQCD:2012aa,Inoue:2011ai} has used their potential method to conclude that there are no bound states (again at higher than physical pion masses). Below we discuss possible sources of discrepancy.

\begin{figure}
\includegraphics[width=\columnwidth]{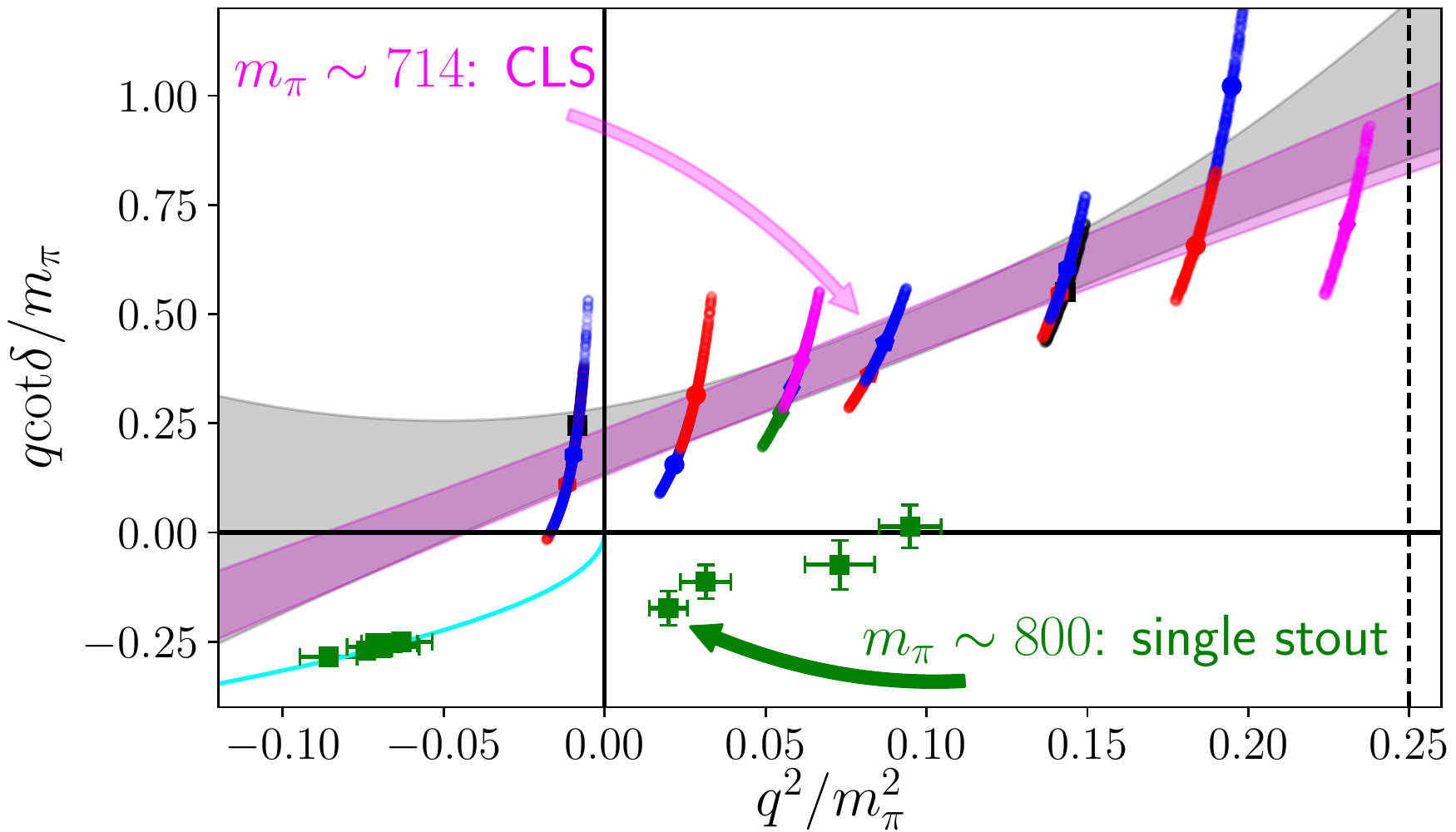}
\caption{\label{fig:qcotd_deuteron_nplqcd}
Values of $q\cot\d$ in the deuteron channel in the present work compared with NPLQCD results at $m_\pi\approx800$~MeV.
While there is some communal expectation that discretization effects should be subdominant, one should be cautious to note that these computations have been performed with different lattice actions at only a single lattice spacing each:
in the present case, the CLS clover-Wilson action~\cite{Bruno:2014jqa} and with NPLQCD, a single-stout smeared~\cite{Morningstar:2003gk}, tadpole improved~\cite{Lepage:1992xa} clover-Wilson action~\cite{Beane:2012vq}.
Assuming the discretizaton effects are relatively small, and that the phase shift does not have a strong pion mass dependence, these results are in conflict.
}
\end{figure}

We will focus our comparison with the results from NPLQCD at $m_\pi\approx800$~MeV~\cite{Beane:2013br,Wagman:2017tmp} as their results are the most similar to ours also being at the SU(3) flavor-symmetric point near the physical strange quark mass.%
% FOOTNOTE
\footnote{NPLQCD also has results at $m_\pi\approx450$~MeV with bound states, however, these results are self-inconsistent as pointed about by HAL QCD~\cite{Iritani:2017rlk} as well as with a low-energy scattering analysis~\cite{Baru:2016evv}, and so we do not compare with them.
The more recent update of the $m_\pi\approx450$~MeV results~\cite{Illa:2020nsi} addresses these issues and leads to larger uncertainties in the constraint of the scattering parameters, though they still identify bound states in the di-nucleon channels.}
%----------------------------------------------------------
They have found that both the deuteron and dineutron channels form bound states with a relatively large binding energy of $B\approx20$~MeV at the SU(3) flavor-symmetric point.  In \figref{fig:qcotd_deuteron_nplqcd} we show our present determination of $q\cot\d$ in the deuteron channel along with the values from Ref.~\cite{Wagman:2017tmp}.%

As is clearly visible from the figure, the results from NPLQCD and the present work are not compatible with each other:
To have a bound state, there must be negative values of $q\cot\d$ at positive values of $q^2$ for the ERE to cross the $-\sqrt{-q^2}$ line with a slope smaller than the tangent to this line~\cite{Iritani:2017rlk}.
Further, we have no evidence of such large negative values of $q^2$ as does NPLQCD (the clustering of (green) points around $q_{\rm cm}^2/m_\pi^2\approx-0.08$).
One should always be cautious comparing results at finite lattice spacing, at least from calculations with different lattice actions.  There is an expectation in the community that discretization effects are a subdominant source of systematic uncertainty.  If this is found to be true (with future work), then there must be another unresolved systematic uncertainty.

%----------------------------------------------------------
%    qcotd deuteron

While our results strongly disfavor the existence of a bound deuteron or dineutron at this pion mass, they are not sufficient to settle the discrepancy in the literature between HAL QCD~\cite{HALQCD:2012aa,Inoue:2011ai}, NPLQCD~\cite{Beane:2012vq,Beane:2013br,Beane:2017edf}, Yamazaki et al.~\cite{Yamazaki:2012hi,Yamazaki:2015asa} and CalLat~\cite{Berkowitz:2015eaa}.
The possible source of the existing discrepancy can be any of the following:
\begin{itemize}[leftmargin=*]
\item There are larger systematic uncertainties in the HAL QCD method and/or the local two-nucleon creation operators typically used by NPLQCD and Yamazaki et al. than are currently understood.  HAL QCD has speculated that these calculations suffer from ``false plateaus'' that arise through an unfortunate linear combination of elastic scattering states~\cite{Iritani:2016jie};

\item Our variational calculation has not utilized a local hexaquark creation/annihilation operator.  It is possible that such a local operator may couple to a deep bound state with a significantly larger overlap such that, without it, the operator basis is not sufficient to identify the state.
If this were the case, then the addition of the hexaquark operator would have to shift the resulting spectrum in all the irreps presented in this work down in a coordinated way that does not spoil the otherwise very smooth $q^2$ dependence observed.

\item None of the calculations have been performed with more than a single lattice spacing and so there could be larger-than-expected discretization effects which prohibit the rigorous identification (or exclusion) of bound states.

\end{itemize}
All of these sources of potential systematic uncertainty may need to be explored in more detail to resolve the discrepancy.  A good first start would be the use of all methods in the literature on the same set of gauge configurations such that one could eliminate all the systematic uncertainties aside from the method in the determination of the spectrum.
Provided the dispersion relation is continuum like in the range of momentum considered, the \luscher method remains valid.
HAL QCD has performed the computation of the $\Xi\Xi$ spectrum and interactions using their potential method and the \luscher method with a local source~\cite{Iritani:2018zbt}, which led them to conclude the local source method has elastic excited state pollution leading to a false plateau.

In a forthcoming publication, we will compare and contrast our present work (with more statistics) to the local source method used by NPLQCD and Yamazaki et al., as well as the displaced nucleons used by CalLat. With both methods, we will implement the HALQCD potential approach such that we can isolate possible sources of systematic uncertainty arising in the method.  We also plan to implement a hexaquark operator into the basis to see how much it shifts the spectrum, if at all.
Resolving this discrepancy is critical if we are to have confidence in the application of LQCD to multinucleon systems, and more importantly for the NP and HEP long-range science goals, to be able to compute the response of few-nucleon systems to SM and BSM currents.
NPLQCD has invested significant effort in computing such matrix elements, see for example the recent review~\cite{Davoudi:2020ngi}, but if the spectrum has been misidentified, it is not clear how much these systematic uncertainties would modify their results and conclusions.

%----------------------------------------------------------
%    ACKNOWLEDGEMENTS
\begin{acknowledgments}
We would like to thank Ra{\'u}l Brice{\~n}o, Evgeny Epelbaum, Daniel Phillips and Bira van Kolck for helpful correspondence regarding the parameterization of the two-nucleon amplitudes.

Computing time for this work was provided through the Innovative and Novel Computational Impact on Theory and Experiment (INCITE) program,  the LLNL Multiprogrammatic and Institutional Computing program for Grand Challenge allocations on the LLNL supercomputers and the Energy Research Computing Allocations Process (ERCAP).
This research utilized the  NVIDIA GPU-accelerated Summit supercomputer at Oak Ridge Leadership Computing Facility at the Oak Ridge National Laboratory, which is supported by the Office of Science of the U.S. Department of Energy under Contract No. DE-AC05-00OR22725,
the Lassen (NVIDIA-GPU), and Vulcan (BG/Q) supercomputers at Lawrence Livermore National Laboratory and resources of the National Energy Research Scientific Computing Center (NERSC), a U.S. Department of Energy Office of Science User Facility located at Lawrence Berkeley National Laboratory, operated under Contract No. DE-AC02-05CH11231.

The computations were performed using the \texttt{chroma\_laph} and \texttt{last\_laph} software suites.
\texttt{chroma\_laph} uses the USQCD \texttt{chroma}~\cite{Edwards:2004sx} library and the \texttt{QDP++} library.
The propagator solves were efficiently performed with \texttt{QUDA}~\cite{Clark:2009wm,Babich:2011np} and much of the sLapH workflow has been ported to \texttt{QUDA} as well.  The contractions were optimized with \texttt{contraction\_optimizer}~\cite{contraction_optimizer}.
The computations were managed with \texttt{METAQ}~\cite{Berkowitz:2017vcp,Berkowitz:2017xna}.
The correlation function analysis was performed with \texttt{lsqfit}~\cite{lsqfit:11.5.1} and \texttt{gvar}~\cite{gvar:11.2} and Sigmond.
The resulting correlation functions will be released in with a future publication that includes a larger number of correlation functions with a full partial-wave analysis.
The phase shift analysis code and resulting bootstrap results of the data presented here are included with the github repository~\cite{laphnn_qcotd}.

This work was supported by the NVIDIA Corporation (MAC), the Alexander von Humboldt Foundation through a Feodor Lynen Research Fellowship (CK), the RIKEN Special Postdoctoral Researcher Program (ER), the U.S. Department of Energy, Office of Science, Office of Nuclear Physics under Award Numbers DE-AC02-05CH11231 (BH, CCC, AWL), DE-AC52-07NA27344 (DH, PV), DE-FG02-93ER-40762 (EB);
the Nuclear Physics Double Beta Decay Topical Collaboration (AN); an LBNL LDRD grant (BH, AWL) and the DOE Early Career Award Program (AWL).
CJM acknowledges support from the U.S.~NSF under award PHY-1913158.

\end{acknowledgments}

\onecolumngrid
\appendix

\section{Energy levels for the deuteron and dineutron \label{app:energy_levels}}

The deuteron and dineutron irreps, extracted energies and processed values of $q^2$ and $q\cot\d$ are provided in \tabref{tab:deuteron} and \tabref{tab:dineutron} respectively.

%-------------------------------------------------------------------------------
% Deuteron Table
\begin{table*}[]
\caption{\label{tab:deuteron}
Energy levels and phase shifts for the deuteron channel.
These are determined with a 2-exponential fit to both the single-nucleon ($t=[5,20]$) and ratio two-nucleon correlation functions ($t=[5,15]$) defined in \eqnref{eq:n_exp} and \eqnref{eq:nn_ratio_exp}.
The total momentum is given by the boost vector $\mathbf{d}$, \eqnref{eq:boost_vec}.
The state indicates the resulting principle correlation function after performing the GEVP.
$\mathbf{d}_{1,2}$ are the squared boost vectors of the individual nucleons used in the ratio correlation function which is fit to determine the interacting energy $\D E_{NN}$ and total energy $E_{NN}$ which is converted to the CoM frame $E^*_{NN}$ and processed to get $q^* \cot \d$.
}
\begin{ruledtabular}
\begin{tabular}{cccccccccccc}
 $\mathbf{d}^2$& irrep& state& $\mathbf{d}_1$& $E_1$& $\mathbf{d}_2$& $E_2$& $\D E_{NN}$& $E_{NN}$& $E_{NN}^{*}$& $q^{*2} / m_\pi^2$& $q^* \cot \delta / m_\pi$\\
\hline
0& T1g& 0& 0& 0.70262(59)& 0& 0.70262(59)& -0.00115(27)& 1.4041(12)& 1.4041(12)& -0.0084(40)& 0.24(18)\\
0& T1g& 1& 1& 0.71459(50)& 1& 0.71459(50)& -0.00439(46)& 1.4248(12)& 1.4248(12)& 0.143(13)& 0.55(16)\\
1&  A2& 0& 0& 0.70272(57)& 1& 0.71462(51)& -0.00306(33)& 1.4143(11)& 1.4082(11)& 0.0217(97)& 0.155(96)\\
1&  A2& 1& 1& 0.71463(50)& 2& 0.72627(49)& -0.00314(40)& 1.4378(11)& 1.4318(11)& 0.195(13)& 1.02(41)\\
1&   E& 0& 0& 0.70271(56)& 1& 0.71458(50)& -0.00208(28)& 1.4152(11)& 1.4091(11)& 0.0284(96)& 0.31(46)\\
1&   E& 1& 1& 0.71454(49)& 2& 0.72617(50)& -0.00446(37)& 1.4362(11)& 1.4303(11)& 0.184(13)& 0.66(17)\\
3&   E& 0& 0& 0.70274(54)& 3& 0.73768(51)& -0.00589(58)& 1.4345(13)& 1.4165(13)& 0.082(13)& 0.362(90)\\
4&   E& 0& 1& 0.71469(51)& 1& 0.71469(51)& -0.00153(23)& 1.4279(11)& 1.4037(11)& -0.012(12)& 0.1(2.5)\\
4&   E& 1& 0& 0.70270(57)& 4& 0.74877(59)& -0.00307(43)& 1.4484(12)& 1.4245(12)& 0.141(11)& 0.55(13)\\
2&  A2& 0& 1& 0.71461(51)& 1& 0.71461(51)& -0.00393(39)& 1.4253(11)& 1.4132(11)& 0.058(11)& 0.33(12)\\
3&  A2& 0& 0& 0.70269(56)& 3& 0.73769(52)& -0.00519(68)& 1.4352(13)& 1.4172(13)& 0.087(14)& 0.43(12)\\
4&  A2& 0& 1& 0.71471(50)& 1& 0.71471(50)& -0.00130(25)& 1.4281(10)& 1.4039(11)& -0.010(12)& 0.2(4.6)\\
4&  A2& 1& 0& 0.70284(56)& 4& 0.74890(57)& -0.00301(44)& 1.4487(12)& 1.4249(12)& 0.144(11)& 0.60(16)\\
2&  B1& 0& 1& 0.71459(51)& 1& 0.71459(51)& -0.00436(37)& 1.4248(12)& 1.4127(12)& 0.055(12)& 0.27(10)\\
2&  B2& 0& 1& 0.71455(52)& 1& 0.71455(52)& -0.00340(40)& 1.4257(12)& 1.4136(12)& 0.061(12)& 0.39(19)\\
2&  B2& 3& 1& 0.71465(50)& 3& 0.73786(50)& -0.00401(57)& 1.4485(12)& 1.4366(12)& 0.231(15)& 0.71(24)\\
\end{tabular}
\end{ruledtabular}
\end{table*}
%-------------------------------------------------------------------------------

%-------------------------------------------------------------------------------
% dineutron Table
\begin{table*}[htp]
\caption{\label{tab:dineutron}
Energy levels and phase shifts for the dineutron channel.  These are determined with a 2-exponential fit to both the single-nucleon ($t=[5,20]$) and ratio two-nucleon correlation functions defined in \eqnref{eq:n_exp} and \eqnref{eq:nn_ratio_exp}.
}
\begin{ruledtabular}
\begin{tabular}{cccccccccccc}
 $P^2$& irrep& state& $N_1$& $E_1$& $N_2$& $E_2$& $\D E_{NN}$& $E_{NN}$& $E_{NN}^{\rm cm}$& $q_{\rm cm}^2 / m_\pi^2$& $q_{\rm cm} \cot \delta / m_\pi$\\
\hline
0& A1g& 0& 0& 0.70259(57)& 0& 0.70259(57)& -0.00161(37)& 1.4036(13)& 1.4036(13)& -0.0117(54)& 0.11(11)\\
0& A1g& 1& 1& 0.71454(51)& 1& 0.71454(51)& -0.00262(53)& 1.4265(13)& 1.4265(13)& 0.156(14)& 0.98(91)\\
1&  A1& 0& 0& 0.70282(56)& 1& 0.71466(51)& -0.00147(27)& 1.4160(11)& 1.4099(11)& 0.0347(95)& 0.7(29.5)\\
1&  A1& 1& 1& 0.71451(51)& 2& 0.72616(51)& -0.00202(47)& 1.4387(12)& 1.4327(12)& 0.202(15)& 1.5(5.5)\\
2&  A1& 0& 1& 0.71452(51)& 1& 0.71452(51)& -0.00271(36)& 1.4263(12)& 1.4143(12)& 0.066(11)& 0.54(27)\\
2&  A1& 3& 1& 0.71461(50)& 3& 0.73784(50)& -0.00226(67)& 1.4502(13)& 1.4383(13)& 0.244(16)& 1.3(4.5)\\
3&  A1& 0& 0& 0.70267(56)& 3& 0.73762(52)& -0.00451(89)& 1.4358(15)& 1.4178(15)& 0.092(17)& 0.52(19)\\
4&  A1& 0& 1& 0.71479(50)& 1& 0.71479(50)& -0.00123(22)& 1.4283(10)& 1.4041(11)& -0.007(11)& 0.3(4.1)\\
4&  A1& 1& 0& 0.70267(57)& 4& 0.74860(60)& -0.00205(48)& 1.4492(13)& 1.4254(14)& 0.148(13)& 0.73(25)\\
\end{tabular}
\end{ruledtabular}
\end{table*}
%-------------------------------------------------------------------------------

\section{Computational and algorithmic optimization\label{app:code}}

Several of the kernels required in the stochastic LapH workflow have been implemented to run on NVIDIA V100 GPUs using the QUDA library~\cite{Clark:2009wm}. We constructed new routines that compute the cross product and contraction of color vectors, as well as specialized routines that compute time-slice reductions. Due to the reduction strategy we employ, the bulk of these contractions is expressed in terms of BLAS3 (matrix-matrix) operations, automatically improving the arithmetic intensity of the computation. These operations take the form
\begin{equation}\label{eq:stridedBatchZGEMM}
C_{i} = A_{0} B_{i}
\end{equation}
for dense matrices $A,B,C$, and batch index $i$. The matrix $A_{0}$ is constant with respect to the batch index. As such, we wrote interfaces in QUDA for the cuBLAS function \texttt{stridedBatchZGEMM} which minimizes data-transfer latency between the host and device by caching the $A_{0}$ matrix. We found that by using the 4 V100 accelerators on a single Lassen (LLNL) node we gained speed-up factors of $\approx$30x over using the host IBM Power-9 CPUs. All the code we constructed specifically for this computation is publicly available in the QUDA GitHub repository. The contraction, time-slice reductions, and cuBLAS interface components are in mainline QUDA.

Overall, correlation function construction requires a large number of tensor contractions, and therefore dedicated computational optimizations are vital. A total of 32,960 correlation functions is computed on each gauge configuration. A strategy to minimize the amount of computational work by optimizing the contraction order as well as re-using common subexpressions is described in Ref.~\cite{Horz:2019rrn}. Following the nomenclature of that reference, 200,370,960 diagrams are left to evaluate after consolidating duplicates in the initial set of 2,052,792,360 diagrams. Eliminating common subexpressions in the set of remaining diagrams reduces the number of computationally dominant contractions with $N_\mathrm{dil}^4$ scaling from 344,163,600 to 9,969,360 for a combined speedup by roughly a factor $350\times$ compared to the naive evaluation of all tensor contractions.

\section{Error propagation for $q\cot(\d)$\label{app:qcot_analysis}}

\subsection{Unbiasedness in Weighted Regression}
In this section we offer a short refresher on the guarantees and assumptions behind classic regression. The statistics regression model specifies the relationship between data $(X, Y)$ and the associated parameters of interest $\beta$ via an additive error:
\begin{eqnarray}\label{eq:linearity}
Y = X\beta + \epsilon
\end{eqnarray}
where $Y$ is a $n \times 1$ vector, $X$ is a $n \times p$ matrix, $\beta$ is a $p \times 1$ vector, and $\epsilon$ is a $n \times 1$ random vector. Intuitively, $n$ is the number of data points and $p$ is the number of coefficients in the regression. The usual inference task is to infer $\beta$ using the noisy observed values of $Y$. Any estimate for $\beta$ is often denoted as $\hat{\beta}$ and an estimate is unbiased if $E(\hat{\beta}) = \beta$.

The weighted regression estimate minimizes the squared loss between the $Y$ values and the inferred line $X\hat{\beta}$, $\hat{\beta}_{reg} = \arg\min \|W^{1/2}(Y - X\hat{\beta})\|^2 = (X^TWX)^{-1}X^TWY$. We will show that $\hat{\beta}_{reg}$ is unbiased if \eqnref{eq:linearity} and the following assumption holds:
\begin{eqnarray}\label{eq:cond_mean0}
E(\epsilon|X) = 0
\end{eqnarray}
The addition of mean 0 error in the $Y$ dimension intuitively justifies why minimizing the squared loss in $Y$ alone would yield the unbiased results. Mathematically the derivation further highlights the assumption on treating $X$ as given implied in both \eqnref{eq:linearity} and \eqnref{eq:cond_mean0}
\begin{align}
E(\hat{\beta}_{\rm reg}|X) &= E((X^TWX)^{-1}X^TWY|X) \\
&= E((X^TWX)^{-1}X^TW(X\beta + \epsilon)|X) \\
&= \beta + (X^TWX)^{-1}X^TW E(\epsilon|X) \\
&= \beta
\end{align}
The most common choice for $W = \Sigma^{-1}$ where ${\rm Cov}(\epsilon|X) = \Sigma$ but its choice affects the variance for $\hat{\beta}$ instead of its unbiasedness. It is also worth pointing out that Normality nor symmetry in $\epsilon$ are necessary for the unbiasedness to hold.

\subsection{Assumptions for unbiasedness are not met}
In this section we explain the possible errors in fitting the classic weighted least squares for our curve fitting exercise at hand.
The first violation is the existence of error in the horizontal axis for each data point. Given the derivations above, if $X$ is measured with error as well, our objective would incorporates both error in $X$ and $Y$ instead of solely minimizing errors in the vertical axis.

The second deviation from the classic setting is the strong nonlinear relationship between the errors in $X$ and $Y$ within each irrep. Although we have errors in both axes, knowing the error in one dimension allows us to infer the error in the other dimension. The seemingly 2 dimensional error is therefore more appropriately modeled as having a single source of variability.

\subsection{Our modified weighted least squares}
\begin{figure}
\includegraphics[width=0.6\textwidth]{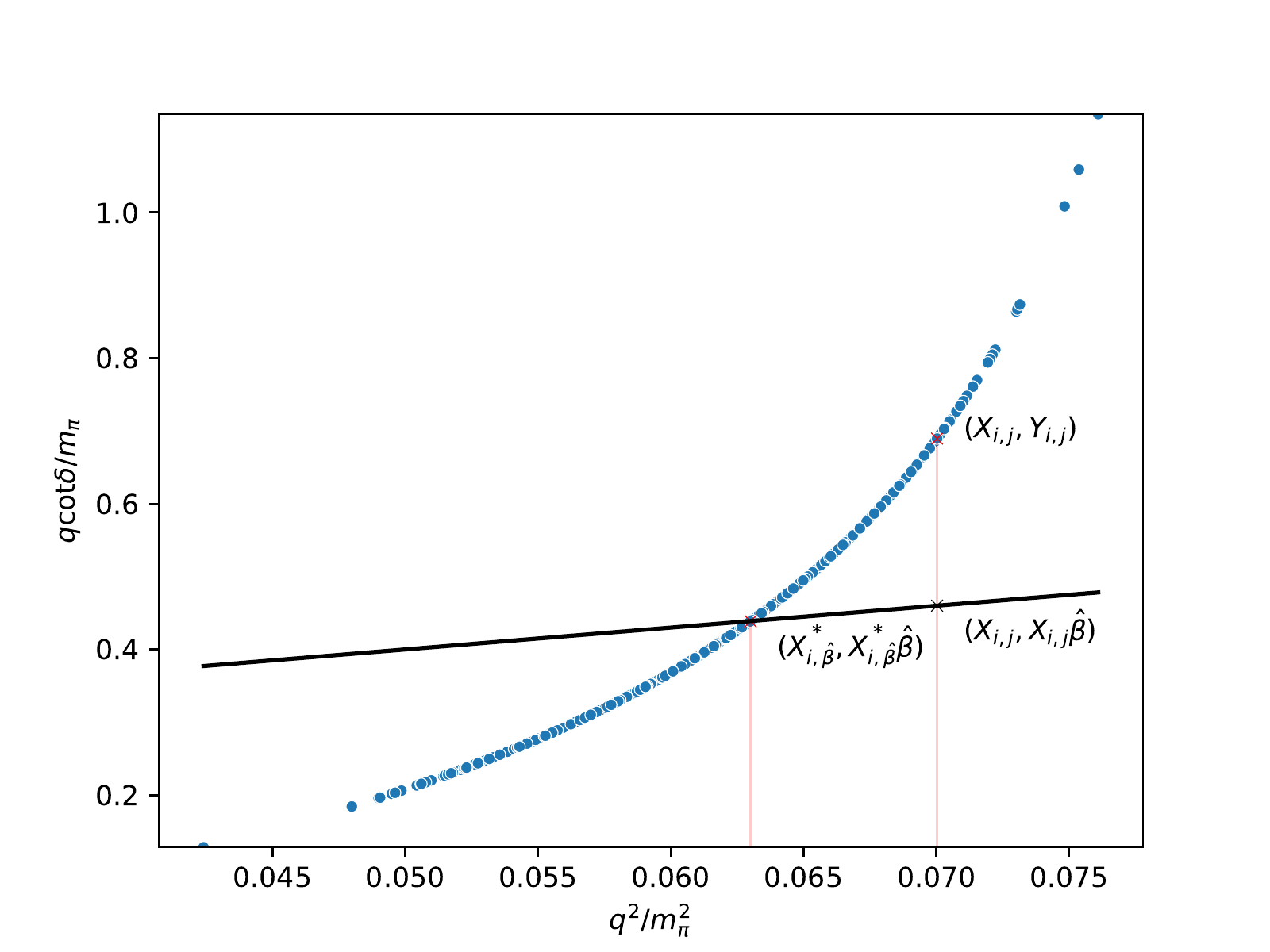}
\caption{\label{fig:mod_wls_demo}
Plotting the bootstrap samples within a single irrep, demonstrating the difference between the modified distance vs the usual regression
}
\end{figure}

Our modification essentially converts the problem at hand into the classical settings by re-defining the error in terms of squared distance along the curve rather than the vertical axis. Fig.~\ref{fig:mod_wls_demo} demonstrates this modified distance between a data point and any candidate regression line, implied by $\hat{\beta}$ is the distance along the blue curve between the red points $X_{i,\hat{\beta}}^*$ to $X_{i,j}$. Classic regression, however, computes the vertical distance between $Y_{i,j}$ and $X_{i,j}\hat{\beta}$ which forces the regression line towards implausible values.

Let $f_i$ denote the functional relationship between $X_{i,\cdot}$ and $Y_{i,\cdot}$ for irrep $i$ so that $Y_{i,\cdot} = f_i(X_{i,\cdot})$. The general equation for the length along a differentiable curve between 2 points is $s(f'_i, a, b) = |\int_a^b \sqrt{1 + f'_i(X)^2} dx|$. Let the point of intersection between $f_i$ and the regression line will be denoted as $(X_{i,\hat{\beta}}^*, X_{i,\hat{\beta}}^* \hat{\beta})$. An unweighted least square penalty would then be $\sum_i s(f'_i, X_{i, \hat{\beta}}, X_{i,j})^2$, where $j$ denotes the index for different bootstrap values.

To weigh the different irreps, we estimate the covariance using a similar approach as in the Delta Method~\cite{bickel2001mathematical}. The Delta Method states that if $(\bar{X} - \mu_X) \Rightarrow N(0, \Sigma_X)$, then for a differentiable $f$ we have $f(\bar{X}) - f(\mu_X) \Rightarrow N(0, \nabla f(\mu_X)^T \Sigma_X \nabla f(\mu_X))$. We recycle the same $f'$ from calculating the lengths before, and we estimate the covariance of $X$ treating each irrep as a different dimension, specifically:
\begin{align}
\hat{\Sigma}_X &= \frac{1}{n-p}\sum_j \begin{bmatrix} X_{1, j} - \bar{X}_1\\ \vdots \\ X_{k,j} - \bar{X}_k \end{bmatrix} \begin{bmatrix} X_{1, j} - \bar{X}_1& \dots & X_{k,j} - \bar{X}_k \end{bmatrix} \\
\nabla f(X_{\cdot, j}) &= \begin{bmatrix} f'_1(X_{1,j}) & \dots & 0 \\ \vdots & \ddots & \vdots \\ 0 & \dots & f'_k(X_{k,j}) \end{bmatrix}\\
W_j &= \left[\nabla f(X_{\cdot, j})^T \hat{\Sigma}_X \nabla f(X_{\cdot, j}) \right]^{-1}
\end{align}
where $\bar{X}_i = \frac{1}{n} \sum_j X_{i,j}$ is the average over the bootstraps, $k$ is the total number of irreps, and $W$ is the weights we will use to scale the modified errors.
It is worth noting that the Delta Method is more practical than directly estimating the covariance empirically because the the distance along the curve between several bootstrap
samples on certain irreps are infinite if they are across the critical point of the $cot$ function. These points make empirically estimating the covariance of $Y$ infeasible.

Our optimization for each bootstrap $j$, is to solve
\begin{eqnarray}\label{eq:mod_obj}
\hat{\beta}_j = \arg\min_{\beta} L(f'_{\cdot}, \hat{\beta}, X_{\cdot, j})^T W_j L(f'_{\cdot}, \hat{\beta}, X_{\cdot, j})
\end{eqnarray}
where $L(f'_{\cdot}, \hat{\beta}, X_{\cdot, j}) = \begin{bmatrix} s(f'_1, X_{1, \hat{\beta}}, X_{1,j}) \\ \vdots \\ s(f'_k, X_{k, \hat{\beta}}, X_{k,j}) \end{bmatrix}$.

The estimation of $f_i$ and $f'_i$ were done using splines which are piece-wise cubic polynomials that ensure the derivative is continuous.

\twocolumngrid

\bibliography{nn}

\end{document}

%% file: preamble.tex
\usepackage{graphicx}  % needed for figures
\usepackage{dcolumn}   % needed for some tables
\usepackage{bm}        % for math
\usepackage{amssymb}   % for math
\usepackage{xspace}
\usepackage{standalone}
\usepackage{enumitem}
\usepackage[pdftex]{color}
\usepackage{xcolor}
\usepackage{slashed}
\usepackage{booktabs}
\usepackage{multirow}
\usepackage{amsmath}
\usepackage{stackrel}
\usepackage{rotating}
\usepackage{CJKutf8}
\usepackage{pifont}
\usepackage{mathtools}
\usepackage{hyperref}
\hypersetup{
    colorlinks=true,       % false: boxed links; true: colored links
    linkcolor=blue,          % color of internal links
    citecolor=blue,        % color of links to bibliography
    filecolor=blue,      % color of file links
    urlcolor=blue           % color of external links
}
\usepackage{simplewick}
\usepackage{float} % Forces placement of figures
\usepackage{tikz} % adds /foreach (looping) command

%% file: def.tex
\def\eqref#1{{(\ref{#1})}}
\newcommand{\eqnref}[1]{Eq.~\eqref{#1}}

\newcommand{\figref}[1]{Fig.~\ref{#1}}

\newcommand{\secref}[1]{Sec.~\ref{#1}}
\newcommand{\appref}[1]{Appendix~\ref{#1}}
\newcommand{\tabref}[1]{Table~\ref{#1}}

\definecolor{kngrey}{HTML}{A6AAA9}
\definecolor{knred}{HTML}{EC5D57}
\definecolor{knorange}{HTML}{F39019}
\definecolor{knyellow}{HTML}{F5D328}
\definecolor{kngreen}{HTML}{70BF41}
\definecolor{knblue}{HTML}{51A7F9}
\definecolor{knpurple}{HTML}{B36AE2}

\def\b{{\beta}}
\def\d{{\delta}}
\def\D{{\Delta}}

\def\g{{\gamma}}

\def\L{{\Lambda}}

\def\O{{\Omega}}
\def\S{{\Sigma}}

\def\luscher{L{\"u}scher\xspace}

%% file: affiliations.tex
\newcommand{\columbia}{
Department of Statistics, Columbia University,
New York, NY 10027, USA
}
\newcommand{\ithems}{
RIKEN iTHEMS,
Wako, Saitama 351-0198, Japan
}
\newcommand{\bochum}{
    Institut f{\"u}r Theoretische Physik II,
    Ruhr-Universit{\"a}t Bochum, D-44780 Bochum, Germany
}
\newcommand{\cmu}{
    Department of Physics, Carnegie Mellon University,
    Pittsburgh, Pennsylvania 15213, USA
}
\newcommand{\cpthree}{
    CP$^3$-Origins \& Dept. of Mathematics and Computer Science,
    University of Southern Denmark Campusvej 55,
    5230 Odense M, Denmark
    }
\newcommand{\arithmer}{
    Arithmer Inc., R\&D Headquarters,
    Minato, Tokyo 106-6040, Japan
}

\newcommand{\lblnsd}{
    Nuclear Science Division,
    Lawrence Berkeley National Laboratory,
	Berkeley, CA 94720, USA
	}

\newcommand{\llnl}{
	Physics Division,
	Lawrence Livermore National Laboratory,
	Livermore, CA 94550, USA
	}

\newcommand{\mainz}{
    Helmholtz Institute Mainz,
    55099 Mainz, Germany;\\
    GSI Helmholtzzentrum f\"{u}r Schwerionenforschung,
    64291 Darmstadt, Germany
    }
\newcommand{\nvidia}{
    NVIDIA Corporation,
    2701 San Tomas Expressway, Santa Clara, CA 95050, USA
    }

\newcommand{\ucb}{
	Department of Physics,
	University of California,
	Berkeley, CA 94720, USA
	}

\newcommand{\mcfp}{
	Maryland Center for Fundamental Physics,
	University of Maryland,
	College Park, MD 20742, USA
}

\newcommand{\unc}{
	Department of Physics and Astronomy,
	University of North Carolina,
	Chapel Hill, NC 27516-3255, USA
	}

%% file: nn_su3_slaph_firstlook.bbl
%merlin.mbs apsrev4-1.bst 2010-07-25 4.21a (PWD, AO, DPC) hacked
%Control: key (0)
%Control: author (0) dotless jnrlst
%Control: editor formatted (1) identically to author
%Control: production of article title (0) allowed
%Control: page (1) range
%Control: year (0) verbatim
%Control: production of eprint (0) enabled
\begin{thebibliography}{96}%
\makeatletter
\providecommand \@ifxundefined [1]{%
 \@ifx{#1\undefined}
}%
\providecommand \@ifnum [1]{%
 \ifnum #1\expandafter \@firstoftwo
 \else \expandafter \@secondoftwo
 \fi
}%
\providecommand \@ifx [1]{%
 \ifx #1\expandafter \@firstoftwo
 \else \expandafter \@secondoftwo
 \fi
}%
\providecommand \natexlab [1]{#1}%
\providecommand \enquote  [1]{``#1''}%
\providecommand \bibnamefont  [1]{#1}%
\providecommand \bibfnamefont [1]{#1}%
\providecommand \citenamefont [1]{#1}%
\providecommand \href@noop [0]{\@secondoftwo}%
\providecommand \href [0]{\begingroup \@sanitize@url \@href}%
\providecommand \@href[1]{\@@startlink{#1}\@@href}%
\providecommand \@@href[1]{\endgroup#1\@@endlink}%
\providecommand \@sanitize@url [0]{\catcode `\\12\catcode `\$12\catcode
  `\&12\catcode `\#12\catcode `\^12\catcode `\_12\catcode `\%12\relax}%
\providecommand \@@startlink[1]{}%
\providecommand \@@endlink[0]{}%
\providecommand \url  [0]{\begingroup\@sanitize@url \@url }%
\providecommand \@url [1]{\endgroup\@href {#1}{\urlprefix }}%
\providecommand \urlprefix  [0]{URL }%
\providecommand \Eprint [0]{\href }%
\providecommand \doibase [0]{http://dx.doi.org/}%
\providecommand \selectlanguage [0]{\@gobble}%
\providecommand \bibinfo  [0]{\@secondoftwo}%
\providecommand \bibfield  [0]{\@secondoftwo}%
\providecommand \translation [1]{[#1]}%
\providecommand \BibitemOpen [0]{}%
\providecommand \bibitemStop [0]{}%
\providecommand \bibitemNoStop [0]{.\EOS\space}%
\providecommand \EOS [0]{\spacefactor3000\relax}%
\providecommand \BibitemShut  [1]{\csname bibitem#1\endcsname}%
\let\auto@bib@innerbib\@empty
%</preamble>
\bibitem [{\citenamefont {Drischler}\ \emph {et~al.}(2019)\citenamefont
  {Drischler}, \citenamefont {Haxton}, \citenamefont {McElvain}, \citenamefont
  {Mereghetti}, \citenamefont {Nicholson}, \citenamefont {Vranas},\ and\
  \citenamefont {Walker-Loud}}]{Drischler:2019xuo}%
  \BibitemOpen
  \bibfield  {author} {\bibinfo {author} {\bibfnamefont {Christian}\
  \bibnamefont {Drischler}}, \bibinfo {author} {\bibfnamefont {Wick}\
  \bibnamefont {Haxton}}, \bibinfo {author} {\bibfnamefont {Kenneth}\
  \bibnamefont {McElvain}}, \bibinfo {author} {\bibfnamefont {Emanuele}\
  \bibnamefont {Mereghetti}}, \bibinfo {author} {\bibfnamefont {Amy}\
  \bibnamefont {Nicholson}}, \bibinfo {author} {\bibfnamefont {Pavlos}\
  \bibnamefont {Vranas}}, \ and\ \bibinfo {author} {\bibfnamefont {Andr{\'e}}\
  \bibnamefont {Walker-Loud}},\ }\bibfield  {title} {\enquote {\bibinfo {title}
  {{Towards grounding nuclear physics in QCD}},}\ }\href@noop {} {\  (\bibinfo
  {year} {2019})},\ \Eprint {http://arxiv.org/abs/1910.07961} {arXiv:1910.07961
  [nucl-th]} \BibitemShut {NoStop}%
\bibitem [{\citenamefont {Tews}\ \emph {et~al.}(2020)\citenamefont {Tews},
  \citenamefont {Davoudi}, \citenamefont {Ekstr\"om}, \citenamefont {Holt},\
  and\ \citenamefont {Lynn}}]{Tews:2020hgp}%
  \BibitemOpen
  \bibfield  {author} {\bibinfo {author} {\bibfnamefont {Ingo}\ \bibnamefont
  {Tews}}, \bibinfo {author} {\bibfnamefont {Zohreh}\ \bibnamefont {Davoudi}},
  \bibinfo {author} {\bibfnamefont {Andreas}\ \bibnamefont {Ekstr\"om}},
  \bibinfo {author} {\bibfnamefont {Jason~D.}\ \bibnamefont {Holt}}, \ and\
  \bibinfo {author} {\bibfnamefont {Joel~E.}\ \bibnamefont {Lynn}},\ }\bibfield
   {title} {\enquote {\bibinfo {title} {{New Ideas in Constraining Nuclear
  Forces}},}\ }\href {\doibase 10.1088/1361-6471/ab9079} {\bibfield  {journal}
  {\bibinfo  {journal} {J. Phys. G}\ }\textbf {\bibinfo {volume} {47}},\
  \bibinfo {pages} {103001} (\bibinfo {year} {2020})},\ \Eprint
  {http://arxiv.org/abs/2001.03334} {arXiv:2001.03334 [nucl-th]} \BibitemShut
  {NoStop}%
\bibitem [{\citenamefont {Cirigliano}\ \emph {et~al.}(2020)\citenamefont
  {Cirigliano}, \citenamefont {Detmold}, \citenamefont {Nicholson},\ and\
  \citenamefont {Shanahan}}]{Cirigliano:2020yhp}%
  \BibitemOpen
  \bibfield  {author} {\bibinfo {author} {\bibfnamefont {Vincenzo}\
  \bibnamefont {Cirigliano}}, \bibinfo {author} {\bibfnamefont {William}\
  \bibnamefont {Detmold}}, \bibinfo {author} {\bibfnamefont {Amy}\ \bibnamefont
  {Nicholson}}, \ and\ \bibinfo {author} {\bibfnamefont {Phiala}\ \bibnamefont
  {Shanahan}},\ }\bibfield  {title} {\enquote {\bibinfo {title} {{Lattice QCD
  Inputs for Nuclear Double Beta Decay}},}\ }\href {\doibase
  10.1016/j.ppnp.2020.103771} {\bibfield  {journal} {\bibinfo  {journal}
  {Progress in Particle and Nuclear Physics}\ }\textbf {\bibinfo {volume}
  {112}},\ \bibinfo {pages} {103771} (\bibinfo {year} {2020})},\ \Eprint
  {http://arxiv.org/abs/2003.08493} {arXiv:2003.08493 [nucl-th]} \BibitemShut
  {NoStop}%
\bibitem [{\citenamefont {Fukugita}\ \emph {et~al.}(1995)\citenamefont
  {Fukugita}, \citenamefont {Kuramashi}, \citenamefont {Okawa}, \citenamefont
  {Mino},\ and\ \citenamefont {Ukawa}}]{Fukugita:1994ve}%
  \BibitemOpen
  \bibfield  {author} {\bibinfo {author} {\bibfnamefont {M.}~\bibnamefont
  {Fukugita}}, \bibinfo {author} {\bibfnamefont {Y.}~\bibnamefont {Kuramashi}},
  \bibinfo {author} {\bibfnamefont {M.}~\bibnamefont {Okawa}}, \bibinfo
  {author} {\bibfnamefont {H.}~\bibnamefont {Mino}}, \ and\ \bibinfo {author}
  {\bibfnamefont {A.}~\bibnamefont {Ukawa}},\ }\bibfield  {title} {\enquote
  {\bibinfo {title} {{Hadron scattering lengths in lattice QCD}},}\ }\href
  {\doibase 10.1103/PhysRevD.52.3003} {\bibfield  {journal} {\bibinfo
  {journal} {Phys. Rev. D}\ }\textbf {\bibinfo {volume} {52}},\ \bibinfo
  {pages} {3003--3023} (\bibinfo {year} {1995})},\ \Eprint
  {http://arxiv.org/abs/hep-lat/9501024} {arXiv:hep-lat/9501024} \BibitemShut
  {NoStop}%
\bibitem [{\citenamefont {Beane}\ \emph {et~al.}(2006)\citenamefont {Beane},
  \citenamefont {Bedaque}, \citenamefont {Orginos},\ and\ \citenamefont
  {Savage}}]{Beane:2006mx}%
  \BibitemOpen
  \bibfield  {author} {\bibinfo {author} {\bibfnamefont {S.R.}\ \bibnamefont
  {Beane}}, \bibinfo {author} {\bibfnamefont {P.F.}\ \bibnamefont {Bedaque}},
  \bibinfo {author} {\bibfnamefont {K.}~\bibnamefont {Orginos}}, \ and\
  \bibinfo {author} {\bibfnamefont {M.J.}\ \bibnamefont {Savage}},\ }\bibfield
  {title} {\enquote {\bibinfo {title} {{Nucleon-nucleon scattering from
  fully-dynamical lattice QCD}},}\ }\href {\doibase
  10.1103/PhysRevLett.97.012001} {\bibfield  {journal} {\bibinfo  {journal}
  {Phys. Rev. Lett.}\ }\textbf {\bibinfo {volume} {97}},\ \bibinfo {pages}
  {012001} (\bibinfo {year} {2006})},\ \Eprint
  {http://arxiv.org/abs/hep-lat/0602010} {arXiv:hep-lat/0602010} \BibitemShut
  {NoStop}%
\bibitem [{\citenamefont {Beane}\ \emph {et~al.}(2011)\citenamefont {Beane}
  \emph {et~al.}}]{Beane:2010hg}%
  \BibitemOpen
  \bibfield  {author} {\bibinfo {author} {\bibfnamefont {S.R.}\ \bibnamefont
  {Beane}} \emph {et~al.} (\bibinfo {collaboration} {NPLQCD}),\ }\bibfield
  {title} {\enquote {\bibinfo {title} {{Evidence for a Bound H-dibaryon from
  Lattice QCD}},}\ }\href {\doibase 10.1103/PhysRevLett.106.162001} {\bibfield
  {journal} {\bibinfo  {journal} {Phys. Rev. Lett.}\ }\textbf {\bibinfo
  {volume} {106}},\ \bibinfo {pages} {162001} (\bibinfo {year} {2011})},\
  \Eprint {http://arxiv.org/abs/1012.3812} {arXiv:1012.3812 [hep-lat]}
  \BibitemShut {NoStop}%
\bibitem [{\citenamefont {Inoue}\ \emph {et~al.}(2011)\citenamefont {Inoue},
  \citenamefont {Ishii}, \citenamefont {Aoki}, \citenamefont {Doi},
  \citenamefont {Hatsuda}, \citenamefont {Ikeda}, \citenamefont {Murano},
  \citenamefont {Nemura},\ and\ \citenamefont {Sasaki}}]{Inoue:2010es}%
  \BibitemOpen
  \bibfield  {author} {\bibinfo {author} {\bibfnamefont {Takashi}\ \bibnamefont
  {Inoue}}, \bibinfo {author} {\bibfnamefont {Noriyoshi}\ \bibnamefont
  {Ishii}}, \bibinfo {author} {\bibfnamefont {Sinya}\ \bibnamefont {Aoki}},
  \bibinfo {author} {\bibfnamefont {Takumi}\ \bibnamefont {Doi}}, \bibinfo
  {author} {\bibfnamefont {Tetsuo}\ \bibnamefont {Hatsuda}}, \bibinfo {author}
  {\bibfnamefont {Yoichi}\ \bibnamefont {Ikeda}}, \bibinfo {author}
  {\bibfnamefont {Keiko}\ \bibnamefont {Murano}}, \bibinfo {author}
  {\bibfnamefont {Hidekatsu}\ \bibnamefont {Nemura}}, \ and\ \bibinfo {author}
  {\bibfnamefont {Kenji}\ \bibnamefont {Sasaki}} (\bibinfo {collaboration} {HAL
  QCD}),\ }\bibfield  {title} {\enquote {\bibinfo {title} {{Bound H-dibaryon in
  Flavor SU(3) Limit of Lattice QCD}},}\ }\href {\doibase
  10.1103/PhysRevLett.106.162002} {\bibfield  {journal} {\bibinfo  {journal}
  {Phys. Rev. Lett.}\ }\textbf {\bibinfo {volume} {106}},\ \bibinfo {pages}
  {162002} (\bibinfo {year} {2011})},\ \Eprint {http://arxiv.org/abs/1012.5928}
  {arXiv:1012.5928 [hep-lat]} \BibitemShut {NoStop}%
\bibitem [{\citenamefont {Lepage}(1989)}]{Lepage:1989hd}%
  \BibitemOpen
  \bibfield  {author} {\bibinfo {author} {\bibfnamefont {G.~Peter}\
  \bibnamefont {Lepage}},\ }\bibfield  {title} {\enquote {\bibinfo {title}
  {{The Analysis of Algorithms for Lattice Field Theory}},}\ }in\ \href@noop {}
  {\emph {\bibinfo {booktitle} {{Boulder ASI 1989:97-120}}}}\ (\bibinfo {year}
  {1989})\ pp.\ \bibinfo {pages} {97--120}\BibitemShut {NoStop}%
%%CITATION = CLNS-89-971;%%
\bibitem [{\citenamefont {Luscher}(1986)}]{Luscher:1986pf}%
  \BibitemOpen
  \bibfield  {author} {\bibinfo {author} {\bibfnamefont {M.}~\bibnamefont
  {Luscher}},\ }\bibfield  {title} {\enquote {\bibinfo {title} {{Volume
  Dependence of the Energy Spectrum in Massive Quantum Field Theories. 2.
  Scattering States}},}\ }\href {\doibase 10.1007/BF01211097} {\bibfield
  {journal} {\bibinfo  {journal} {Commun. Math. Phys.}\ }\textbf {\bibinfo
  {volume} {105}},\ \bibinfo {pages} {153--188} (\bibinfo {year}
  {1986})}\BibitemShut {NoStop}%
\bibitem [{\citenamefont {Luscher}(1991)}]{Luscher:1990ux}%
  \BibitemOpen
  \bibfield  {author} {\bibinfo {author} {\bibfnamefont {Martin}\ \bibnamefont
  {Luscher}},\ }\bibfield  {title} {\enquote {\bibinfo {title} {{Two particle
  states on a torus and their relation to the scattering matrix}},}\ }\href
  {\doibase 10.1016/0550-3213(91)90366-6} {\bibfield  {journal} {\bibinfo
  {journal} {Nucl. Phys. B}\ }\textbf {\bibinfo {volume} {354}},\ \bibinfo
  {pages} {531--578} (\bibinfo {year} {1991})}\BibitemShut {NoStop}%
\bibitem [{\citenamefont {Yamazaki}\ \emph {et~al.}(2015)\citenamefont
  {Yamazaki}, \citenamefont {Ishikawa}, \citenamefont {Kuramashi},\ and\
  \citenamefont {Ukawa}}]{Yamazaki:2015asa}%
  \BibitemOpen
  \bibfield  {author} {\bibinfo {author} {\bibfnamefont {Takeshi}\ \bibnamefont
  {Yamazaki}}, \bibinfo {author} {\bibfnamefont {Ken-ichi}\ \bibnamefont
  {Ishikawa}}, \bibinfo {author} {\bibfnamefont {Yoshinobu}\ \bibnamefont
  {Kuramashi}}, \ and\ \bibinfo {author} {\bibfnamefont {Akira}\ \bibnamefont
  {Ukawa}},\ }\bibfield  {title} {\enquote {\bibinfo {title} {{Study of quark
  mass dependence of binding energy for light nuclei in 2+1 flavor lattice
  QCD}},}\ }\href {\doibase 10.1103/PhysRevD.92.014501} {\bibfield  {journal}
  {\bibinfo  {journal} {Phys. Rev. D}\ }\textbf {\bibinfo {volume} {92}},\
  \bibinfo {pages} {014501} (\bibinfo {year} {2015})},\ \Eprint
  {http://arxiv.org/abs/1502.04182} {arXiv:1502.04182 [hep-lat]} \BibitemShut
  {NoStop}%
\bibitem [{\citenamefont {Beane}\ \emph {et~al.}(2009)\citenamefont {Beane},
  \citenamefont {Detmold}, \citenamefont {Luu}, \citenamefont {Orginos},
  \citenamefont {Parreno}, \citenamefont {Savage}, \citenamefont {Torok},\ and\
  \citenamefont {Walker-Loud}}]{Beane:2009gs}%
  \BibitemOpen
  \bibfield  {author} {\bibinfo {author} {\bibfnamefont {Silas~R.}\
  \bibnamefont {Beane}}, \bibinfo {author} {\bibfnamefont {William}\
  \bibnamefont {Detmold}}, \bibinfo {author} {\bibfnamefont {Thomas~C}\
  \bibnamefont {Luu}}, \bibinfo {author} {\bibfnamefont {Kostas}\ \bibnamefont
  {Orginos}}, \bibinfo {author} {\bibfnamefont {Assumpta}\ \bibnamefont
  {Parreno}}, \bibinfo {author} {\bibfnamefont {Martin~J.}\ \bibnamefont
  {Savage}}, \bibinfo {author} {\bibfnamefont {Aaron}\ \bibnamefont {Torok}}, \
  and\ \bibinfo {author} {\bibfnamefont {Andre}\ \bibnamefont {Walker-Loud}},\
  }\bibfield  {title} {\enquote {\bibinfo {title} {{High Statistics Analysis
  using Anisotropic Clover Lattices. II. Three-Baryon Systems}},}\ }\href
  {\doibase 10.1103/PhysRevD.80.074501} {\bibfield  {journal} {\bibinfo
  {journal} {Phys. Rev. D}\ }\textbf {\bibinfo {volume} {80}},\ \bibinfo
  {pages} {074501} (\bibinfo {year} {2009})},\ \Eprint
  {http://arxiv.org/abs/0905.0466} {arXiv:0905.0466 [hep-lat]} \BibitemShut
  {NoStop}%
\bibitem [{\citenamefont {Beane}\ \emph
  {et~al.}(2013{\natexlab{a}})\citenamefont {Beane}, \citenamefont {Chang},
  \citenamefont {Cohen}, \citenamefont {Detmold}, \citenamefont {Lin},
  \citenamefont {Luu}, \citenamefont {Orginos}, \citenamefont {Parreno},
  \citenamefont {Savage},\ and\ \citenamefont {Walker-Loud}}]{Beane:2012vq}%
  \BibitemOpen
  \bibfield  {author} {\bibinfo {author} {\bibfnamefont {S.R.}\ \bibnamefont
  {Beane}}, \bibinfo {author} {\bibfnamefont {E.}~\bibnamefont {Chang}},
  \bibinfo {author} {\bibfnamefont {S.D.}\ \bibnamefont {Cohen}}, \bibinfo
  {author} {\bibfnamefont {William}\ \bibnamefont {Detmold}}, \bibinfo {author}
  {\bibfnamefont {H.W.}\ \bibnamefont {Lin}}, \bibinfo {author} {\bibfnamefont
  {T.C.}\ \bibnamefont {Luu}}, \bibinfo {author} {\bibfnamefont
  {K.}~\bibnamefont {Orginos}}, \bibinfo {author} {\bibfnamefont
  {A.}~\bibnamefont {Parreno}}, \bibinfo {author} {\bibfnamefont {M.J.}\
  \bibnamefont {Savage}}, \ and\ \bibinfo {author} {\bibfnamefont
  {A.}~\bibnamefont {Walker-Loud}} (\bibinfo {collaboration} {NPLQCD}),\
  }\bibfield  {title} {\enquote {\bibinfo {title} {{Light Nuclei and
  Hypernuclei from Quantum Chromodynamics in the Limit of SU(3) Flavor
  Symmetry}},}\ }\href {\doibase 10.1103/PhysRevD.87.034506} {\bibfield
  {journal} {\bibinfo  {journal} {Phys. Rev. D}\ }\textbf {\bibinfo {volume}
  {87}},\ \bibinfo {pages} {034506} (\bibinfo {year} {2013}{\natexlab{a}})},\
  \Eprint {http://arxiv.org/abs/1206.5219} {arXiv:1206.5219 [hep-lat]}
  \BibitemShut {NoStop}%
\bibitem [{\citenamefont {Barnea}\ \emph {et~al.}(2015)\citenamefont {Barnea},
  \citenamefont {Contessi}, \citenamefont {Gazit}, \citenamefont {Pederiva},\
  and\ \citenamefont {van Kolck}}]{Barnea:2013uqa}%
  \BibitemOpen
  \bibfield  {author} {\bibinfo {author} {\bibfnamefont {N.}~\bibnamefont
  {Barnea}}, \bibinfo {author} {\bibfnamefont {L.}~\bibnamefont {Contessi}},
  \bibinfo {author} {\bibfnamefont {D.}~\bibnamefont {Gazit}}, \bibinfo
  {author} {\bibfnamefont {F.}~\bibnamefont {Pederiva}}, \ and\ \bibinfo
  {author} {\bibfnamefont {U.}~\bibnamefont {van Kolck}},\ }\bibfield  {title}
  {\enquote {\bibinfo {title} {{Effective Field Theory for Lattice Nuclei}},}\
  }\href {\doibase 10.1103/PhysRevLett.114.052501} {\bibfield  {journal}
  {\bibinfo  {journal} {Phys. Rev. Lett.}\ }\textbf {\bibinfo {volume} {114}},\
  \bibinfo {pages} {052501} (\bibinfo {year} {2015})},\ \Eprint
  {http://arxiv.org/abs/1311.4966} {arXiv:1311.4966 [nucl-th]} \BibitemShut
  {NoStop}%
\bibitem [{\citenamefont {Ishii}\ \emph {et~al.}(2007)\citenamefont {Ishii},
  \citenamefont {Aoki},\ and\ \citenamefont {Hatsuda}}]{Ishii:2006ec}%
  \BibitemOpen
  \bibfield  {author} {\bibinfo {author} {\bibfnamefont {N.}~\bibnamefont
  {Ishii}}, \bibinfo {author} {\bibfnamefont {S.}~\bibnamefont {Aoki}}, \ and\
  \bibinfo {author} {\bibfnamefont {T.}~\bibnamefont {Hatsuda}},\ }\bibfield
  {title} {\enquote {\bibinfo {title} {{The Nuclear Force from Lattice QCD}},}\
  }\href {\doibase 10.1103/PhysRevLett.99.022001} {\bibfield  {journal}
  {\bibinfo  {journal} {Phys. Rev. Lett.}\ }\textbf {\bibinfo {volume} {99}},\
  \bibinfo {pages} {022001} (\bibinfo {year} {2007})},\ \Eprint
  {http://arxiv.org/abs/nucl-th/0611096} {arXiv:nucl-th/0611096} \BibitemShut
  {NoStop}%
\bibitem [{\citenamefont {Aoki}\ \emph {et~al.}(2010)\citenamefont {Aoki},
  \citenamefont {Hatsuda},\ and\ \citenamefont {Ishii}}]{Aoki:2009ji}%
  \BibitemOpen
  \bibfield  {author} {\bibinfo {author} {\bibfnamefont {Sinya}\ \bibnamefont
  {Aoki}}, \bibinfo {author} {\bibfnamefont {Tetsuo}\ \bibnamefont {Hatsuda}},
  \ and\ \bibinfo {author} {\bibfnamefont {Noriyoshi}\ \bibnamefont {Ishii}},\
  }\bibfield  {title} {\enquote {\bibinfo {title} {{Theoretical Foundation of
  the Nuclear Force in QCD and its applications to Central and Tensor Forces in
  Quenched Lattice QCD Simulations}},}\ }\href {\doibase 10.1143/PTP.123.89}
  {\bibfield  {journal} {\bibinfo  {journal} {Prog. Theor. Phys.}\ }\textbf
  {\bibinfo {volume} {123}},\ \bibinfo {pages} {89--128} (\bibinfo {year}
  {2010})},\ \Eprint {http://arxiv.org/abs/0909.5585} {arXiv:0909.5585
  [hep-lat]} \BibitemShut {NoStop}%
\bibitem [{\citenamefont {Ishii}\ \emph {et~al.}(2012)\citenamefont {Ishii},
  \citenamefont {Aoki}, \citenamefont {Doi}, \citenamefont {Hatsuda},
  \citenamefont {Ikeda}, \citenamefont {Inoue}, \citenamefont {Murano},
  \citenamefont {Nemura},\ and\ \citenamefont {Sasaki}}]{HALQCD:2012aa}%
  \BibitemOpen
  \bibfield  {author} {\bibinfo {author} {\bibfnamefont {Noriyoshi}\
  \bibnamefont {Ishii}}, \bibinfo {author} {\bibfnamefont {Sinya}\ \bibnamefont
  {Aoki}}, \bibinfo {author} {\bibfnamefont {Takumi}\ \bibnamefont {Doi}},
  \bibinfo {author} {\bibfnamefont {Tetsuo}\ \bibnamefont {Hatsuda}}, \bibinfo
  {author} {\bibfnamefont {Yoichi}\ \bibnamefont {Ikeda}}, \bibinfo {author}
  {\bibfnamefont {Takashi}\ \bibnamefont {Inoue}}, \bibinfo {author}
  {\bibfnamefont {Keiko}\ \bibnamefont {Murano}}, \bibinfo {author}
  {\bibfnamefont {Hidekatsu}\ \bibnamefont {Nemura}}, \ and\ \bibinfo {author}
  {\bibfnamefont {Kenji}\ \bibnamefont {Sasaki}} (\bibinfo {collaboration} {HAL
  QCD}),\ }\bibfield  {title} {\enquote {\bibinfo {title} {{Hadron-hadron
  interactions from imaginary-time Nambu-Bethe-Salpeter wave function on the
  lattice}},}\ }\href {\doibase 10.1016/j.physletb.2012.04.076} {\bibfield
  {journal} {\bibinfo  {journal} {Phys. Lett. B}\ }\textbf {\bibinfo {volume}
  {712}},\ \bibinfo {pages} {437--441} (\bibinfo {year} {2012})},\ \Eprint
  {http://arxiv.org/abs/1203.3642} {arXiv:1203.3642 [hep-lat]} \BibitemShut
  {NoStop}%
\bibitem [{\citenamefont {Aoki}\ \emph {et~al.}(2012)\citenamefont {Aoki},
  \citenamefont {Doi}, \citenamefont {Hatsuda}, \citenamefont {Ikeda},
  \citenamefont {Inoue}, \citenamefont {Ishii}, \citenamefont {Murano},
  \citenamefont {Nemura},\ and\ \citenamefont {Sasaki}}]{Aoki:2012tk}%
  \BibitemOpen
  \bibfield  {author} {\bibinfo {author} {\bibfnamefont {Sinya}\ \bibnamefont
  {Aoki}}, \bibinfo {author} {\bibfnamefont {Takumi}\ \bibnamefont {Doi}},
  \bibinfo {author} {\bibfnamefont {Tetsuo}\ \bibnamefont {Hatsuda}}, \bibinfo
  {author} {\bibfnamefont {Yoichi}\ \bibnamefont {Ikeda}}, \bibinfo {author}
  {\bibfnamefont {Takashi}\ \bibnamefont {Inoue}}, \bibinfo {author}
  {\bibfnamefont {Noriyoshi}\ \bibnamefont {Ishii}}, \bibinfo {author}
  {\bibfnamefont {Keiko}\ \bibnamefont {Murano}}, \bibinfo {author}
  {\bibfnamefont {Hidekatsu}\ \bibnamefont {Nemura}}, \ and\ \bibinfo {author}
  {\bibfnamefont {Kenji}\ \bibnamefont {Sasaki}} (\bibinfo {collaboration} {HAL
  QCD}),\ }\bibfield  {title} {\enquote {\bibinfo {title} {{Lattice QCD
  approach to Nuclear Physics}},}\ }\href {\doibase 10.1093/ptep/pts010}
  {\bibfield  {journal} {\bibinfo  {journal} {PTEP}\ }\textbf {\bibinfo
  {volume} {2012}},\ \bibinfo {pages} {01A105} (\bibinfo {year} {2012})},\
  \Eprint {http://arxiv.org/abs/1206.5088} {arXiv:1206.5088 [hep-lat]}
  \BibitemShut {NoStop}%
\bibitem [{\citenamefont {Aoki}\ \emph {et~al.}(2013)\citenamefont {Aoki},
  \citenamefont {Charron}, \citenamefont {Doi}, \citenamefont {Hatsuda},
  \citenamefont {Inoue},\ and\ \citenamefont {Ishii}}]{Aoki:2012bb}%
  \BibitemOpen
  \bibfield  {author} {\bibinfo {author} {\bibfnamefont {Sinya}\ \bibnamefont
  {Aoki}}, \bibinfo {author} {\bibfnamefont {Bruno}\ \bibnamefont {Charron}},
  \bibinfo {author} {\bibfnamefont {Takumi}\ \bibnamefont {Doi}}, \bibinfo
  {author} {\bibfnamefont {Tetsuo}\ \bibnamefont {Hatsuda}}, \bibinfo {author}
  {\bibfnamefont {Takashi}\ \bibnamefont {Inoue}}, \ and\ \bibinfo {author}
  {\bibfnamefont {Noriyoshi}\ \bibnamefont {Ishii}},\ }\bibfield  {title}
  {\enquote {\bibinfo {title} {{Construction of energy-independent potentials
  above inelastic thresholds in quantum field theories}},}\ }\href {\doibase
  10.1103/PhysRevD.87.034512} {\bibfield  {journal} {\bibinfo  {journal} {Phys.
  Rev. D}\ }\textbf {\bibinfo {volume} {87}},\ \bibinfo {pages} {034512}
  (\bibinfo {year} {2013})},\ \Eprint {http://arxiv.org/abs/1212.4896}
  {arXiv:1212.4896 [hep-lat]} \BibitemShut {NoStop}%
\bibitem [{\citenamefont {Yamazaki}\ \emph {et~al.}(2012)\citenamefont
  {Yamazaki}, \citenamefont {Ishikawa}, \citenamefont {Kuramashi},\ and\
  \citenamefont {Ukawa}}]{Yamazaki:2012hi}%
  \BibitemOpen
  \bibfield  {author} {\bibinfo {author} {\bibfnamefont {Takeshi}\ \bibnamefont
  {Yamazaki}}, \bibinfo {author} {\bibfnamefont {Ken-ichi}\ \bibnamefont
  {Ishikawa}}, \bibinfo {author} {\bibfnamefont {Yoshinobu}\ \bibnamefont
  {Kuramashi}}, \ and\ \bibinfo {author} {\bibfnamefont {Akira}\ \bibnamefont
  {Ukawa}},\ }\bibfield  {title} {\enquote {\bibinfo {title} {{Helium nuclei,
  deuteron and dineutron in 2+1 flavor lattice QCD}},}\ }\href {\doibase
  10.1103/PhysRevD.86.074514} {\bibfield  {journal} {\bibinfo  {journal} {Phys.
  Rev. D}\ }\textbf {\bibinfo {volume} {86}},\ \bibinfo {pages} {074514}
  (\bibinfo {year} {2012})},\ \Eprint {http://arxiv.org/abs/1207.4277}
  {arXiv:1207.4277 [hep-lat]} \BibitemShut {NoStop}%
\bibitem [{\citenamefont {Beane}\ \emph
  {et~al.}(2013{\natexlab{b}})\citenamefont {Beane} \emph
  {et~al.}}]{Beane:2013br}%
  \BibitemOpen
  \bibfield  {author} {\bibinfo {author} {\bibfnamefont {S.R.}\ \bibnamefont
  {Beane}} \emph {et~al.} (\bibinfo {collaboration} {NPLQCD}),\ }\bibfield
  {title} {\enquote {\bibinfo {title} {{Nucleon-Nucleon Scattering Parameters
  in the Limit of SU(3) Flavor Symmetry}},}\ }\href {\doibase
  10.1103/PhysRevC.88.024003} {\bibfield  {journal} {\bibinfo  {journal} {Phys.
  Rev. C}\ }\textbf {\bibinfo {volume} {88}},\ \bibinfo {pages} {024003}
  (\bibinfo {year} {2013}{\natexlab{b}})},\ \Eprint
  {http://arxiv.org/abs/1301.5790} {arXiv:1301.5790 [hep-lat]} \BibitemShut
  {NoStop}%
\bibitem [{\citenamefont {Berkowitz}\ \emph {et~al.}(2017)\citenamefont
  {Berkowitz}, \citenamefont {Kurth}, \citenamefont {Nicholson}, \citenamefont
  {Joo}, \citenamefont {Rinaldi}, \citenamefont {Strother}, \citenamefont
  {Vranas},\ and\ \citenamefont {Walker-Loud}}]{Berkowitz:2015eaa}%
  \BibitemOpen
  \bibfield  {author} {\bibinfo {author} {\bibfnamefont {Evan}\ \bibnamefont
  {Berkowitz}}, \bibinfo {author} {\bibfnamefont {Thorsten}\ \bibnamefont
  {Kurth}}, \bibinfo {author} {\bibfnamefont {Amy}\ \bibnamefont {Nicholson}},
  \bibinfo {author} {\bibfnamefont {Balint}\ \bibnamefont {Joo}}, \bibinfo
  {author} {\bibfnamefont {Enrico}\ \bibnamefont {Rinaldi}}, \bibinfo {author}
  {\bibfnamefont {Mark}\ \bibnamefont {Strother}}, \bibinfo {author}
  {\bibfnamefont {Pavlos~M.}\ \bibnamefont {Vranas}}, \ and\ \bibinfo {author}
  {\bibfnamefont {Andre}\ \bibnamefont {Walker-Loud}},\ }\bibfield  {title}
  {\enquote {\bibinfo {title} {{Two-Nucleon Higher Partial-Wave Scattering from
  Lattice QCD}},}\ }\href {\doibase 10.1016/j.physletb.2016.12.024} {\bibfield
  {journal} {\bibinfo  {journal} {Phys. Lett. B}\ }\textbf {\bibinfo {volume}
  {765}},\ \bibinfo {pages} {285--292} (\bibinfo {year} {2017})},\ \Eprint
  {http://arxiv.org/abs/1508.00886} {arXiv:1508.00886 [hep-lat]} \BibitemShut
  {NoStop}%
\bibitem [{\citenamefont {Orginos}\ \emph {et~al.}(2015)\citenamefont
  {Orginos}, \citenamefont {Parreno}, \citenamefont {Savage}, \citenamefont
  {Beane}, \citenamefont {Chang},\ and\ \citenamefont
  {Detmold}}]{Orginos:2015aya}%
  \BibitemOpen
  \bibfield  {author} {\bibinfo {author} {\bibfnamefont {Kostas}\ \bibnamefont
  {Orginos}}, \bibinfo {author} {\bibfnamefont {Assumpta}\ \bibnamefont
  {Parreno}}, \bibinfo {author} {\bibfnamefont {Martin~J.}\ \bibnamefont
  {Savage}}, \bibinfo {author} {\bibfnamefont {Silas~R.}\ \bibnamefont
  {Beane}}, \bibinfo {author} {\bibfnamefont {Emmanuel}\ \bibnamefont {Chang}},
  \ and\ \bibinfo {author} {\bibfnamefont {William}\ \bibnamefont {Detmold}},\
  }\bibfield  {title} {\enquote {\bibinfo {title} {{Two nucleon systems at
  $m_\pi\sim 450~{\rm MeV}$ from lattice QCD}},}\ }\href {\doibase
  10.1103/PhysRevD.92.114512} {\bibfield  {journal} {\bibinfo  {journal} {Phys.
  Rev. D}\ }\textbf {\bibinfo {volume} {92}},\ \bibinfo {pages} {114512}
  (\bibinfo {year} {2015})},\ \Eprint {http://arxiv.org/abs/1508.07583}
  {arXiv:1508.07583 [hep-lat]} \BibitemShut {NoStop}%
\bibitem [{\citenamefont {Wagman}\ \emph {et~al.}(2017)\citenamefont {Wagman},
  \citenamefont {Winter}, \citenamefont {Chang}, \citenamefont {Davoudi},
  \citenamefont {Detmold}, \citenamefont {Orginos}, \citenamefont {Savage},\
  and\ \citenamefont {Shanahan}}]{Wagman:2017tmp}%
  \BibitemOpen
  \bibfield  {author} {\bibinfo {author} {\bibfnamefont {Michael~L.}\
  \bibnamefont {Wagman}}, \bibinfo {author} {\bibfnamefont {Frank}\
  \bibnamefont {Winter}}, \bibinfo {author} {\bibfnamefont {Emmanuel}\
  \bibnamefont {Chang}}, \bibinfo {author} {\bibfnamefont {Zohreh}\
  \bibnamefont {Davoudi}}, \bibinfo {author} {\bibfnamefont {William}\
  \bibnamefont {Detmold}}, \bibinfo {author} {\bibfnamefont {Kostas}\
  \bibnamefont {Orginos}}, \bibinfo {author} {\bibfnamefont {Martin~J.}\
  \bibnamefont {Savage}}, \ and\ \bibinfo {author} {\bibfnamefont {Phiala~E.}\
  \bibnamefont {Shanahan}},\ }\bibfield  {title} {\enquote {\bibinfo {title}
  {{Baryon-Baryon Interactions and Spin-Flavor Symmetry from Lattice Quantum
  Chromodynamics}},}\ }\href {\doibase 10.1103/PhysRevD.96.114510} {\bibfield
  {journal} {\bibinfo  {journal} {Phys. Rev. D}\ }\textbf {\bibinfo {volume}
  {96}},\ \bibinfo {pages} {114510} (\bibinfo {year} {2017})},\ \Eprint
  {http://arxiv.org/abs/1706.06550} {arXiv:1706.06550 [hep-lat]} \BibitemShut
  {NoStop}%
\bibitem [{\citenamefont {Francis}\ \emph {et~al.}(2019)\citenamefont
  {Francis}, \citenamefont {Green}, \citenamefont {Junnarkar}, \citenamefont
  {Miao}, \citenamefont {Rae},\ and\ \citenamefont {Wittig}}]{Francis:2018qch}%
  \BibitemOpen
  \bibfield  {author} {\bibinfo {author} {\bibfnamefont {A.}~\bibnamefont
  {Francis}}, \bibinfo {author} {\bibfnamefont {J.~R.}\ \bibnamefont {Green}},
  \bibinfo {author} {\bibfnamefont {P.~M.}\ \bibnamefont {Junnarkar}}, \bibinfo
  {author} {\bibfnamefont {Ch.}\ \bibnamefont {Miao}}, \bibinfo {author}
  {\bibfnamefont {T.~D.}\ \bibnamefont {Rae}}, \ and\ \bibinfo {author}
  {\bibfnamefont {H.}~\bibnamefont {Wittig}},\ }\bibfield  {title} {\enquote
  {\bibinfo {title} {{Lattice QCD study of the $H$ dibaryon using hexaquark and
  two-baryon interpolators}},}\ }\href {\doibase 10.1103/PhysRevD.99.074505}
  {\bibfield  {journal} {\bibinfo  {journal} {Phys. Rev.}\ }\textbf {\bibinfo
  {volume} {D99}},\ \bibinfo {pages} {074505} (\bibinfo {year} {2019})},\
  \Eprint {http://arxiv.org/abs/1805.03966} {arXiv:1805.03966 [hep-lat]}
  \BibitemShut {NoStop}%
%%CITATION = ARXIV:1805.03966;%%
\bibitem [{\citenamefont {Inoue}\ \emph {et~al.}(2012)\citenamefont {Inoue},
  \citenamefont {Aoki}, \citenamefont {Doi}, \citenamefont {Hatsuda},
  \citenamefont {Ikeda}, \citenamefont {Ishii}, \citenamefont {Murano},
  \citenamefont {Nemura},\ and\ \citenamefont {Sasaki}}]{Inoue:2011ai}%
  \BibitemOpen
  \bibfield  {author} {\bibinfo {author} {\bibfnamefont {Takashi}\ \bibnamefont
  {Inoue}}, \bibinfo {author} {\bibfnamefont {Sinya}\ \bibnamefont {Aoki}},
  \bibinfo {author} {\bibfnamefont {Takumi}\ \bibnamefont {Doi}}, \bibinfo
  {author} {\bibfnamefont {Tetsuo}\ \bibnamefont {Hatsuda}}, \bibinfo {author}
  {\bibfnamefont {Yoichi}\ \bibnamefont {Ikeda}}, \bibinfo {author}
  {\bibfnamefont {Noriyoshi}\ \bibnamefont {Ishii}}, \bibinfo {author}
  {\bibfnamefont {Keiko}\ \bibnamefont {Murano}}, \bibinfo {author}
  {\bibfnamefont {Hidekatsu}\ \bibnamefont {Nemura}}, \ and\ \bibinfo {author}
  {\bibfnamefont {Kanji}\ \bibnamefont {Sasaki}} (\bibinfo {collaboration} {HAL
  QCD}),\ }\bibfield  {title} {\enquote {\bibinfo {title} {{Two-Baryon
  Potentials and H-Dibaryon from 3-flavor Lattice QCD Simulations}},}\
  }\bibfield  {booktitle} {\emph {\bibinfo {booktitle} {{Progress in
  strangeness nuclear physics. Proceedings, ECT Workshop on Strange Hadronic
  Matter, Trento, Italy, September 26-30, 2011}}},\ }\href {\doibase
  10.1016/j.nuclphysa.2012.02.008} {\bibfield  {journal} {\bibinfo  {journal}
  {Nucl. Phys.}\ }\textbf {\bibinfo {volume} {A881}},\ \bibinfo {pages}
  {28--43} (\bibinfo {year} {2012})},\ \Eprint {http://arxiv.org/abs/1112.5926}
  {arXiv:1112.5926 [hep-lat]} \BibitemShut {NoStop}%
%%CITATION = ARXIV:1112.5926;%%
\bibitem [{\citenamefont {Iritani}\ \emph {et~al.}(2017)\citenamefont
  {Iritani}, \citenamefont {Aoki}, \citenamefont {Doi}, \citenamefont
  {Hatsuda}, \citenamefont {Ikeda}, \citenamefont {Inoue}, \citenamefont
  {Ishii}, \citenamefont {Nemura},\ and\ \citenamefont
  {Sasaki}}]{Iritani:2017rlk}%
  \BibitemOpen
  \bibfield  {author} {\bibinfo {author} {\bibfnamefont {Takumi}\ \bibnamefont
  {Iritani}}, \bibinfo {author} {\bibfnamefont {Sinya}\ \bibnamefont {Aoki}},
  \bibinfo {author} {\bibfnamefont {Takumi}\ \bibnamefont {Doi}}, \bibinfo
  {author} {\bibfnamefont {Testuo}\ \bibnamefont {Hatsuda}}, \bibinfo {author}
  {\bibfnamefont {Yoichi}\ \bibnamefont {Ikeda}}, \bibinfo {author}
  {\bibfnamefont {Takashi}\ \bibnamefont {Inoue}}, \bibinfo {author}
  {\bibfnamefont {Noriyoshi}\ \bibnamefont {Ishii}}, \bibinfo {author}
  {\bibfnamefont {Hidekatsu}\ \bibnamefont {Nemura}}, \ and\ \bibinfo {author}
  {\bibfnamefont {Kenji}\ \bibnamefont {Sasaki}},\ }\bibfield  {title}
  {\enquote {\bibinfo {title} {{Are two nucleons bound in lattice QCD for heavy
  quark masses? Consistency check with Lüscher's finite volume formula}},}\
  }\href {\doibase 10.1103/PhysRevD.96.034521} {\bibfield  {journal} {\bibinfo
  {journal} {Phys. Rev. D}\ }\textbf {\bibinfo {volume} {96}},\ \bibinfo
  {pages} {034521} (\bibinfo {year} {2017})},\ \Eprint
  {http://arxiv.org/abs/1703.07210} {arXiv:1703.07210 [hep-lat]} \BibitemShut
  {NoStop}%
\bibitem [{\citenamefont {Iritani}\ \emph {et~al.}(2016)\citenamefont {Iritani}
  \emph {et~al.}}]{Iritani:2016jie}%
  \BibitemOpen
  \bibfield  {author} {\bibinfo {author} {\bibfnamefont {Takumi}\ \bibnamefont
  {Iritani}} \emph {et~al.},\ }\bibfield  {title} {\enquote {\bibinfo {title}
  {{Mirage in Temporal Correlation functions for Baryon-Baryon Interactions in
  Lattice QCD}},}\ }\href {\doibase 10.1007/JHEP10(2016)101} {\bibfield
  {journal} {\bibinfo  {journal} {JHEP}\ }\textbf {\bibinfo {volume} {10}},\
  \bibinfo {pages} {101} (\bibinfo {year} {2016})},\ \Eprint
  {http://arxiv.org/abs/1607.06371} {arXiv:1607.06371 [hep-lat]} \BibitemShut
  {NoStop}%
\bibitem [{\citenamefont {Yamazaki}\ \emph {et~al.}(2017)\citenamefont
  {Yamazaki}, \citenamefont {Ishikawa}, \citenamefont {Kuramashi},\ and\
  \citenamefont {Ukawa}}]{Yamazaki:2017euu}%
  \BibitemOpen
  \bibfield  {author} {\bibinfo {author} {\bibfnamefont {Takeshi}\ \bibnamefont
  {Yamazaki}}, \bibinfo {author} {\bibfnamefont {Ken-Ichi}\ \bibnamefont
  {Ishikawa}}, \bibinfo {author} {\bibfnamefont {Yoshinobu}\ \bibnamefont
  {Kuramashi}}, \ and\ \bibinfo {author} {\bibfnamefont {Akira}\ \bibnamefont
  {Ukawa}} (\bibinfo {collaboration} {PACS}),\ }\bibfield  {title} {\enquote
  {\bibinfo {title} {{Systematic study of operator dependence in nucleus
  calculation at large quark mass}},}\ }\href {\doibase 10.22323/1.256.0108}
  {\bibfield  {journal} {\bibinfo  {journal} {PoS}\ }\textbf {\bibinfo {volume}
  {LATTICE2016}},\ \bibinfo {pages} {108} (\bibinfo {year} {2017})},\ \Eprint
  {http://arxiv.org/abs/1702.00541} {arXiv:1702.00541 [hep-lat]} \BibitemShut
  {NoStop}%
\bibitem [{\citenamefont {Yamazaki}\ \emph {et~al.}(2018)\citenamefont
  {Yamazaki}, \citenamefont {Ishikawa},\ and\ \citenamefont
  {Kuramashi}}]{Yamazaki:2017jfh}%
  \BibitemOpen
  \bibfield  {author} {\bibinfo {author} {\bibfnamefont {Takeshi}\ \bibnamefont
  {Yamazaki}}, \bibinfo {author} {\bibfnamefont {Ken-ichi}\ \bibnamefont
  {Ishikawa}}, \ and\ \bibinfo {author} {\bibfnamefont {Yoshinobu}\
  \bibnamefont {Kuramashi}} (\bibinfo {collaboration} {PACS}),\ }\bibfield
  {title} {\enquote {\bibinfo {title} {{Comparison of different source
  calculations in two-nucleon channel at large quark mass}},}\ }\href {\doibase
  10.1051/epjconf/201817505019} {\bibfield  {journal} {\bibinfo  {journal} {EPJ
  Web Conf.}\ }\textbf {\bibinfo {volume} {175}},\ \bibinfo {pages} {05019}
  (\bibinfo {year} {2018})},\ \Eprint {http://arxiv.org/abs/1710.08066}
  {arXiv:1710.08066 [hep-lat]} \BibitemShut {NoStop}%
\bibitem [{\citenamefont {Beane}\ \emph {et~al.}(2017)\citenamefont {Beane}
  \emph {et~al.}}]{Beane:2017edf}%
  \BibitemOpen
  \bibfield  {author} {\bibinfo {author} {\bibfnamefont {Silas~R.}\
  \bibnamefont {Beane}} \emph {et~al.},\ }\bibfield  {title} {\enquote
  {\bibinfo {title} {{Comment on "Are two nucleons bound in lattice QCD for
  heavy quark masses? - Sanity check with L\"uscher's finite volume formula
  -"}},}\ }\href@noop {} {\  (\bibinfo {year} {2017})},\ \Eprint
  {http://arxiv.org/abs/1705.09239} {arXiv:1705.09239 [hep-lat]} \BibitemShut
  {NoStop}%
\bibitem [{\citenamefont {Beane}\ \emph
  {et~al.}(2012{\natexlab{a}})\citenamefont {Beane}, \citenamefont {Chang},
  \citenamefont {Detmold}, \citenamefont {Lin}, \citenamefont {Luu},
  \citenamefont {Orginos}, \citenamefont {Parreno}, \citenamefont {Savage},
  \citenamefont {Torok},\ and\ \citenamefont {Walker-Loud}}]{Beane:2011sc}%
  \BibitemOpen
  \bibfield  {author} {\bibinfo {author} {\bibfnamefont {S.R.}\ \bibnamefont
  {Beane}}, \bibinfo {author} {\bibfnamefont {E.}~\bibnamefont {Chang}},
  \bibinfo {author} {\bibfnamefont {W.}~\bibnamefont {Detmold}}, \bibinfo
  {author} {\bibfnamefont {H.W.}\ \bibnamefont {Lin}}, \bibinfo {author}
  {\bibfnamefont {T.C.}\ \bibnamefont {Luu}}, \bibinfo {author} {\bibfnamefont
  {K.}~\bibnamefont {Orginos}}, \bibinfo {author} {\bibfnamefont
  {A.}~\bibnamefont {Parreno}}, \bibinfo {author} {\bibfnamefont {M.J.}\
  \bibnamefont {Savage}}, \bibinfo {author} {\bibfnamefont {A.}~\bibnamefont
  {Torok}}, \ and\ \bibinfo {author} {\bibfnamefont {A.}~\bibnamefont
  {Walker-Loud}} (\bibinfo {collaboration} {NPLQCD}),\ }\bibfield  {title}
  {\enquote {\bibinfo {title} {{The I=2 pipi S-wave Scattering Phase Shift from
  Lattice QCD}},}\ }\href {\doibase 10.1103/PhysRevD.85.034505} {\bibfield
  {journal} {\bibinfo  {journal} {Phys. Rev. D}\ }\textbf {\bibinfo {volume}
  {85}},\ \bibinfo {pages} {034505} (\bibinfo {year} {2012}{\natexlab{a}})},\
  \Eprint {http://arxiv.org/abs/1107.5023} {arXiv:1107.5023 [hep-lat]}
  \BibitemShut {NoStop}%
\bibitem [{\citenamefont {Dudek}\ \emph {et~al.}(2011)\citenamefont {Dudek},
  \citenamefont {Edwards}, \citenamefont {Peardon}, \citenamefont {Richards},\
  and\ \citenamefont {Thomas}}]{Dudek:2010ew}%
  \BibitemOpen
  \bibfield  {author} {\bibinfo {author} {\bibfnamefont {Jozef~J.}\
  \bibnamefont {Dudek}}, \bibinfo {author} {\bibfnamefont {Robert~G.}\
  \bibnamefont {Edwards}}, \bibinfo {author} {\bibfnamefont {Michael~J.}\
  \bibnamefont {Peardon}}, \bibinfo {author} {\bibfnamefont {David~G.}\
  \bibnamefont {Richards}}, \ and\ \bibinfo {author} {\bibfnamefont
  {Christopher~E.}\ \bibnamefont {Thomas}},\ }\bibfield  {title} {\enquote
  {\bibinfo {title} {{The phase-shift of isospin-2 pi-pi scattering from
  lattice QCD}},}\ }\href {\doibase 10.1103/PhysRevD.83.071504} {\bibfield
  {journal} {\bibinfo  {journal} {Phys. Rev. D}\ }\textbf {\bibinfo {volume}
  {83}},\ \bibinfo {pages} {071504} (\bibinfo {year} {2011})},\ \Eprint
  {http://arxiv.org/abs/1011.6352} {arXiv:1011.6352 [hep-ph]} \BibitemShut
  {NoStop}%
\bibitem [{\citenamefont {Dudek}\ \emph {et~al.}(2013)\citenamefont {Dudek},
  \citenamefont {Edwards},\ and\ \citenamefont {Thomas}}]{Dudek:2012xn}%
  \BibitemOpen
  \bibfield  {author} {\bibinfo {author} {\bibfnamefont {Jozef~J.}\
  \bibnamefont {Dudek}}, \bibinfo {author} {\bibfnamefont {Robert~G.}\
  \bibnamefont {Edwards}}, \ and\ \bibinfo {author} {\bibfnamefont
  {Christopher~E.}\ \bibnamefont {Thomas}} (\bibinfo {collaboration} {Hadron
  Spectrum}),\ }\bibfield  {title} {\enquote {\bibinfo {title} {{Energy
  dependence of the $\rho$ resonance in $\pi\pi$ elastic scattering from
  lattice QCD}},}\ }\href {\doibase 10.1103/PhysRevD.87.034505} {\bibfield
  {journal} {\bibinfo  {journal} {Phys. Rev. D}\ }\textbf {\bibinfo {volume}
  {87}},\ \bibinfo {pages} {034505} (\bibinfo {year} {2013})},\ \bibinfo {note}
  {[Erratum: Phys.Rev.D 90, 099902 (2014)]},\ \Eprint
  {http://arxiv.org/abs/1212.0830} {arXiv:1212.0830 [hep-ph]} \BibitemShut
  {NoStop}%
\bibitem [{\citenamefont {Lang}\ and\ \citenamefont
  {Verduci}(2013)}]{Lang:2012db}%
  \BibitemOpen
  \bibfield  {author} {\bibinfo {author} {\bibfnamefont {C.~B.}\ \bibnamefont
  {Lang}}\ and\ \bibinfo {author} {\bibfnamefont {V.}~\bibnamefont {Verduci}},\
  }\bibfield  {title} {\enquote {\bibinfo {title} {{Scattering in the $\pi$N
  negative parity channel in lattice QCD}},}\ }\href {\doibase
  10.1103/PhysRevD.87.054502} {\bibfield  {journal} {\bibinfo  {journal} {Phys.
  Rev.}\ }\textbf {\bibinfo {volume} {D87}},\ \bibinfo {pages} {054502}
  (\bibinfo {year} {2013})},\ \Eprint {http://arxiv.org/abs/1212.5055}
  {arXiv:1212.5055 [hep-lat]} \BibitemShut {NoStop}%
%%CITATION = ARXIV:1212.5055;%%
\bibitem [{\citenamefont {Hanlon}\ \emph {et~al.}(2018)\citenamefont {Hanlon},
  \citenamefont {Francis}, \citenamefont {Green}, \citenamefont {Junnarkar},\
  and\ \citenamefont {Wittig}}]{Hanlon:2018yfv}%
  \BibitemOpen
  \bibfield  {author} {\bibinfo {author} {\bibfnamefont {Andrew}\ \bibnamefont
  {Hanlon}}, \bibinfo {author} {\bibfnamefont {Anthony}\ \bibnamefont
  {Francis}}, \bibinfo {author} {\bibfnamefont {Jeremy}\ \bibnamefont {Green}},
  \bibinfo {author} {\bibfnamefont {Parikshit}\ \bibnamefont {Junnarkar}}, \
  and\ \bibinfo {author} {\bibfnamefont {Hartmut}\ \bibnamefont {Wittig}},\
  }\bibfield  {title} {\enquote {\bibinfo {title} {{The $H$ dibaryon from
  lattice QCD with SU(3) flavor symmetry}},}\ }\href {\doibase
  10.22323/1.334.0081} {\bibfield  {journal} {\bibinfo  {journal} {PoS}\
  }\textbf {\bibinfo {volume} {LATTICE2018}},\ \bibinfo {pages} {081} (\bibinfo
  {year} {2018})},\ \Eprint {http://arxiv.org/abs/1810.13282} {arXiv:1810.13282
  [hep-lat]} \BibitemShut {NoStop}%
\bibitem [{\citenamefont {Morningstar}\ \emph {et~al.}(2011)\citenamefont
  {Morningstar}, \citenamefont {Bulava}, \citenamefont {Foley}, \citenamefont
  {Juge}, \citenamefont {Lenkner}, \citenamefont {Peardon},\ and\ \citenamefont
  {Wong}}]{Morningstar:2011ka}%
  \BibitemOpen
  \bibfield  {author} {\bibinfo {author} {\bibfnamefont {Colin}\ \bibnamefont
  {Morningstar}}, \bibinfo {author} {\bibfnamefont {John}\ \bibnamefont
  {Bulava}}, \bibinfo {author} {\bibfnamefont {Justin}\ \bibnamefont {Foley}},
  \bibinfo {author} {\bibfnamefont {Keisuke~J.}\ \bibnamefont {Juge}}, \bibinfo
  {author} {\bibfnamefont {David}\ \bibnamefont {Lenkner}}, \bibinfo {author}
  {\bibfnamefont {Mike}\ \bibnamefont {Peardon}}, \ and\ \bibinfo {author}
  {\bibfnamefont {Chik~Him}\ \bibnamefont {Wong}},\ }\bibfield  {title}
  {\enquote {\bibinfo {title} {{Improved stochastic estimation of quark
  propagation with Laplacian Heaviside smearing in lattice QCD}},}\ }\href
  {\doibase 10.1103/PhysRevD.83.114505} {\bibfield  {journal} {\bibinfo
  {journal} {Phys. Rev. D}\ }\textbf {\bibinfo {volume} {83}},\ \bibinfo
  {pages} {114505} (\bibinfo {year} {2011})},\ \Eprint
  {http://arxiv.org/abs/1104.3870} {arXiv:1104.3870 [hep-lat]} \BibitemShut
  {NoStop}%
\bibitem [{\citenamefont {Peardon}\ \emph {et~al.}(2009)\citenamefont
  {Peardon}, \citenamefont {Bulava}, \citenamefont {Foley}, \citenamefont
  {Morningstar}, \citenamefont {Dudek}, \citenamefont {Edwards}, \citenamefont
  {Joo}, \citenamefont {Lin}, \citenamefont {Richards},\ and\ \citenamefont
  {Juge}}]{Peardon:2009gh}%
  \BibitemOpen
  \bibfield  {author} {\bibinfo {author} {\bibfnamefont {Michael}\ \bibnamefont
  {Peardon}}, \bibinfo {author} {\bibfnamefont {John}\ \bibnamefont {Bulava}},
  \bibinfo {author} {\bibfnamefont {Justin}\ \bibnamefont {Foley}}, \bibinfo
  {author} {\bibfnamefont {Colin}\ \bibnamefont {Morningstar}}, \bibinfo
  {author} {\bibfnamefont {Jozef}\ \bibnamefont {Dudek}}, \bibinfo {author}
  {\bibfnamefont {Robert~G.}\ \bibnamefont {Edwards}}, \bibinfo {author}
  {\bibfnamefont {Balint}\ \bibnamefont {Joo}}, \bibinfo {author}
  {\bibfnamefont {Huey-Wen}\ \bibnamefont {Lin}}, \bibinfo {author}
  {\bibfnamefont {David~G.}\ \bibnamefont {Richards}}, \ and\ \bibinfo {author}
  {\bibfnamefont {Keisuke~Jimmy}\ \bibnamefont {Juge}} (\bibinfo
  {collaboration} {Hadron Spectrum}),\ }\bibfield  {title} {\enquote {\bibinfo
  {title} {{A Novel quark-field creation operator construction for hadronic
  physics in lattice QCD}},}\ }\href {\doibase 10.1103/PhysRevD.80.054506}
  {\bibfield  {journal} {\bibinfo  {journal} {Phys. Rev. D}\ }\textbf {\bibinfo
  {volume} {80}},\ \bibinfo {pages} {054506} (\bibinfo {year} {2009})},\
  \Eprint {http://arxiv.org/abs/0905.2160} {arXiv:0905.2160 [hep-lat]}
  \BibitemShut {NoStop}%
\bibitem [{\citenamefont {Bulava}\ \emph {et~al.}(2016)\citenamefont {Bulava},
  \citenamefont {Fahy}, \citenamefont {H\"orz}, \citenamefont {Juge},
  \citenamefont {Morningstar},\ and\ \citenamefont {Wong}}]{Bulava:2016mks}%
  \BibitemOpen
  \bibfield  {author} {\bibinfo {author} {\bibfnamefont {John}\ \bibnamefont
  {Bulava}}, \bibinfo {author} {\bibfnamefont {Brendan}\ \bibnamefont {Fahy}},
  \bibinfo {author} {\bibfnamefont {Ben}\ \bibnamefont {H\"orz}}, \bibinfo
  {author} {\bibfnamefont {Keisuke~J.}\ \bibnamefont {Juge}}, \bibinfo {author}
  {\bibfnamefont {Colin}\ \bibnamefont {Morningstar}}, \ and\ \bibinfo {author}
  {\bibfnamefont {Chik~Him}\ \bibnamefont {Wong}},\ }\bibfield  {title}
  {\enquote {\bibinfo {title} {{$I=1$ and $I=2$ $\pi-\pi$ scattering phase
  shifts from $N_{\mathrm{f}} = 2+1$ lattice QCD}},}\ }\href {\doibase
  10.1016/j.nuclphysb.2016.07.024} {\bibfield  {journal} {\bibinfo  {journal}
  {Nucl. Phys. B}\ }\textbf {\bibinfo {volume} {910}},\ \bibinfo {pages}
  {842--867} (\bibinfo {year} {2016})},\ \Eprint
  {http://arxiv.org/abs/1604.05593} {arXiv:1604.05593 [hep-lat]} \BibitemShut
  {NoStop}%
\bibitem [{\citenamefont {Brett}\ \emph {et~al.}(2018)\citenamefont {Brett},
  \citenamefont {Bulava}, \citenamefont {Fallica}, \citenamefont {Hanlon},
  \citenamefont {H\"orz},\ and\ \citenamefont {Morningstar}}]{Brett:2018jqw}%
  \BibitemOpen
  \bibfield  {author} {\bibinfo {author} {\bibfnamefont {Ruair\'\i{}}\
  \bibnamefont {Brett}}, \bibinfo {author} {\bibfnamefont {John}\ \bibnamefont
  {Bulava}}, \bibinfo {author} {\bibfnamefont {Jacob}\ \bibnamefont {Fallica}},
  \bibinfo {author} {\bibfnamefont {Andrew}\ \bibnamefont {Hanlon}}, \bibinfo
  {author} {\bibfnamefont {Ben}\ \bibnamefont {H\"orz}}, \ and\ \bibinfo
  {author} {\bibfnamefont {Colin}\ \bibnamefont {Morningstar}},\ }\bibfield
  {title} {\enquote {\bibinfo {title} {{Determination of $s$- and $p$-wave
  $I=1/2$ $K\pi$ scattering amplitudes in $N_{\mathrm{f}}=2+1$ lattice QCD}},}\
  }\href {\doibase 10.1016/j.nuclphysb.2018.05.008} {\bibfield  {journal}
  {\bibinfo  {journal} {Nucl. Phys. B}\ }\textbf {\bibinfo {volume} {932}},\
  \bibinfo {pages} {29--51} (\bibinfo {year} {2018})},\ \Eprint
  {http://arxiv.org/abs/1802.03100} {arXiv:1802.03100 [hep-lat]} \BibitemShut
  {NoStop}%
\bibitem [{\citenamefont {Andersen}\ \emph {et~al.}(2018)\citenamefont
  {Andersen}, \citenamefont {Bulava}, \citenamefont {H\"orz},\ and\
  \citenamefont {Morningstar}}]{Andersen:2017una}%
  \BibitemOpen
  \bibfield  {author} {\bibinfo {author} {\bibfnamefont {Christian~Walther}\
  \bibnamefont {Andersen}}, \bibinfo {author} {\bibfnamefont {John}\
  \bibnamefont {Bulava}}, \bibinfo {author} {\bibfnamefont {Ben}\ \bibnamefont
  {H\"orz}}, \ and\ \bibinfo {author} {\bibfnamefont {Colin}\ \bibnamefont
  {Morningstar}},\ }\bibfield  {title} {\enquote {\bibinfo {title} {{Elastic
  $I=3/2 p$-wave nucleon-pion scattering amplitude and the $\Delta$(1232)
  resonance from N$_f$=2+1 lattice QCD}},}\ }\href {\doibase
  10.1103/PhysRevD.97.014506} {\bibfield  {journal} {\bibinfo  {journal} {Phys.
  Rev. D}\ }\textbf {\bibinfo {volume} {97}},\ \bibinfo {pages} {014506}
  (\bibinfo {year} {2018})},\ \Eprint {http://arxiv.org/abs/1710.01557}
  {arXiv:1710.01557 [hep-lat]} \BibitemShut {NoStop}%
\bibitem [{\citenamefont {Foley}\ \emph {et~al.}(2005)\citenamefont {Foley},
  \citenamefont {Jimmy~Juge}, \citenamefont {O'Cais}, \citenamefont {Peardon},
  \citenamefont {Ryan},\ and\ \citenamefont {Skullerud}}]{Foley:2005ac}%
  \BibitemOpen
  \bibfield  {author} {\bibinfo {author} {\bibfnamefont {Justin}\ \bibnamefont
  {Foley}}, \bibinfo {author} {\bibfnamefont {K.}~\bibnamefont {Jimmy~Juge}},
  \bibinfo {author} {\bibfnamefont {Alan}\ \bibnamefont {O'Cais}}, \bibinfo
  {author} {\bibfnamefont {Mike}\ \bibnamefont {Peardon}}, \bibinfo {author}
  {\bibfnamefont {Sinead~M.}\ \bibnamefont {Ryan}}, \ and\ \bibinfo {author}
  {\bibfnamefont {Jon-Ivar}\ \bibnamefont {Skullerud}},\ }\bibfield  {title}
  {\enquote {\bibinfo {title} {{Practical all-to-all propagators for lattice
  QCD}},}\ }\href {\doibase 10.1016/j.cpc.2005.06.008} {\bibfield  {journal}
  {\bibinfo  {journal} {Comput. Phys. Commun.}\ }\textbf {\bibinfo {volume}
  {172}},\ \bibinfo {pages} {145--162} (\bibinfo {year} {2005})},\ \Eprint
  {http://arxiv.org/abs/hep-lat/0505023} {arXiv:hep-lat/0505023} \BibitemShut
  {NoStop}%
\bibitem [{\citenamefont {Clark}\ \emph {et~al.}(2010)\citenamefont {Clark},
  \citenamefont {Babich}, \citenamefont {Barros}, \citenamefont {Brower},\ and\
  \citenamefont {Rebbi}}]{Clark:2009wm}%
  \BibitemOpen
  \bibfield  {author} {\bibinfo {author} {\bibfnamefont {M.A.}\ \bibnamefont
  {Clark}}, \bibinfo {author} {\bibfnamefont {R.}~\bibnamefont {Babich}},
  \bibinfo {author} {\bibfnamefont {K.}~\bibnamefont {Barros}}, \bibinfo
  {author} {\bibfnamefont {R.C.}\ \bibnamefont {Brower}}, \ and\ \bibinfo
  {author} {\bibfnamefont {C.}~\bibnamefont {Rebbi}},\ }\bibfield  {title}
  {\enquote {\bibinfo {title} {{Solving Lattice QCD systems of equations using
  mixed precision solvers on GPUs}},}\ }\href {\doibase
  10.1016/j.cpc.2010.05.002} {\bibfield  {journal} {\bibinfo  {journal}
  {Comput.Phys.Commun.}\ }\textbf {\bibinfo {volume} {181}},\ \bibinfo {pages}
  {1517--1528} (\bibinfo {year} {2010})},\ \bibinfo {note}
  {\url{https://github.com/lattice/quda}},\ \Eprint
  {http://arxiv.org/abs/0911.3191} {arXiv:0911.3191 [hep-lat]} \BibitemShut
  {NoStop}%
%%CITATION = ARXIV:0911.3191;%%
\bibitem [{\citenamefont {Babich}\ \emph {et~al.}(2011)\citenamefont {Babich},
  \citenamefont {Clark}, \citenamefont {Joo}, \citenamefont {Shi},
  \citenamefont {Brower} \emph {et~al.}}]{Babich:2011np}%
  \BibitemOpen
  \bibfield  {author} {\bibinfo {author} {\bibfnamefont {R.}~\bibnamefont
  {Babich}}, \bibinfo {author} {\bibfnamefont {M.A.}\ \bibnamefont {Clark}},
  \bibinfo {author} {\bibfnamefont {B.}~\bibnamefont {Joo}}, \bibinfo {author}
  {\bibfnamefont {G.}~\bibnamefont {Shi}}, \bibinfo {author} {\bibfnamefont
  {R.C.}\ \bibnamefont {Brower}},  \emph {et~al.},\ }\bibfield  {title}
  {\enquote {\bibinfo {title} {{Scaling Lattice QCD beyond 100 GPUs}},}\
  }\href@noop {} {\  (\bibinfo {year} {2011})},\ \Eprint
  {http://arxiv.org/abs/1109.2935} {arXiv:1109.2935 [hep-lat]} \BibitemShut
  {NoStop}%
%%CITATION = ARXIV:1109.2935;%%
\bibitem [{\citenamefont {Morningstar}\ \emph {et~al.}(2013)\citenamefont
  {Morningstar}, \citenamefont {Bulava}, \citenamefont {Fahy}, \citenamefont
  {Foley}, \citenamefont {Jhang}, \citenamefont {Juge}, \citenamefont
  {Lenkner},\ and\ \citenamefont {Wong}}]{Morningstar:2013bda}%
  \BibitemOpen
  \bibfield  {author} {\bibinfo {author} {\bibfnamefont {C.}~\bibnamefont
  {Morningstar}}, \bibinfo {author} {\bibfnamefont {J.}~\bibnamefont {Bulava}},
  \bibinfo {author} {\bibfnamefont {B.}~\bibnamefont {Fahy}}, \bibinfo {author}
  {\bibfnamefont {J.}~\bibnamefont {Foley}}, \bibinfo {author} {\bibfnamefont
  {Y.C.}\ \bibnamefont {Jhang}}, \bibinfo {author} {\bibfnamefont {K.J.}\
  \bibnamefont {Juge}}, \bibinfo {author} {\bibfnamefont {D.}~\bibnamefont
  {Lenkner}}, \ and\ \bibinfo {author} {\bibfnamefont {C.H.}\ \bibnamefont
  {Wong}},\ }\bibfield  {title} {\enquote {\bibinfo {title} {{Extended hadron
  and two-hadron operators of definite momentum for spectrum calculations in
  lattice QCD}},}\ }\href {\doibase 10.1103/PhysRevD.88.014511} {\bibfield
  {journal} {\bibinfo  {journal} {Phys. Rev. D}\ }\textbf {\bibinfo {volume}
  {88}},\ \bibinfo {pages} {014511} (\bibinfo {year} {2013})},\ \Eprint
  {http://arxiv.org/abs/1303.6816} {arXiv:1303.6816 [hep-lat]} \BibitemShut
  {NoStop}%
\bibitem [{\citenamefont {Bulava}\ \emph {et~al.}(2018)\citenamefont {Bulava},
  \citenamefont {Hörz},\ and\ \citenamefont {Morningstar}}]{Bulava:2017stw}%
  \BibitemOpen
  \bibfield  {author} {\bibinfo {author} {\bibfnamefont {John}\ \bibnamefont
  {Bulava}}, \bibinfo {author} {\bibfnamefont {Ben}\ \bibnamefont {Hörz}}, \
  and\ \bibinfo {author} {\bibfnamefont {Colin}\ \bibnamefont {Morningstar}},\
  }\bibfield  {title} {\enquote {\bibinfo {title} {{Multi-hadron spectroscopy
  in a large physical volume}},}\ }\href {\doibase
  10.1051/epjconf/201817505026} {\bibfield  {journal} {\bibinfo  {journal} {EPJ
  Web Conf.}\ }\textbf {\bibinfo {volume} {175}},\ \bibinfo {pages} {05026}
  (\bibinfo {year} {2018})},\ \Eprint {http://arxiv.org/abs/1710.04545}
  {arXiv:1710.04545 [hep-lat]} \BibitemShut {NoStop}%
\bibitem [{\citenamefont {Bruno}\ \emph {et~al.}(2015)\citenamefont {Bruno}
  \emph {et~al.}}]{Bruno:2014jqa}%
  \BibitemOpen
  \bibfield  {author} {\bibinfo {author} {\bibfnamefont {Mattia}\ \bibnamefont
  {Bruno}} \emph {et~al.},\ }\bibfield  {title} {\enquote {\bibinfo {title}
  {{Simulation of QCD with N$_{f} =$ 2 $+$ 1 flavors of non-perturbatively
  improved Wilson fermions}},}\ }\href {\doibase 10.1007/JHEP02(2015)043}
  {\bibfield  {journal} {\bibinfo  {journal} {JHEP}\ }\textbf {\bibinfo
  {volume} {02}},\ \bibinfo {pages} {043} (\bibinfo {year} {2015})},\ \Eprint
  {http://arxiv.org/abs/1411.3982} {arXiv:1411.3982 [hep-lat]} \BibitemShut
  {NoStop}%
\bibitem [{\citenamefont {L\"uscher}()}]{openQCD}%
  \BibitemOpen
  \bibfield  {author} {\bibinfo {author} {\bibfnamefont {M.}~\bibnamefont
  {L\"uscher}},\ }\href@noop {} {\enquote {\bibinfo {title} {\texttt{openQCD}:
  simulation programs for lattice qcd},}\ }\bibinfo {note}
  {\url{http://luscher.web.cern.ch/luscher/openQCD/}}\BibitemShut {NoStop}%
\bibitem [{\citenamefont {Bruno}\ \emph {et~al.}(2017)\citenamefont {Bruno},
  \citenamefont {Korzec},\ and\ \citenamefont {Schaefer}}]{Bruno:2016plf}%
  \BibitemOpen
  \bibfield  {author} {\bibinfo {author} {\bibfnamefont {Mattia}\ \bibnamefont
  {Bruno}}, \bibinfo {author} {\bibfnamefont {Tomasz}\ \bibnamefont {Korzec}},
  \ and\ \bibinfo {author} {\bibfnamefont {Stefan}\ \bibnamefont {Schaefer}},\
  }\bibfield  {title} {\enquote {\bibinfo {title} {{Setting the scale for the
  CLS $2 + 1$ flavor ensembles}},}\ }\href {\doibase
  10.1103/PhysRevD.95.074504} {\bibfield  {journal} {\bibinfo  {journal} {Phys.
  Rev. D}\ }\textbf {\bibinfo {volume} {95}},\ \bibinfo {pages} {074504}
  (\bibinfo {year} {2017})},\ \Eprint {http://arxiv.org/abs/1608.08900}
  {arXiv:1608.08900 [hep-lat]} \BibitemShut {NoStop}%
\bibitem [{\citenamefont {Gusken}\ \emph {et~al.}(1989)\citenamefont {Gusken},
  \citenamefont {Low}, \citenamefont {Mutter}, \citenamefont {Sommer},
  \citenamefont {Patel},\ and\ \citenamefont {Schilling}}]{Gusken:1989ad}%
  \BibitemOpen
  \bibfield  {author} {\bibinfo {author} {\bibfnamefont {S.}~\bibnamefont
  {Gusken}}, \bibinfo {author} {\bibfnamefont {U.}~\bibnamefont {Low}},
  \bibinfo {author} {\bibfnamefont {K.H.}\ \bibnamefont {Mutter}}, \bibinfo
  {author} {\bibfnamefont {R.}~\bibnamefont {Sommer}}, \bibinfo {author}
  {\bibfnamefont {A.}~\bibnamefont {Patel}}, \ and\ \bibinfo {author}
  {\bibfnamefont {K.}~\bibnamefont {Schilling}},\ }\bibfield  {title} {\enquote
  {\bibinfo {title} {{Nonsinglet Axial Vector Couplings of the Baryon Octet in
  Lattice {QCD}}},}\ }\href {\doibase 10.1016/S0370-2693(89)80034-6} {\bibfield
   {journal} {\bibinfo  {journal} {Phys. Lett. B}\ }\textbf {\bibinfo {volume}
  {227}},\ \bibinfo {pages} {266--269} (\bibinfo {year} {1989})}\BibitemShut
  {NoStop}%
\bibitem [{\citenamefont {Alexandrou}\ \emph {et~al.}(1991)\citenamefont
  {Alexandrou}, \citenamefont {Jegerlehner}, \citenamefont {Gusken},
  \citenamefont {Schilling},\ and\ \citenamefont {Sommer}}]{Alexandrou:1990dq}%
  \BibitemOpen
  \bibfield  {author} {\bibinfo {author} {\bibfnamefont {C.}~\bibnamefont
  {Alexandrou}}, \bibinfo {author} {\bibfnamefont {F.}~\bibnamefont
  {Jegerlehner}}, \bibinfo {author} {\bibfnamefont {S.}~\bibnamefont {Gusken}},
  \bibinfo {author} {\bibfnamefont {K.}~\bibnamefont {Schilling}}, \ and\
  \bibinfo {author} {\bibfnamefont {R.}~\bibnamefont {Sommer}},\ }\bibfield
  {title} {\enquote {\bibinfo {title} {{B meson properties from lattice
  QCD}},}\ }\href {\doibase 10.1016/0370-2693(91)90219-G} {\bibfield  {journal}
  {\bibinfo  {journal} {Phys. Lett. B}\ }\textbf {\bibinfo {volume} {256}},\
  \bibinfo {pages} {60--67} (\bibinfo {year} {1991})}\BibitemShut {NoStop}%
\bibitem [{\citenamefont {Beane}\ \emph {et~al.}(2010)\citenamefont {Beane},
  \citenamefont {Detmold}, \citenamefont {Lin}, \citenamefont {Luu},
  \citenamefont {Orginos}, \citenamefont {Savage}, \citenamefont {Torok},\ and\
  \citenamefont {Walker-Loud}}]{Beane:2009py}%
  \BibitemOpen
  \bibfield  {author} {\bibinfo {author} {\bibfnamefont {Silas~R.}\
  \bibnamefont {Beane}}, \bibinfo {author} {\bibfnamefont {William}\
  \bibnamefont {Detmold}}, \bibinfo {author} {\bibfnamefont {Huey-Wen}\
  \bibnamefont {Lin}}, \bibinfo {author} {\bibfnamefont {Thomas~C.}\
  \bibnamefont {Luu}}, \bibinfo {author} {\bibfnamefont {Kostas}\ \bibnamefont
  {Orginos}}, \bibinfo {author} {\bibfnamefont {Martin~J.}\ \bibnamefont
  {Savage}}, \bibinfo {author} {\bibfnamefont {Aaron}\ \bibnamefont {Torok}}, \
  and\ \bibinfo {author} {\bibfnamefont {Andre}\ \bibnamefont {Walker-Loud}}
  (\bibinfo {collaboration} {NPLQCD}),\ }\bibfield  {title} {\enquote {\bibinfo
  {title} {{High Statistics Analysis using Anisotropic Clover Lattices: (III)
  Baryon-Baryon Interactions}},}\ }\href {\doibase 10.1103/PhysRevD.81.054505}
  {\bibfield  {journal} {\bibinfo  {journal} {Phys. Rev. D}\ }\textbf {\bibinfo
  {volume} {81}},\ \bibinfo {pages} {054505} (\bibinfo {year} {2010})},\
  \Eprint {http://arxiv.org/abs/0912.4243} {arXiv:0912.4243 [hep-lat]}
  \BibitemShut {NoStop}%
\bibitem [{\citenamefont {Beane}\ \emph
  {et~al.}(2012{\natexlab{b}})\citenamefont {Beane}, \citenamefont {Chang},
  \citenamefont {Detmold}, \citenamefont {Lin}, \citenamefont {Luu},
  \citenamefont {Orginos}, \citenamefont {Parreno}, \citenamefont {Savage},
  \citenamefont {Torok},\ and\ \citenamefont {Walker-Loud}}]{Beane:2011iw}%
  \BibitemOpen
  \bibfield  {author} {\bibinfo {author} {\bibfnamefont {S.R.}\ \bibnamefont
  {Beane}}, \bibinfo {author} {\bibfnamefont {E.}~\bibnamefont {Chang}},
  \bibinfo {author} {\bibfnamefont {W.}~\bibnamefont {Detmold}}, \bibinfo
  {author} {\bibfnamefont {H.W.}\ \bibnamefont {Lin}}, \bibinfo {author}
  {\bibfnamefont {T.C.}\ \bibnamefont {Luu}}, \bibinfo {author} {\bibfnamefont
  {K.}~\bibnamefont {Orginos}}, \bibinfo {author} {\bibfnamefont
  {A.}~\bibnamefont {Parreno}}, \bibinfo {author} {\bibfnamefont {M.J.}\
  \bibnamefont {Savage}}, \bibinfo {author} {\bibfnamefont {A.}~\bibnamefont
  {Torok}}, \ and\ \bibinfo {author} {\bibfnamefont {A.}~\bibnamefont
  {Walker-Loud}} (\bibinfo {collaboration} {NPLQCD}),\ }\bibfield  {title}
  {\enquote {\bibinfo {title} {{The Deuteron and Exotic Two-Body Bound States
  from Lattice QCD}},}\ }\href {\doibase 10.1103/PhysRevD.85.054511} {\bibfield
   {journal} {\bibinfo  {journal} {Phys. Rev. D}\ }\textbf {\bibinfo {volume}
  {85}},\ \bibinfo {pages} {054511} (\bibinfo {year} {2012}{\natexlab{b}})},\
  \Eprint {http://arxiv.org/abs/1109.2889} {arXiv:1109.2889 [hep-lat]}
  \BibitemShut {NoStop}%
\bibitem [{\citenamefont {Lepage}\ \emph {et~al.}(2002)\citenamefont {Lepage},
  \citenamefont {Clark}, \citenamefont {Davies}, \citenamefont {Hornbostel},
  \citenamefont {Mackenzie}, \citenamefont {Morningstar},\ and\ \citenamefont
  {Trottier}}]{Lepage:2001ym}%
  \BibitemOpen
  \bibfield  {author} {\bibinfo {author} {\bibfnamefont {G.P.}\ \bibnamefont
  {Lepage}}, \bibinfo {author} {\bibfnamefont {B.}~\bibnamefont {Clark}},
  \bibinfo {author} {\bibfnamefont {C.T.H.}\ \bibnamefont {Davies}}, \bibinfo
  {author} {\bibfnamefont {K.}~\bibnamefont {Hornbostel}}, \bibinfo {author}
  {\bibfnamefont {P.B.}\ \bibnamefont {Mackenzie}}, \bibinfo {author}
  {\bibfnamefont {C.}~\bibnamefont {Morningstar}}, \ and\ \bibinfo {author}
  {\bibfnamefont {H.}~\bibnamefont {Trottier}},\ }\bibfield  {title} {\enquote
  {\bibinfo {title} {{Constrained curve fitting}},}\ }\href {\doibase
  10.1016/S0920-5632(01)01638-3} {\bibfield  {journal} {\bibinfo  {journal}
  {Nucl. Phys. B Proc. Suppl.}\ }\textbf {\bibinfo {volume} {106}},\ \bibinfo
  {pages} {12--20} (\bibinfo {year} {2002})},\ \Eprint
  {http://arxiv.org/abs/hep-lat/0110175} {arXiv:hep-lat/0110175} \BibitemShut
  {NoStop}%
\bibitem [{\citenamefont {sLapHnn}(2020)}]{laphnn_qcotd}%
  \BibitemOpen
  \bibfield  {author} {\bibinfo {author} {\bibnamefont {sLapHnn}},\ }\href@noop
  {} {\enquote {\bibinfo {title} {Phase shift analysis code for {C103} $s$-wave
  $nn$},}\ } (\bibinfo {year} {2020}),\ \bibinfo {note}
  {\url{https://github.com/laphnn/nn_c103_qcotd_swave_only}}\BibitemShut
  {NoStop}%
\bibitem [{\citenamefont {Rummukainen}\ and\ \citenamefont
  {Gottlieb}(1995)}]{Rummukainen:1995vs}%
  \BibitemOpen
  \bibfield  {author} {\bibinfo {author} {\bibfnamefont {K.}~\bibnamefont
  {Rummukainen}}\ and\ \bibinfo {author} {\bibfnamefont {Steven~A.}\
  \bibnamefont {Gottlieb}},\ }\bibfield  {title} {\enquote {\bibinfo {title}
  {{Resonance scattering phase shifts on a nonrest frame lattice}},}\ }\href
  {\doibase 10.1016/0550-3213(95)00313-H} {\bibfield  {journal} {\bibinfo
  {journal} {Nucl. Phys. B}\ }\textbf {\bibinfo {volume} {450}},\ \bibinfo
  {pages} {397--436} (\bibinfo {year} {1995})},\ \Eprint
  {http://arxiv.org/abs/hep-lat/9503028} {arXiv:hep-lat/9503028} \BibitemShut
  {NoStop}%
\bibitem [{\citenamefont {Kim}\ \emph {et~al.}(2005)\citenamefont {Kim},
  \citenamefont {Sachrajda},\ and\ \citenamefont {Sharpe}}]{Kim:2005gf}%
  \BibitemOpen
  \bibfield  {author} {\bibinfo {author} {\bibfnamefont {C.h.}\ \bibnamefont
  {Kim}}, \bibinfo {author} {\bibfnamefont {C.T.}\ \bibnamefont {Sachrajda}}, \
  and\ \bibinfo {author} {\bibfnamefont {Stephen~R.}\ \bibnamefont {Sharpe}},\
  }\bibfield  {title} {\enquote {\bibinfo {title} {{Finite-volume effects for
  two-hadron states in moving frames}},}\ }\href {\doibase
  10.1016/j.nuclphysb.2005.08.029} {\bibfield  {journal} {\bibinfo  {journal}
  {Nucl. Phys. B}\ }\textbf {\bibinfo {volume} {727}},\ \bibinfo {pages}
  {218--243} (\bibinfo {year} {2005})},\ \Eprint
  {http://arxiv.org/abs/hep-lat/0507006} {arXiv:hep-lat/0507006} \BibitemShut
  {NoStop}%
\bibitem [{\citenamefont {Luu}\ and\ \citenamefont
  {Savage}(2011)}]{PhysRevD.83.114508}%
  \BibitemOpen
  \bibfield  {author} {\bibinfo {author} {\bibfnamefont {Thomas}\ \bibnamefont
  {Luu}}\ and\ \bibinfo {author} {\bibfnamefont {Martin~J.}\ \bibnamefont
  {Savage}},\ }\bibfield  {title} {\enquote {\bibinfo {title} {Extracting
  scattering phase shifts in higher partial waves from lattice qcd
  calculations},}\ }\href {\doibase 10.1103/PhysRevD.83.114508} {\bibfield
  {journal} {\bibinfo  {journal} {Phys. Rev. D}\ }\textbf {\bibinfo {volume}
  {83}},\ \bibinfo {pages} {114508} (\bibinfo {year} {2011})}\BibitemShut
  {NoStop}%
\bibitem [{\citenamefont {Fu}(2012)}]{Fu:2011xz}%
  \BibitemOpen
  \bibfield  {author} {\bibinfo {author} {\bibfnamefont {Ziwen}\ \bibnamefont
  {Fu}},\ }\bibfield  {title} {\enquote {\bibinfo {title}
  {{Rummukainen-Gottlieb's formula on two-particle system with different
  mass}},}\ }\href {\doibase 10.1103/PhysRevD.85.014506} {\bibfield  {journal}
  {\bibinfo  {journal} {Phys. Rev. D}\ }\textbf {\bibinfo {volume} {85}},\
  \bibinfo {pages} {014506} (\bibinfo {year} {2012})},\ \Eprint
  {http://arxiv.org/abs/1110.0319} {arXiv:1110.0319 [hep-lat]} \BibitemShut
  {NoStop}%
\bibitem [{\citenamefont {Leskovec}\ and\ \citenamefont
  {Prelovsek}(2012)}]{Leskovec:2012gb}%
  \BibitemOpen
  \bibfield  {author} {\bibinfo {author} {\bibfnamefont {Luka}\ \bibnamefont
  {Leskovec}}\ and\ \bibinfo {author} {\bibfnamefont {Sasa}\ \bibnamefont
  {Prelovsek}},\ }\bibfield  {title} {\enquote {\bibinfo {title} {{Scattering
  phase shifts for two particles of different mass and non-zero total momentum
  in lattice QCD}},}\ }\href {\doibase 10.1103/PhysRevD.85.114507} {\bibfield
  {journal} {\bibinfo  {journal} {Phys. Rev. D}\ }\textbf {\bibinfo {volume}
  {85}},\ \bibinfo {pages} {114507} (\bibinfo {year} {2012})},\ \Eprint
  {http://arxiv.org/abs/1202.2145} {arXiv:1202.2145 [hep-lat]} \BibitemShut
  {NoStop}%
\bibitem [{\citenamefont {Hansen}\ and\ \citenamefont
  {Sharpe}(2012)}]{Hansen:2012tf}%
  \BibitemOpen
  \bibfield  {author} {\bibinfo {author} {\bibfnamefont {Maxwell~T.}\
  \bibnamefont {Hansen}}\ and\ \bibinfo {author} {\bibfnamefont {Stephen~R.}\
  \bibnamefont {Sharpe}},\ }\bibfield  {title} {\enquote {\bibinfo {title}
  {{Multiple-channel generalization of Lellouch-Luscher formula}},}\ }\href
  {\doibase 10.1103/PhysRevD.86.016007} {\bibfield  {journal} {\bibinfo
  {journal} {Phys. Rev. D}\ }\textbf {\bibinfo {volume} {86}},\ \bibinfo
  {pages} {016007} (\bibinfo {year} {2012})},\ \Eprint
  {http://arxiv.org/abs/1204.0826} {arXiv:1204.0826 [hep-lat]} \BibitemShut
  {NoStop}%
\bibitem [{\citenamefont {Gockeler}\ \emph {et~al.}(2012)\citenamefont
  {Gockeler}, \citenamefont {Horsley}, \citenamefont {Lage}, \citenamefont
  {Meissner}, \citenamefont {Rakow}, \citenamefont {Rusetsky}, \citenamefont
  {Schierholz},\ and\ \citenamefont {Zanotti}}]{Gockeler:2012yj}%
  \BibitemOpen
  \bibfield  {author} {\bibinfo {author} {\bibfnamefont {M.}~\bibnamefont
  {Gockeler}}, \bibinfo {author} {\bibfnamefont {R.}~\bibnamefont {Horsley}},
  \bibinfo {author} {\bibfnamefont {M.}~\bibnamefont {Lage}}, \bibinfo {author}
  {\bibfnamefont {U.-G.}\ \bibnamefont {Meissner}}, \bibinfo {author}
  {\bibfnamefont {P.E.L.}\ \bibnamefont {Rakow}}, \bibinfo {author}
  {\bibfnamefont {A.}~\bibnamefont {Rusetsky}}, \bibinfo {author}
  {\bibfnamefont {G.}~\bibnamefont {Schierholz}}, \ and\ \bibinfo {author}
  {\bibfnamefont {J.M.}\ \bibnamefont {Zanotti}},\ }\bibfield  {title}
  {\enquote {\bibinfo {title} {{Scattering phases for meson and baryon
  resonances on general moving-frame lattices}},}\ }\href {\doibase
  10.1103/PhysRevD.86.094513} {\bibfield  {journal} {\bibinfo  {journal} {Phys.
  Rev. D}\ }\textbf {\bibinfo {volume} {86}},\ \bibinfo {pages} {094513}
  (\bibinfo {year} {2012})},\ \Eprint {http://arxiv.org/abs/1206.4141}
  {arXiv:1206.4141 [hep-lat]} \BibitemShut {NoStop}%
\bibitem [{\citenamefont {Brice{\~n}o}\ \emph {et~al.}(2013)\citenamefont
  {Brice{\~n}o}, \citenamefont {Davoudi},\ and\ \citenamefont
  {Luu}}]{Briceno:2013lba}%
  \BibitemOpen
  \bibfield  {author} {\bibinfo {author} {\bibfnamefont {Ra{\'u}l~A.}\
  \bibnamefont {Brice{\~n}o}}, \bibinfo {author} {\bibfnamefont {Zohreh}\
  \bibnamefont {Davoudi}}, \ and\ \bibinfo {author} {\bibfnamefont {Thomas~C.}\
  \bibnamefont {Luu}},\ }\bibfield  {title} {\enquote {\bibinfo {title}
  {{Two-Nucleon Systems in a Finite Volume: (I) Quantization Conditions}},}\
  }\href {\doibase 10.1103/PhysRevD.88.034502} {\bibfield  {journal} {\bibinfo
  {journal} {Phys. Rev. D}\ }\textbf {\bibinfo {volume} {88}},\ \bibinfo
  {pages} {034502} (\bibinfo {year} {2013})},\ \Eprint
  {http://arxiv.org/abs/1305.4903} {arXiv:1305.4903 [hep-lat]} \BibitemShut
  {NoStop}%
\bibitem [{\citenamefont {Briceno}(2014)}]{Briceno:2014oea}%
  \BibitemOpen
  \bibfield  {author} {\bibinfo {author} {\bibfnamefont {Raul~A.}\ \bibnamefont
  {Briceno}},\ }\bibfield  {title} {\enquote {\bibinfo {title} {{Two-particle
  multichannel systems in a finite volume with arbitrary spin}},}\ }\href
  {\doibase 10.1103/PhysRevD.89.074507} {\bibfield  {journal} {\bibinfo
  {journal} {Phys. Rev. D}\ }\textbf {\bibinfo {volume} {89}},\ \bibinfo
  {pages} {074507} (\bibinfo {year} {2014})},\ \Eprint
  {http://arxiv.org/abs/1401.3312} {arXiv:1401.3312 [hep-lat]} \BibitemShut
  {NoStop}%
\bibitem [{\citenamefont {Morningstar}\ \emph {et~al.}(2017)\citenamefont
  {Morningstar}, \citenamefont {Bulava}, \citenamefont {Singha}, \citenamefont
  {Brett}, \citenamefont {Fallica}, \citenamefont {Hanlon},\ and\ \citenamefont
  {H\"orz}}]{Morningstar:2017spu}%
  \BibitemOpen
  \bibfield  {author} {\bibinfo {author} {\bibfnamefont {Colin}\ \bibnamefont
  {Morningstar}}, \bibinfo {author} {\bibfnamefont {John}\ \bibnamefont
  {Bulava}}, \bibinfo {author} {\bibfnamefont {Bijit}\ \bibnamefont {Singha}},
  \bibinfo {author} {\bibfnamefont {Ruair\'\i{}}\ \bibnamefont {Brett}},
  \bibinfo {author} {\bibfnamefont {Jacob}\ \bibnamefont {Fallica}}, \bibinfo
  {author} {\bibfnamefont {Andrew}\ \bibnamefont {Hanlon}}, \ and\ \bibinfo
  {author} {\bibfnamefont {Ben}\ \bibnamefont {H\"orz}},\ }\bibfield  {title}
  {\enquote {\bibinfo {title} {{Estimating the two-particle $K$-matrix for
  multiple partial waves and decay channels from finite-volume energies}},}\
  }\href {\doibase 10.1016/j.nuclphysb.2017.09.014} {\bibfield  {journal}
  {\bibinfo  {journal} {Nucl. Phys. B}\ }\textbf {\bibinfo {volume} {924}},\
  \bibinfo {pages} {477--507} (\bibinfo {year} {2017})},\ \Eprint
  {http://arxiv.org/abs/1707.05817} {arXiv:1707.05817 [hep-lat]} \BibitemShut
  {NoStop}%
\bibitem [{\citenamefont {Bickel}\ and\ \citenamefont
  {Doksum}(2001)}]{bickel2001mathematical}%
  \BibitemOpen
  \bibfield  {author} {\bibinfo {author} {\bibfnamefont {P.J.}\ \bibnamefont
  {Bickel}}\ and\ \bibinfo {author} {\bibfnamefont {K.A.}\ \bibnamefont
  {Doksum}},\ }\href {https://books.google.com/books?id=8poZAQAAIAAJ} {\emph
  {\bibinfo {title} {Mathematical Statistics: Basic Ideas and Selected
  Topics}}},\ \bibinfo {series} {Mathematical Statistics: Basic Ideas and
  Selected Topics}\ No.\ \bibinfo {number} {v. 1}\ (\bibinfo  {publisher}
  {Prentice Hall},\ \bibinfo {year} {2001})\BibitemShut {NoStop}%
\bibitem [{\citenamefont {Steiner}(2018)}]{steiner2018multidimensional}%
  \BibitemOpen
  \bibfield  {author} {\bibinfo {author} {\bibfnamefont {Andrew~W.}\
  \bibnamefont {Steiner}},\ }\href@noop {} {\enquote {\bibinfo {title} {Two-
  and multi-dimensional curve fitting using bayesian inference},}\ } (\bibinfo
  {year} {2018}),\ \Eprint {http://arxiv.org/abs/1802.05339} {arXiv:1802.05339
  [physics.data-an]} \BibitemShut {NoStop}%
\bibitem [{\citenamefont {Bernard}\ \emph {et~al.}(2008)\citenamefont
  {Bernard}, \citenamefont {Meissner},\ and\ \citenamefont
  {Rusetsky}}]{Bernard:2007cm}%
  \BibitemOpen
  \bibfield  {author} {\bibinfo {author} {\bibfnamefont {Veronique}\
  \bibnamefont {Bernard}}, \bibinfo {author} {\bibfnamefont {Ulf-G.}\
  \bibnamefont {Meissner}}, \ and\ \bibinfo {author} {\bibfnamefont {Akaki}\
  \bibnamefont {Rusetsky}},\ }\bibfield  {title} {\enquote {\bibinfo {title}
  {{The Delta-resonance in a finite volume}},}\ }\href {\doibase
  10.1016/j.nuclphysb.2007.07.030} {\bibfield  {journal} {\bibinfo  {journal}
  {Nucl. Phys. B}\ }\textbf {\bibinfo {volume} {788}},\ \bibinfo {pages}
  {1--20} (\bibinfo {year} {2008})},\ \Eprint
  {http://arxiv.org/abs/hep-lat/0702012} {arXiv:hep-lat/0702012} \BibitemShut
  {NoStop}%
\bibitem [{\citenamefont {Beane}\ \emph
  {et~al.}(2012{\natexlab{c}})\citenamefont {Beane}, \citenamefont {Chang},
  \citenamefont {Cohen}, \citenamefont {Detmold}, \citenamefont {Lin},
  \citenamefont {Luu}, \citenamefont {Orginos}, \citenamefont {Parreno},
  \citenamefont {Savage},\ and\ \citenamefont {Walker-Loud}}]{Beane:2012ey}%
  \BibitemOpen
  \bibfield  {author} {\bibinfo {author} {\bibfnamefont {S.R.}\ \bibnamefont
  {Beane}}, \bibinfo {author} {\bibfnamefont {E.}~\bibnamefont {Chang}},
  \bibinfo {author} {\bibfnamefont {S.D.}\ \bibnamefont {Cohen}}, \bibinfo
  {author} {\bibfnamefont {W.}~\bibnamefont {Detmold}}, \bibinfo {author}
  {\bibfnamefont {H.-W.}\ \bibnamefont {Lin}}, \bibinfo {author} {\bibfnamefont
  {T.C.}\ \bibnamefont {Luu}}, \bibinfo {author} {\bibfnamefont
  {K.}~\bibnamefont {Orginos}}, \bibinfo {author} {\bibfnamefont
  {A.}~\bibnamefont {Parreno}}, \bibinfo {author} {\bibfnamefont {M.J.}\
  \bibnamefont {Savage}}, \ and\ \bibinfo {author} {\bibfnamefont
  {A.}~\bibnamefont {Walker-Loud}},\ }\bibfield  {title} {\enquote {\bibinfo
  {title} {{Hyperon-Nucleon Interactions and the Composition of Dense Nuclear
  Matter from Quantum Chromodynamics}},}\ }\href {\doibase
  10.1103/PhysRevLett.109.172001} {\bibfield  {journal} {\bibinfo  {journal}
  {Phys. Rev. Lett.}\ }\textbf {\bibinfo {volume} {109}},\ \bibinfo {pages}
  {172001} (\bibinfo {year} {2012}{\natexlab{c}})},\ \Eprint
  {http://arxiv.org/abs/1204.3606} {arXiv:1204.3606 [hep-lat]} \BibitemShut
  {NoStop}%
\bibitem [{\citenamefont {Hall}\ \emph {et~al.}(2013)\citenamefont {Hall},
  \citenamefont {Hsu}, \citenamefont {Leinweber}, \citenamefont {Thomas},\ and\
  \citenamefont {Young}}]{Hall:2013qba}%
  \BibitemOpen
  \bibfield  {author} {\bibinfo {author} {\bibfnamefont {J.M.M.}\ \bibnamefont
  {Hall}}, \bibinfo {author} {\bibfnamefont {A.~C.~P.}\ \bibnamefont {Hsu}},
  \bibinfo {author} {\bibfnamefont {D.B.}\ \bibnamefont {Leinweber}}, \bibinfo
  {author} {\bibfnamefont {A.W.}\ \bibnamefont {Thomas}}, \ and\ \bibinfo
  {author} {\bibfnamefont {R.D.}\ \bibnamefont {Young}},\ }\bibfield  {title}
  {\enquote {\bibinfo {title} {{Finite-volume matrix Hamiltonian model for a
  $\Delta \to N\pi$ system}},}\ }\href {\doibase 10.1103/PhysRevD.87.094510}
  {\bibfield  {journal} {\bibinfo  {journal} {Phys. Rev. D}\ }\textbf {\bibinfo
  {volume} {87}},\ \bibinfo {pages} {094510} (\bibinfo {year} {2013})},\
  \Eprint {http://arxiv.org/abs/1303.4157} {arXiv:1303.4157 [hep-lat]}
  \BibitemShut {NoStop}%
\bibitem [{\citenamefont {Wu}\ \emph {et~al.}(2014)\citenamefont {Wu},
  \citenamefont {Lee}, \citenamefont {Thomas},\ and\ \citenamefont
  {Young}}]{Wu:2014vma}%
  \BibitemOpen
  \bibfield  {author} {\bibinfo {author} {\bibfnamefont {Jia-Jun}\ \bibnamefont
  {Wu}}, \bibinfo {author} {\bibfnamefont {T.-S.H.}\ \bibnamefont {Lee}},
  \bibinfo {author} {\bibfnamefont {A.W.}\ \bibnamefont {Thomas}}, \ and\
  \bibinfo {author} {\bibfnamefont {R.D.}\ \bibnamefont {Young}},\ }\bibfield
  {title} {\enquote {\bibinfo {title} {{Finite-volume Hamiltonian method for
  coupled-channels interactions in lattice QCD}},}\ }\href {\doibase
  10.1103/PhysRevC.90.055206} {\bibfield  {journal} {\bibinfo  {journal} {Phys.
  Rev. C}\ }\textbf {\bibinfo {volume} {90}},\ \bibinfo {pages} {055206}
  (\bibinfo {year} {2014})},\ \Eprint {http://arxiv.org/abs/1402.4868}
  {arXiv:1402.4868 [hep-lat]} \BibitemShut {NoStop}%
\bibitem [{\citenamefont {Wilson}\ \emph
  {et~al.}(2015{\natexlab{a}})\citenamefont {Wilson}, \citenamefont {Dudek},
  \citenamefont {Edwards},\ and\ \citenamefont {Thomas}}]{Wilson:2014cna}%
  \BibitemOpen
  \bibfield  {author} {\bibinfo {author} {\bibfnamefont {David~J.}\
  \bibnamefont {Wilson}}, \bibinfo {author} {\bibfnamefont {Jozef~J.}\
  \bibnamefont {Dudek}}, \bibinfo {author} {\bibfnamefont {Robert~G.}\
  \bibnamefont {Edwards}}, \ and\ \bibinfo {author} {\bibfnamefont
  {Christopher~E.}\ \bibnamefont {Thomas}},\ }\bibfield  {title} {\enquote
  {\bibinfo {title} {{Resonances in coupled $\pi K, \eta K$ scattering from
  lattice QCD}},}\ }\href {\doibase 10.1103/PhysRevD.91.054008} {\bibfield
  {journal} {\bibinfo  {journal} {Phys. Rev. D}\ }\textbf {\bibinfo {volume}
  {91}},\ \bibinfo {pages} {054008} (\bibinfo {year} {2015}{\natexlab{a}})},\
  \Eprint {http://arxiv.org/abs/1411.2004} {arXiv:1411.2004 [hep-ph]}
  \BibitemShut {NoStop}%
\bibitem [{\citenamefont {Wilson}\ \emph
  {et~al.}(2015{\natexlab{b}})\citenamefont {Wilson}, \citenamefont {Briceno},
  \citenamefont {Dudek}, \citenamefont {Edwards},\ and\ \citenamefont
  {Thomas}}]{Wilson:2015dqa}%
  \BibitemOpen
  \bibfield  {author} {\bibinfo {author} {\bibfnamefont {David~J.}\
  \bibnamefont {Wilson}}, \bibinfo {author} {\bibfnamefont {Raul~A.}\
  \bibnamefont {Briceno}}, \bibinfo {author} {\bibfnamefont {Jozef~J.}\
  \bibnamefont {Dudek}}, \bibinfo {author} {\bibfnamefont {Robert~G.}\
  \bibnamefont {Edwards}}, \ and\ \bibinfo {author} {\bibfnamefont
  {Christopher~E.}\ \bibnamefont {Thomas}},\ }\bibfield  {title} {\enquote
  {\bibinfo {title} {{Coupled $\pi\pi, K\bar{K}$ scattering in $P$-wave and the
  $\rho$ resonance from lattice QCD}},}\ }\href {\doibase
  10.1103/PhysRevD.92.094502} {\bibfield  {journal} {\bibinfo  {journal} {Phys.
  Rev. D}\ }\textbf {\bibinfo {volume} {92}},\ \bibinfo {pages} {094502}
  (\bibinfo {year} {2015}{\natexlab{b}})},\ \Eprint
  {http://arxiv.org/abs/1507.02599} {arXiv:1507.02599 [hep-ph]} \BibitemShut
  {NoStop}%
\bibitem [{\citenamefont {Dudek}\ \emph {et~al.}(2016)\citenamefont {Dudek},
  \citenamefont {Edwards},\ and\ \citenamefont {Wilson}}]{Dudek:2016cru}%
  \BibitemOpen
  \bibfield  {author} {\bibinfo {author} {\bibfnamefont {Jozef~J.}\
  \bibnamefont {Dudek}}, \bibinfo {author} {\bibfnamefont {Robert~G.}\
  \bibnamefont {Edwards}}, \ and\ \bibinfo {author} {\bibfnamefont {David~J.}\
  \bibnamefont {Wilson}} (\bibinfo {collaboration} {Hadron Spectrum}),\
  }\bibfield  {title} {\enquote {\bibinfo {title} {{An $a_0$ resonance in
  strongly coupled $\pi \eta$, $K\overline{K}$ scattering from lattice QCD}},}\
  }\href {\doibase 10.1103/PhysRevD.93.094506} {\bibfield  {journal} {\bibinfo
  {journal} {Phys. Rev. D}\ }\textbf {\bibinfo {volume} {93}},\ \bibinfo
  {pages} {094506} (\bibinfo {year} {2016})},\ \Eprint
  {http://arxiv.org/abs/1602.05122} {arXiv:1602.05122 [hep-ph]} \BibitemShut
  {NoStop}%
\bibitem [{\citenamefont {Guo}\ \emph {et~al.}(2016)\citenamefont {Guo},
  \citenamefont {Alexandru}, \citenamefont {Molina},\ and\ \citenamefont
  {D\"oring}}]{Guo:2016zos}%
  \BibitemOpen
  \bibfield  {author} {\bibinfo {author} {\bibfnamefont {Dehua}\ \bibnamefont
  {Guo}}, \bibinfo {author} {\bibfnamefont {Andrei}\ \bibnamefont {Alexandru}},
  \bibinfo {author} {\bibfnamefont {Raquel}\ \bibnamefont {Molina}}, \ and\
  \bibinfo {author} {\bibfnamefont {Michael}\ \bibnamefont {D\"oring}},\
  }\bibfield  {title} {\enquote {\bibinfo {title} {{Rho resonance parameters
  from lattice QCD}},}\ }\href {\doibase 10.1103/PhysRevD.94.034501} {\bibfield
   {journal} {\bibinfo  {journal} {Phys. Rev. D}\ }\textbf {\bibinfo {volume}
  {94}},\ \bibinfo {pages} {034501} (\bibinfo {year} {2016})},\ \Eprint
  {http://arxiv.org/abs/1605.03993} {arXiv:1605.03993 [hep-lat]} \BibitemShut
  {NoStop}%
\bibitem [{\citenamefont {Woss}\ \emph {et~al.}(2020)\citenamefont {Woss},
  \citenamefont {Wilson},\ and\ \citenamefont {Dudek}}]{Woss:2020cmp}%
  \BibitemOpen
  \bibfield  {author} {\bibinfo {author} {\bibfnamefont {Antoni~J.}\
  \bibnamefont {Woss}}, \bibinfo {author} {\bibfnamefont {David~J.}\
  \bibnamefont {Wilson}}, \ and\ \bibinfo {author} {\bibfnamefont {Jozef~J.}\
  \bibnamefont {Dudek}} (\bibinfo {collaboration} {Hadron Spectrum}),\
  }\bibfield  {title} {\enquote {\bibinfo {title} {{Efficient solution of the
  multichannel L\"uscher determinant condition through eigenvalue
  decomposition}},}\ }\href {\doibase 10.1103/PhysRevD.101.114505} {\bibfield
  {journal} {\bibinfo  {journal} {Phys. Rev. D}\ }\textbf {\bibinfo {volume}
  {101}},\ \bibinfo {pages} {114505} (\bibinfo {year} {2020})},\ \Eprint
  {http://arxiv.org/abs/2001.08474} {arXiv:2001.08474 [hep-lat]} \BibitemShut
  {NoStop}%
\bibitem [{\citenamefont {Jay}\ and\ \citenamefont {Neil}(2020)}]{Jay:2020jkz}%
  \BibitemOpen
  \bibfield  {author} {\bibinfo {author} {\bibfnamefont {William~I.}\
  \bibnamefont {Jay}}\ and\ \bibinfo {author} {\bibfnamefont {Ethan~T.}\
  \bibnamefont {Neil}},\ }\bibfield  {title} {\enquote {\bibinfo {title}
  {{Bayesian model averaging for analysis of lattice field theory results}},}\
  }\href@noop {} {\  (\bibinfo {year} {2020})},\ \Eprint
  {http://arxiv.org/abs/2008.01069} {arXiv:2008.01069 [stat.ME]} \BibitemShut
  {NoStop}%
\bibitem [{\citenamefont {Dudek}\ \emph {et~al.}(2014)\citenamefont {Dudek},
  \citenamefont {Edwards}, \citenamefont {Thomas},\ and\ \citenamefont
  {Wilson}}]{Dudek:2014qha}%
  \BibitemOpen
  \bibfield  {author} {\bibinfo {author} {\bibfnamefont {Jozef~J.}\
  \bibnamefont {Dudek}}, \bibinfo {author} {\bibfnamefont {Robert~G.}\
  \bibnamefont {Edwards}}, \bibinfo {author} {\bibfnamefont {Christopher~E.}\
  \bibnamefont {Thomas}}, \ and\ \bibinfo {author} {\bibfnamefont {David~J.}\
  \bibnamefont {Wilson}} (\bibinfo {collaboration} {Hadron Spectrum}),\
  }\bibfield  {title} {\enquote {\bibinfo {title} {{Resonances in coupled $\pi
  K -\eta K$ scattering from quantum chromodynamics}},}\ }\href {\doibase
  10.1103/PhysRevLett.113.182001} {\bibfield  {journal} {\bibinfo  {journal}
  {Phys. Rev. Lett.}\ }\textbf {\bibinfo {volume} {113}},\ \bibinfo {pages}
  {182001} (\bibinfo {year} {2014})},\ \Eprint {http://arxiv.org/abs/1406.4158}
  {arXiv:1406.4158 [hep-ph]} \BibitemShut {NoStop}%
\bibitem [{\citenamefont {Wilson}\ \emph {et~al.}(2019)\citenamefont {Wilson},
  \citenamefont {Briceno}, \citenamefont {Dudek}, \citenamefont {Edwards},\
  and\ \citenamefont {Thomas}}]{Wilson:2019wfr}%
  \BibitemOpen
  \bibfield  {author} {\bibinfo {author} {\bibfnamefont {David~J.}\
  \bibnamefont {Wilson}}, \bibinfo {author} {\bibfnamefont {Raul~A.}\
  \bibnamefont {Briceno}}, \bibinfo {author} {\bibfnamefont {Jozef~J.}\
  \bibnamefont {Dudek}}, \bibinfo {author} {\bibfnamefont {Robert~G.}\
  \bibnamefont {Edwards}}, \ and\ \bibinfo {author} {\bibfnamefont
  {Christopher~E.}\ \bibnamefont {Thomas}},\ }\bibfield  {title} {\enquote
  {\bibinfo {title} {{The quark-mass dependence of elastic $\pi K$ scattering
  from QCD}},}\ }\href {\doibase 10.1103/PhysRevLett.123.042002} {\bibfield
  {journal} {\bibinfo  {journal} {Phys. Rev. Lett.}\ }\textbf {\bibinfo
  {volume} {123}},\ \bibinfo {pages} {042002} (\bibinfo {year} {2019})},\
  \Eprint {http://arxiv.org/abs/1904.03188} {arXiv:1904.03188 [hep-lat]}
  \BibitemShut {NoStop}%
\bibitem [{\citenamefont {Walker-Loud}\ \emph {et~al.}(2009)\citenamefont
  {Walker-Loud} \emph {et~al.}}]{WalkerLoud:2008bp}%
  \BibitemOpen
  \bibfield  {author} {\bibinfo {author} {\bibfnamefont {A.}~\bibnamefont
  {Walker-Loud}} \emph {et~al.},\ }\bibfield  {title} {\enquote {\bibinfo
  {title} {{Light hadron spectroscopy using domain wall valence quarks on an
  Asqtad sea}},}\ }\href {\doibase 10.1103/PhysRevD.79.054502} {\bibfield
  {journal} {\bibinfo  {journal} {Phys. Rev. D}\ }\textbf {\bibinfo {volume}
  {79}},\ \bibinfo {pages} {054502} (\bibinfo {year} {2009})},\ \Eprint
  {http://arxiv.org/abs/0806.4549} {arXiv:0806.4549 [hep-lat]} \BibitemShut
  {NoStop}%
\bibitem [{\citenamefont {Wigner}(1955)}]{Wigner:1955zz}%
  \BibitemOpen
  \bibfield  {author} {\bibinfo {author} {\bibfnamefont {Eugene~P.}\
  \bibnamefont {Wigner}},\ }\bibfield  {title} {\enquote {\bibinfo {title}
  {{Lower Limit for the Energy Derivative of the Scattering Phase Shift}},}\
  }\href {\doibase 10.1103/PhysRev.98.145} {\bibfield  {journal} {\bibinfo
  {journal} {Phys. Rev.}\ }\textbf {\bibinfo {volume} {98}},\ \bibinfo {pages}
  {145--147} (\bibinfo {year} {1955})}\BibitemShut {NoStop}%
\bibitem [{\citenamefont {Phillips}\ and\ \citenamefont
  {Cohen}(1997)}]{Phillips:1996ae}%
  \BibitemOpen
  \bibfield  {author} {\bibinfo {author} {\bibfnamefont {Daniel~R.}\
  \bibnamefont {Phillips}}\ and\ \bibinfo {author} {\bibfnamefont {Thomas~D.}\
  \bibnamefont {Cohen}},\ }\bibfield  {title} {\enquote {\bibinfo {title} {{How
  short is too short? Constraining contact interactions in nucleon-nucleon
  scattering}},}\ }\href {\doibase 10.1016/S0370-2693(96)01411-6} {\bibfield
  {journal} {\bibinfo  {journal} {Phys. Lett. B}\ }\textbf {\bibinfo {volume}
  {390}},\ \bibinfo {pages} {7--12} (\bibinfo {year} {1997})},\ \Eprint
  {http://arxiv.org/abs/nucl-th/9607048} {arXiv:nucl-th/9607048} \BibitemShut
  {NoStop}%
\bibitem [{\citenamefont {Scaldeferri}\ \emph {et~al.}(1997)\citenamefont
  {Scaldeferri}, \citenamefont {Phillips}, \citenamefont {Kao},\ and\
  \citenamefont {Cohen}}]{Scaldeferri:1996nx}%
  \BibitemOpen
  \bibfield  {author} {\bibinfo {author} {\bibfnamefont {K.A.}\ \bibnamefont
  {Scaldeferri}}, \bibinfo {author} {\bibfnamefont {Daniel~R.}\ \bibnamefont
  {Phillips}}, \bibinfo {author} {\bibfnamefont {C.W.}\ \bibnamefont {Kao}}, \
  and\ \bibinfo {author} {\bibfnamefont {T.D.}\ \bibnamefont {Cohen}},\
  }\bibfield  {title} {\enquote {\bibinfo {title} {{Short range interactions in
  an effective field theory approach for nucleon-nucleon scattering}},}\ }\href
  {\doibase 10.1103/PhysRevC.56.679} {\bibfield  {journal} {\bibinfo  {journal}
  {Phys. Rev. C}\ }\textbf {\bibinfo {volume} {56}},\ \bibinfo {pages}
  {679--688} (\bibinfo {year} {1997})},\ \Eprint
  {http://arxiv.org/abs/nucl-th/9610049} {arXiv:nucl-th/9610049} \BibitemShut
  {NoStop}%
\bibitem [{\citenamefont {Baru}\ \emph {et~al.}(2016)\citenamefont {Baru},
  \citenamefont {Epelbaum},\ and\ \citenamefont {Filin}}]{Baru:2016evv}%
  \BibitemOpen
  \bibfield  {author} {\bibinfo {author} {\bibfnamefont {V.}~\bibnamefont
  {Baru}}, \bibinfo {author} {\bibfnamefont {E.}~\bibnamefont {Epelbaum}}, \
  and\ \bibinfo {author} {\bibfnamefont {A.A.}\ \bibnamefont {Filin}},\
  }\bibfield  {title} {\enquote {\bibinfo {title} {{Low-energy theorems for
  nucleon-nucleon scattering at $M_\pi=450$ MeV}},}\ }\href {\doibase
  10.1103/PhysRevC.94.014001} {\bibfield  {journal} {\bibinfo  {journal} {Phys.
  Rev. C}\ }\textbf {\bibinfo {volume} {94}},\ \bibinfo {pages} {014001}
  (\bibinfo {year} {2016})},\ \Eprint {http://arxiv.org/abs/1604.02551}
  {arXiv:1604.02551 [nucl-th]} \BibitemShut {NoStop}%
\bibitem [{\citenamefont {Illa}\ \emph {et~al.}(2020)\citenamefont {Illa} \emph
  {et~al.}}]{Illa:2020nsi}%
  \BibitemOpen
  \bibfield  {author} {\bibinfo {author} {\bibfnamefont {Marc}\ \bibnamefont
  {Illa}} \emph {et~al.},\ }\bibfield  {title} {\enquote {\bibinfo {title}
  {{Low-energy Scattering and Effective Interactions of Two Baryons at
  $m_{\pi}\sim$ 450 MeV from Lattice Quantum Chromodynamics}},}\ }\href@noop {}
  {\  (\bibinfo {year} {2020})},\ \Eprint {http://arxiv.org/abs/2009.12357}
  {arXiv:2009.12357 [hep-lat]} \BibitemShut {NoStop}%
\bibitem [{\citenamefont {Morningstar}\ and\ \citenamefont
  {Peardon}(2004)}]{Morningstar:2003gk}%
  \BibitemOpen
  \bibfield  {author} {\bibinfo {author} {\bibfnamefont {Colin}\ \bibnamefont
  {Morningstar}}\ and\ \bibinfo {author} {\bibfnamefont {Mike~J.}\ \bibnamefont
  {Peardon}},\ }\bibfield  {title} {\enquote {\bibinfo {title} {{Analytic
  smearing of SU(3) link variables in lattice QCD}},}\ }\href {\doibase
  10.1103/PhysRevD.69.054501} {\bibfield  {journal} {\bibinfo  {journal} {Phys.
  Rev. D}\ }\textbf {\bibinfo {volume} {69}},\ \bibinfo {pages} {054501}
  (\bibinfo {year} {2004})},\ \Eprint {http://arxiv.org/abs/hep-lat/0311018}
  {arXiv:hep-lat/0311018} \BibitemShut {NoStop}%
\bibitem [{\citenamefont {Lepage}\ and\ \citenamefont
  {Mackenzie}(1993)}]{Lepage:1992xa}%
  \BibitemOpen
  \bibfield  {author} {\bibinfo {author} {\bibfnamefont {G.Peter}\ \bibnamefont
  {Lepage}}\ and\ \bibinfo {author} {\bibfnamefont {Paul~B.}\ \bibnamefont
  {Mackenzie}},\ }\bibfield  {title} {\enquote {\bibinfo {title} {{On the
  viability of lattice perturbation theory}},}\ }\href {\doibase
  10.1103/PhysRevD.48.2250} {\bibfield  {journal} {\bibinfo  {journal} {Phys.
  Rev. D}\ }\textbf {\bibinfo {volume} {48}},\ \bibinfo {pages} {2250--2264}
  (\bibinfo {year} {1993})},\ \Eprint {http://arxiv.org/abs/hep-lat/9209022}
  {arXiv:hep-lat/9209022} \BibitemShut {NoStop}%
\bibitem [{\citenamefont {Iritani}\ \emph {et~al.}(2019)\citenamefont
  {Iritani}, \citenamefont {Aoki}, \citenamefont {Doi}, \citenamefont {Gongyo},
  \citenamefont {Hatsuda}, \citenamefont {Ikeda}, \citenamefont {Inoue},
  \citenamefont {Ishii}, \citenamefont {Nemura},\ and\ \citenamefont
  {Sasaki}}]{Iritani:2018zbt}%
  \BibitemOpen
  \bibfield  {author} {\bibinfo {author} {\bibfnamefont {Takumi}\ \bibnamefont
  {Iritani}}, \bibinfo {author} {\bibfnamefont {Sinya}\ \bibnamefont {Aoki}},
  \bibinfo {author} {\bibfnamefont {Takumi}\ \bibnamefont {Doi}}, \bibinfo
  {author} {\bibfnamefont {Shinya}\ \bibnamefont {Gongyo}}, \bibinfo {author}
  {\bibfnamefont {Tetsuo}\ \bibnamefont {Hatsuda}}, \bibinfo {author}
  {\bibfnamefont {Yoichi}\ \bibnamefont {Ikeda}}, \bibinfo {author}
  {\bibfnamefont {Takashi}\ \bibnamefont {Inoue}}, \bibinfo {author}
  {\bibfnamefont {Noriyoshi}\ \bibnamefont {Ishii}}, \bibinfo {author}
  {\bibfnamefont {Hidekatsu}\ \bibnamefont {Nemura}}, \ and\ \bibinfo {author}
  {\bibfnamefont {Kenji}\ \bibnamefont {Sasaki}} (\bibinfo {collaboration} {HAL
  QCD}),\ }\bibfield  {title} {\enquote {\bibinfo {title} {{Systematics of the
  HAL QCD Potential at Low Energies in Lattice QCD}},}\ }\href {\doibase
  10.1103/PhysRevD.99.014514} {\bibfield  {journal} {\bibinfo  {journal} {Phys.
  Rev. D}\ }\textbf {\bibinfo {volume} {99}},\ \bibinfo {pages} {014514}
  (\bibinfo {year} {2019})},\ \Eprint {http://arxiv.org/abs/1805.02365}
  {arXiv:1805.02365 [hep-lat]} \BibitemShut {NoStop}%
\bibitem [{\citenamefont {Davoudi}\ \emph {et~al.}(2020)\citenamefont
  {Davoudi}, \citenamefont {Detmold}, \citenamefont {Orginos}, \citenamefont
  {Parre\~no}, \citenamefont {Savage}, \citenamefont {Shanahan},\ and\
  \citenamefont {Wagman}}]{Davoudi:2020ngi}%
  \BibitemOpen
  \bibfield  {author} {\bibinfo {author} {\bibfnamefont {Zohreh}\ \bibnamefont
  {Davoudi}}, \bibinfo {author} {\bibfnamefont {William}\ \bibnamefont
  {Detmold}}, \bibinfo {author} {\bibfnamefont {Kostas}\ \bibnamefont
  {Orginos}}, \bibinfo {author} {\bibfnamefont {Assumpta}\ \bibnamefont
  {Parre\~no}}, \bibinfo {author} {\bibfnamefont {Martin~J.}\ \bibnamefont
  {Savage}}, \bibinfo {author} {\bibfnamefont {Phiala}\ \bibnamefont
  {Shanahan}}, \ and\ \bibinfo {author} {\bibfnamefont {Michael~L.}\
  \bibnamefont {Wagman}},\ }\bibfield  {title} {\enquote {\bibinfo {title}
  {{Nuclear matrix elements from lattice QCD for electroweak and
  beyond-Standard-Model processes}},}\ }\href@noop {} {\  (\bibinfo {year}
  {2020})},\ \Eprint {http://arxiv.org/abs/2008.11160} {arXiv:2008.11160
  [hep-lat]} \BibitemShut {NoStop}%
\bibitem [{\citenamefont {Edwards}\ and\ \citenamefont
  {Joo}(2005)}]{Edwards:2004sx}%
  \BibitemOpen
  \bibfield  {author} {\bibinfo {author} {\bibfnamefont {Robert~G.}\
  \bibnamefont {Edwards}}\ and\ \bibinfo {author} {\bibfnamefont {Balint}\
  \bibnamefont {Joo}} (\bibinfo {collaboration} {SciDAC, LHPC, UKQCD}),\
  }\bibfield  {title} {\enquote {\bibinfo {title} {{The Chroma software system
  for lattice QCD}},}\ }\href {\doibase 10.1016/j.nuclphysbps.2004.11.254}
  {\bibfield  {journal} {\bibinfo  {journal} {Nucl. Phys. B Proc. Suppl.}\
  }\textbf {\bibinfo {volume} {140}},\ \bibinfo {pages} {832} (\bibinfo {year}
  {2005})},\ \Eprint {http://arxiv.org/abs/hep-lat/0409003}
  {arXiv:hep-lat/0409003} \BibitemShut {NoStop}%
\bibitem [{\citenamefont {H{\"o}rz}(2019)}]{contraction_optimizer}%
  \BibitemOpen
  \bibfield  {author} {\bibinfo {author} {\bibfnamefont {Ben}\ \bibnamefont
  {H{\"o}rz}},\ }\href@noop {} {\enquote {\bibinfo {title} {Contraction
  optimizer},}\ } (\bibinfo {year} {2019}),\ \bibinfo {note}
  {\url{https://github.com/laphnn/contraction_optimizer}}\BibitemShut {NoStop}%
\bibitem [{\citenamefont {Berkowitz}(2017)}]{Berkowitz:2017vcp}%
  \BibitemOpen
  \bibfield  {author} {\bibinfo {author} {\bibfnamefont {Evan}\ \bibnamefont
  {Berkowitz}},\ }\bibfield  {title} {\enquote {\bibinfo {title} {{METAQ:
  Bundle Supercomputing Tasks}},}\ }\href@noop {} {\  (\bibinfo {year}
  {2017})},\ \bibinfo {note} {\url{https://github.com/evanberkowitz/metaq}},\
  \Eprint {http://arxiv.org/abs/1702.06122} {arXiv:1702.06122
  [physics.comp-ph]} \BibitemShut {NoStop}%
%%CITATION = ARXIV:1702.06122;%%
\bibitem [{\citenamefont {Berkowitz}\ \emph {et~al.}(2018)\citenamefont
  {Berkowitz}, \citenamefont {Jansen}, \citenamefont {McElvain},\ and\
  \citenamefont {Walker-Loud}}]{Berkowitz:2017xna}%
  \BibitemOpen
  \bibfield  {author} {\bibinfo {author} {\bibfnamefont {Evan}\ \bibnamefont
  {Berkowitz}}, \bibinfo {author} {\bibfnamefont {Gustav~R.}\ \bibnamefont
  {Jansen}}, \bibinfo {author} {\bibfnamefont {Kenneth}\ \bibnamefont
  {McElvain}}, \ and\ \bibinfo {author} {\bibfnamefont {André}\ \bibnamefont
  {Walker-Loud}},\ }\bibfield  {title} {\enquote {\bibinfo {title} {{Job
  Management and Task Bundling}},}\ }\href {\doibase
  10.1051/epjconf/201817509007} {\bibfield  {journal} {\bibinfo  {journal} {EPJ
  Web Conf.}\ }\textbf {\bibinfo {volume} {175}},\ \bibinfo {pages} {09007}
  (\bibinfo {year} {2018})},\ \Eprint {http://arxiv.org/abs/1710.01986}
  {arXiv:1710.01986 [hep-lat]} \BibitemShut {NoStop}%
\bibitem [{\citenamefont {Lepage}(2020{\natexlab{a}})}]{lsqfit:11.5.1}%
  \BibitemOpen
  \bibfield  {author} {\bibinfo {author} {\bibfnamefont {Peter}\ \bibnamefont
  {Lepage}},\ }\href@noop {} {\enquote {\bibinfo {title} {gplepage/lsqfit:
  lsqfit version 11.5.1},}\ } (\bibinfo {year} {2020}{\natexlab{a}}),\ \bibinfo
  {note} {\url{https://github.com/gplepage/lsqfit}}\BibitemShut {NoStop}%
\bibitem [{\citenamefont {Lepage}(2020{\natexlab{b}})}]{gvar:11.2}%
  \BibitemOpen
  \bibfield  {author} {\bibinfo {author} {\bibfnamefont {Peter}\ \bibnamefont
  {Lepage}},\ }\href@noop {} {\enquote {\bibinfo {title} {gplepage/gvar: gvar
  version 11.2},}\ } (\bibinfo {year} {2020}{\natexlab{b}}),\ \bibinfo {note}
  {\url{https://github.com/gplepage/gvar}}\BibitemShut {NoStop}%
\bibitem [{\citenamefont {Hörz}\ and\ \citenamefont
  {Hanlon}(2019)}]{Horz:2019rrn}%
  \BibitemOpen
  \bibfield  {author} {\bibinfo {author} {\bibfnamefont {Ben}\ \bibnamefont
  {Hörz}}\ and\ \bibinfo {author} {\bibfnamefont {Andrew}\ \bibnamefont
  {Hanlon}},\ }\bibfield  {title} {\enquote {\bibinfo {title} {{Two- and
  three-pion finite-volume spectra at maximal isospin from lattice QCD}},}\
  }\href {\doibase 10.1103/PhysRevLett.123.142002} {\bibfield  {journal}
  {\bibinfo  {journal} {Phys. Rev. Lett.}\ }\textbf {\bibinfo {volume} {123}},\
  \bibinfo {pages} {142002} (\bibinfo {year} {2019})},\ \Eprint
  {http://arxiv.org/abs/1905.04277} {arXiv:1905.04277 [hep-lat]} \BibitemShut
  {NoStop}%
\end{thebibliography}%
